\documentclass[a4paper,11pt]{article}
\pdfoutput=1 

\usepackage{jcappub} 

\usepackage[T1]{fontenc} 
\usepackage{amsmath}
\usepackage{amsfonts}
\usepackage{amsthm}
\usepackage{amssymb}
\usepackage{latexsym, array,multirow,  verbatim, enumerate}
\usepackage{slashed}
\usepackage{booktabs}
\usepackage{tabularx}
\usepackage{cancel}
\usepackage{dcolumn}
\usepackage{bm}
\usepackage{graphicx, caption, subcaption} 
\usepackage{url}
\usepackage[compat=1.0.0]{tikz-feynman}
\usepackage{pifont}
\usepackage{empheq}
\usepackage[utf8]{inputenc}
\usepackage{bm}
\usepackage{blkarray}
\usepackage[font=footnotesize,labelfont=sf]{caption}
\usepackage{cleveref}
\usepackage{float}
\usepackage{multirow}
\usepackage{physics}
\graphicspath{{./Figures/}}
\usepackage{tabularx}
\usepackage{ulem}
\usepackage{tikz-feynman}

\title{Bootstrapping LSS perturbation theory beyond third order}

\author[a]{Arhum Ansari,}
\author[a]{Arka Banerjee,}
\author[a]{Sachin Jain,}
\author[b]{and Sahil Lalsodagar}

\affiliation[a]{Department of Physics, Indian Institute of Science Education and Research Pune, Pune, 411008, India}
\affiliation[b]{Department of Physical Sciences, Indian Institute of Science Education and Research Berhampur, Ganjam, Odisha, 760003, India}
\emailAdd{ansari.arhum@students.iiserpune.ac.in}
\emailAdd{arka@iiserpune.ac.in}
\emailAdd{sachin.jain@iiserpune.ac.in}
\emailAdd{sahild20@iiserbpr.ac.in}

\abstract{Bootstrap techniques relying on the constraints imposed by Extended Galilean Invariance (EGI), have proved to be very useful in the context of perturbation theory of the Large Scale Structure (LSS). It has been formulated in both the Eulerian as well as Lagrangian space. While the Eulerian bootstrap formalism has been successfully applied to both tracer and matter kernels, the application of bootstrap methods in Lagrangian space has so far been restricted to matter. Up to third order, it has been shown that implementing EGI constraints in Eulerian space fully reproduces the bias expansion for tracers. Previous studies have demonstrated that time non-locality affects the bias expansion in a non-trivial way starting from fifth order. Motivated by this fact, we extend the bootstrap approach upto fifth order in both Eulerian and Lagrangian space and demonstrate that it fully captures the time non-local effects. For this, we generalize the Lagrangian bootstrap for tracers, and found that it agrees with the corresponding results obtained in Eulerian space. One of the major challenges in implementing EGI constraints in Eulerian space, is to systematically find out all the "spurious poles" and make them vanish. We have proposed a method that bypasses this difficulty making the procedure tractable at higher orders. From Lagrangian perspective, we have identified coefficients in the tracer kernel whose ratios are independent of tracer properties and may serve as direct probes of the underlying cosmology.}

\begin{document}
\maketitle
\flushbottom

\section{Introduction}\label{sec:intro}
The distribution of matter at very large scales holds information about the initial perturbations seeded during inflation\,\cite{Dodelson:2003ft,Mukhanov:2005sc,Baumann:2022mni}. Studying the properties of correlation of matter overdensities in large scale structure, in principle, can give us a glimpse of pre-inflationary physics and even provide information about the initial state of the Universe. Therefore, understanding the distribution of matter is of utmost importance for the fundamental understanding of the Universe. This distribution is largely governed by the evolution and clustering of dark matter. 

Dark matter clustering on short distances is a highly non-linear process. Therefore, one often uses N-body simulations\,\cite{Angulo:2021kes,hockney2021computer,Springel:2006vs,Vogelsberger:2019ynw,Chisari:2011iq,Adamek:2013wja,Adamek:2016zes,Adamek:2017grt,Ali-Haimoud:2012fzp,Angulo:2016hjd,Bayer:2020tko,Chudaykin:2020hbf,Foreman:2015lca} to study the properties of dark matter evolution and clustering. These N-body simulations work very well down to the non-linear scales. On the other hand, as a theory model, dark matter is described very well as an effective fluid made of collisionless particles interacting only gravitationally\,\cite{Baumann:2010tm}. The framework employed to study such a fluid is known as the Effective Field Theory of Large Scale Structure (EFTofLSS)\,\cite{Carrasco:2012cv}. It is an improvement over Standard Perturbation Theory (SPT) and takes care of the contribution from short scale physics in a systematic manner through an effective stress tensor. The EFTofLSS has been investigated extensively and applied to describe dark matter clustering away from the linear scale where SPT breaks down. The EFTofLSS has been shown to work very well when compared with the N-body simulations. Owing to this robustness, it has been applied in the data analysis of various cosmological surveys\,\cite{Baumann:2010tm,Carrasco:2012cv,Porto:2013qua,Senatore:2014via,Carrasco:2013mua,Angulo:2014tfa,Baldauf:2014qfa,Konstandin:2019bay,Bertolini:2016bmt,Abolhasani:2015mra}.

We do not observe dark matter distribution directly. But there are generic tracers of dark matter, such as galaxies, which are easier to observe and whose distribution is highly correlated with the distribution of dark matter at large scales. Tracers, on the other hand are difficult to model. They do not evolve through an equation of motion and are described, in general, by a bias expansion in terms of dark matter fields. The bias expansion for tracers has been the object of study for many investigations\,\cite{McDonald:2009dh,Senatore:2014eva,Mirbabayi:2014zca,Donath:2023sav,Ansari:2024efj}. SPT plays a crucial role in determining the bias expansion for tracers and for counterterms in the EFT\,\cite{Abolhasani:2015mra,Ansari:2024efj}.

The SPT deals with the perturbative solution to collisionless Boltzmann equation i.e. Vlasov-Poisson equation.
The fluid equations, in particular, Euler and continuity equations are shown to be invariant under certain spacetime transformation\,\cite{Peloso:2013zw,Kehagias:2013yd}. These are termed as Extended Galilean Invariance (EGI) and are the reason behind consistency relations in LSS\,\cite{Peloso:2013zw,Kehagias:2013yd,Creminelli:2013mca,Creminelli:2013poa,Creminelli:2013nua,Horn:2014rta}. 

Due to EGI, the solution of SPT is highly constrained. Tracers inherit these symmetries through bias expansion. Since the restrictions imposed by EGI are very strict, one can try to bootstrap the bias expansion for tracers. In fact, it has been shown that one can actually bootstrap the tracer kernel up to third order, just by imposing constraints from EGI\,\cite{DAmico:2021rdb}. Imposing mass and momentum conservation further restricts the form of the kernel and gives the kernel for dark matter. 

In this study, our objective is to investigate whether one can use EGI constraints alone to bootstrap the kernel for biased tracers at fourth and fifth order. Recently\,\cite{Donath:2023sav}, it has been found that, at fifth order, time non-locality affects the basis expansion for tracers in a non trivial way. It has been shown that time local and non-local basis is qualitatively different at fifth order. Therefore, another motivation for our work is to see whether one can capture these time non-local effects\,\cite{Donath:2023sav} through the bootstrap approach. 

The paper is organized in the following manner. In Sec.\,\ref{sec:eulerian_boot}, we review the ideas of Eulerian bootstrap\,\cite{DAmico:2021rdb} and discuss the constraints imposed by EGI on the kernel for tracers as well as dark matter. We also use this section to set up notations and conventions used throughout the paper. Then in Sec.\,\ref{sec:Eulerian_boot}, we implement the Eulerian bootstrap to get the kernel for biased tracers at fourth and fifth order given in Sec.\,\ref{sec:ker_4_boot} and Sec.\,\ref{sec:fifth_order} respectively. We close this section with some comments on time non-locality, which is particularly relevant at the fifth order. In Sec.\,\ref{sec:lag_boot}, we discuss the bootstrap procedure for tracers in Lagrangian space. This is a generalisation of the Lagrangian bootstrap for matter proposed recently\,\cite{Marinucci:2024add}. In this section, we also discuss the limitation of EGI constraints as implemented in the Eulerian bootstrap procedure and provide a resolution to remedy that. We conclude the discussion with some remarks and future directions in Sec.\,\ref{sec:conc}.
\section{Eulerian Bootstrap and consistency with EGI}\label{sec:eulerian_boot}
In this section, we review the method of Eulerian bootstrap developed in\,\cite{DAmico:2021rdb}. At the same time, we outline a diagrammatic approach to writing kernels and implementing constraints which proves to be useful beyond third order. The idea of Eulerian bootstrap emerges from the symmetry properties of the fluid equations of motion in standard perturbation theory (SPT). It has been shown that dark matter behaves as an effective fluid at large scales\,\cite{Baumann:2010tm,Carrasco:2012cv}. The fluid equations governing the large scale dynamics of dark matter are invariant under the coordinate transformation\,\cite{Peloso:2013zw,Kehagias:2013yd},
\begin{align}\label{eq:EGI_trans}
\eta\rightarrow\tilde{\eta}=\eta\hspace{0.2cm},\hspace{0.2cm}\mathbf{x}\rightarrow \tilde{\mathbf{x}}=\mathbf{x}+\mathbf{d}(\eta)\;,
\end{align}
 where $\eta$ is the conformal time and $\mathbf{x}$ is the comoving distance. Here $\mathbf{d}(\eta)$ can be any arbitrary function of the time variable. Under the transformation given in Eq.\eqref{eq:EGI_trans}, the fluid equations which are given as,
 \begin{gather}
     \delta^{\prime}(\mathbf{x},\eta)+\mathbf{\nabla}\cdot[(1+\delta(\mathbf{x},\eta))\mathbf{v}(\mathbf{x},\eta)]=0 \;,\nonumber\\[5pt]
     \mathbf{v}^{\prime}(\mathbf{x},\eta)+\mathcal{H}(\eta)\mathbf{v}(\mathbf{x},\eta)+\mathbf{v}(\mathbf{x},\eta)\cdot\nabla\mathbf{v}(\mathbf{x},\eta)=-\mathbf{\nabla}\Phi(\mathbf{x},\eta)\;,\nonumber\\[5pt]
     \nabla^2 \Phi(\mathbf{x},\eta) = \frac{3}{2}\Omega_{m}(\eta)\mathcal{H}^2(\eta) \delta(\mathbf{x},\eta)\;,
 \end{gather}
remain invariant, provided that the full non-linear overdensity field $\delta(\mathbf{x},\eta)$, velocity field $\mathbf{v}(\mathbf{x},\eta)$ and the gravitational potential transform as,
 \begin{align}\label{eq:del_v_phi_trans_EGI}
 \delta(\mathbf{x},\eta)\rightarrow\tilde{\delta}(\tilde{\mathbf{x}},\tilde{\eta})&=\delta(\mathbf{x},\eta)\;,\nonumber\\    
 \mathbf{v}(\mathbf{x},\eta)\rightarrow\tilde{\mathbf{v}}(\tilde{\mathbf{x}},\tilde{\eta})&= \mathbf{v}(\mathbf{x},\eta)+\mathcal{H}\partial_\eta\mathbf{d}(\eta)\;,\nonumber\\
 \Phi\rightarrow\tilde{\Phi}=\Phi-[\mathcal{H}\partial_\eta&(\mathcal{H}\mathbf{d})+\mathcal{H}^2\partial_\eta\mathbf{d}]\cdot\mathbf{x}\;.
 \end{align}
This invariance is termed as "Extended Galilean invariance", henceforth denoted as EGI. Here $d(\eta)$, is treated as a non-linear field, a property that is evident from Eq.\eqref{eq:del_v_phi_trans_EGI}, where $\partial_\eta d$ is related to the non-linear velocity field. Therefore, $d(\eta)$ is assumed to admit a power series expansion as follows\,\cite{DAmico:2021rdb},
\begin{align}\label{eq:d_exp}
d(\eta)=\sum_{n=1}^{\infty}d^{(n)}(\eta)\;.
\end{align}
Each term in Eq.\eqref{eq:d_exp} leads to different constraints on matter overdensity as we will see in Sec.\,\ref{ssec:EGI_const}.
In SPT, both $\delta$ and $\theta\equiv\nabla\cdot\mathbf{v}$ admit a power series solution in terms of the linear overdensity as follows,
\begin{align}\label{eq:series_exp_del_v}
\delta(\mathbf{x},\eta)=\sum_{n=1}^{\infty}\delta^{(n)}(\mathbf{x},\eta)\;,\hspace{1cm}\theta(\mathbf{x},\eta)=\sum_{n=1}^{\infty}\theta^{(n)}(\mathbf{x},\eta)\;,
 \end{align}
where $\delta^{(n)}$ and $\theta^{(n)}$ are composed of $n$ linear overdensities $\delta^{(1)}$.

Tracers of dark matter such as galaxies or halos do not have an equation of motion. But it is assumed that the tracer overdensity, denoted by $\delta_t$, is a function of tidal forces and velocity gradient generated by dark matter fields. Hence, $\delta_t$ also admits a power series expansion in linear overdensity fields as,
\begin{align}\label{eq:exp_tracer}
\delta_t(\mathbf{x},\eta)=\sum_{n=1}^\infty\delta_t^{(n)}(\mathbf{x},\eta)\;,
\end{align}
where  $\delta_t^{(n)}$ are composed of $n$ linear overdensities $\delta^{(1)}$. 

In SPT, each term in the perturbative solution given in Eq.\eqref{eq:series_exp_del_v} is solved order by order in momentum space. This gives us the $n^{th}$ order fields in momentum space as a convolution over linear fields with a well defined kernel at each order as\,\cite{DAmico:2021rdb},
\begin{align}\label{eq:ker_def}
&\delta^{(n)}_{\mathbf{k}}(\eta)=\mathcal{I}_{\mathbf{k};\mathbf{q}_1\mathbf{q}_2..\mathbf{q}_n}F_n(\mathbf{q}_1,\mathbf{q}_2,..\mathbf{q}_n,\eta)\delta^{(1)}_{\mathbf{q}_1}(\eta)\delta^{(1)}_{\mathbf{q}_2}(\eta)...\delta^{(1)}_{\mathbf{q}_3}(\eta)\;,\nonumber\\
&\theta^{(n)}_{\mathbf{k}}(\eta)=\mathcal{I}_{\mathbf{k};\mathbf{q}_1\mathbf{q}_2..\mathbf{q}_n}G_n(\mathbf{q}_1,\mathbf{q}_2,..\mathbf{q}_n,\eta)\delta^{(1)}_{\mathbf{q}_1}(\eta)\delta^{(1)}_{\mathbf{q}_2}(\eta)...\delta^{(1)}_{\mathbf{q}_3}(\eta)\;,\nonumber\\
&\delta^{(n)}_{t,\mathbf{k}}(\eta)=\mathcal{I}_{\mathbf{k};\mathbf{q}_1\mathbf{q}_2..\mathbf{q}_n}K_n(\mathbf{q}_1,\mathbf{q}_2,..\mathbf{q}_n,\eta)\delta^{(1)}_{\mathbf{q}_1}(\eta)\delta^{(1)}_{\mathbf{q}_2}(\eta)...\delta^{(1)}_{\mathbf{q}_3}(\eta)\;,
\end{align}
where
\begin{align}\label{eq:Ikq1q2}  \mathcal{I}_{\mathbf{k};\mathbf{q}_1\mathbf{q}_2..\mathbf{q}_n}=\frac{1}{n!}\int\frac{d^3\mathbf{q}_1}{(2\pi)^3}\frac{d^3\mathbf{q}_2}{(2\pi)^3}..\frac{d^3\mathbf{q}_n}{(2\pi)^3}\delta_{D}(\mathbf{k}-\sum_i^n\mathbf{q}_i)\;.
\end{align}
The kernel for dark matter is denoted by $F_n(G_n)$ while the tracer kernel is denoted by $K_n$. The kernel $K_n$ for tracers and it's properties have been investigated extensively in the literature, see \,\cite{DAmico:2022osl,Donath:2023sav} and references therein. This is the main object of study for this paper as well. EGI imposes certain constraints on the form of the kernels given in Eq.\eqref{eq:ker_def}. In the next subsection, we are going to discuss these constraints more explicitly.
 \subsection{Constraints from EGI}\label{ssec:EGI_const}
In this section, we state the constraints imposed by EGI on the tracer and matter kernels denoted by $K_n$ and $F_n$ in Eq.\eqref{eq:ker_def}. We also discuss the EGI constraints coming from "spurious poles" and describe the difficulty in imposing them. Then we discuss the mass and momentum constraints which are to be satisfied by matter kernel but not necessarily by tracers. 

Let us start with describing the constraints coming from EGI. The transformation shown in Eq.\eqref{eq:EGI_trans}, in momentum space implies,
\begin{align}\label{eq:delta_mom_space_trans}
    \delta_{t,\mathbf{k}}\rightarrow e^{i \mathbf{k}\cdot\mathbf{d}}\delta_{t,\mathbf{k}}=\sum_{m=0}^\infty\frac{(i \mathbf{k}\cdot\mathbf{d})^m}{m!}\delta_{t,\mathbf{k}}\;.
\end{align}
Eq.\eqref{eq:delta_mom_space_trans} reveals that the kernels as given in Eq.\eqref{eq:ker_def} should take a particular form. This fact is not apparent from Eq.\eqref{eq:ker_def} and can only be seen by looking at how the kernel transforms as shown in Eq.\eqref{eq:delta_mom_space_trans}. The field $d(\eta)$ is non linear and given in Eq.\eqref{eq:d_exp}. The contribution proportional to $d^{(1)}(\eta)$ is termed as "leading order" while higher order contributions are termed as "next-to-leading order". Below we are going to state the constraints coming from leading and next-to-leading order. For details of the derivation of the constraints, see\,\cite{DAmico:2021rdb}.
\vspace{5pt}\\
\underline{\textbf{Leading Order (LO)}}
\vspace{5pt}\\
The leading order (LO) constraints restricts the structure of the kernel when a subset of the momenta are taken to be soft i.e. $\mathbf{q}_i\rightarrow 0$. The LO constraints state that,
\begin{align}\label{eq:Lo_const}
\lim_{\mathbf{q}_1,\dots,\mathbf{q}_m\rightarrow 0} K_n \left(\mathbf{q}_1, \ldots, \mathbf{q}_m, \mathbf{q}_{m+1}, \ldots, \mathbf{q}_n \right) &= \frac{\mathbf{q}_1\cdot\mathbf{Q}_{n,m}}{\mathbf{q}_1^2} \dots \frac{\mathbf{q}_m\cdot\mathbf{Q}_{n,m}}{\mathbf{q}_m^2} K_{n-m}\left(\mathbf{q}_{m+1}, \ldots, \mathbf{q}_n \right)\nonumber\\
     &+ \mathcal{O}\left(\left(1/q\right)^{m-1}\right)\;,
\end{align}
where $\mathbf{Q}_{n,m} \equiv \sum_{i=m+1}^n \mathbf{q}_i$. We can see from Eq.\eqref{eq:Lo_const}, that the leading pole of the kernel $K_n$ is fixed in terms of the lower order kernels when one or more momenta goes soft. The pole comes with a universal soft factor times a lower order kernel. This is similar to soft theorems in gauge theories.

One can work with the algebraic relation given in Eq.\eqref{eq:Lo_const}. But proceeding to higher order, we find that a diagrammatic representation is more intuitive and beneficial. Therefore, let us represent $K_n$ with a diagram as,
\begin{equation}\label{eq:K_n_diag}
    K_n(\mathbf{q}_1 \dots \mathbf{q}_n, \eta) \equiv \begin{tikzpicture}[scale=1.0, 
            baseline={([xshift=-5cm,yshift=-\the\dimexpr\fontdimen22\textfont2\relax]
                current bounding box.center)}]
\begin{feynman}
    \vertex (b) ;
    \vertex [above left=of b] (i1) {$\mathbf{q}_1$};
    \vertex (i2) at (-1.8,+0.5){$\mathbf{q}_{2}$};
    \vertex [below left=1 and 1 of b] (in) {${\mathbf{q}_n}$};
    \vertex [right=of b] (o1) {};

    \draw[thick] (b) -- node [below, xshift=10pt] {$\mathbf{Q}_{n,0}$} (o1);
    \draw[thick] (i1) -- node [left=4pt] {} (b);
    \draw[thick] (i2) -- node [right=4pt] {} (b);
    \draw[thick] (in) -- node [right=4pt] {} (b);

    \draw[black, thick, fill=gray] (b) circle (0.5cm);

    \node[fill=black,inner sep=0.75pt,circle] at ($(i2)!0.25!(in)$) {};
    \node[fill=black,inner sep=0.75pt,circle] at ($(i2)!0.5!(in)$) {};
    \node[fill=black,inner sep=0.75pt,circle] at ($(i2)!0.75!(in)$) {};
    
\end{feynman}
\end{tikzpicture}\;,
\end{equation}
then we can express Eq.\eqref{eq:Lo_const} in a diagrammatic manner as follows,
\begin{equation}\label{eq:LO_diag_here}
\lim_{\mathbf{q}_1,\dots,\mathbf{q}_m \rightarrow 0}
\begin{tikzpicture}[scale=1.0, 
            baseline={([xshift=-5cm,yshift=-\the\dimexpr\fontdimen22\textfont2\relax]
                current bounding box.center)}]
\begin{feynman}
    \vertex (b) ;
    \vertex [above left=of b] (i1) {$\mathbf{q}_1$};
    \vertex (i2) at (-1.8,+0.5){$\mathbf{q}_{2}$};
    \vertex [below left=1 and 1 of b] (in) {${\mathbf{q}_n}$};
    \vertex [right=of b] (o1) {};

    \draw[thick] (b) -- node [below, xshift=10pt] {$\mathbf{Q}_{n,0}$} (o1);
    \draw[thick] (i1) -- node [left=4pt] {} (b);
    \draw[thick] (i2) -- node [right=4pt] {} (b);
    \draw[thick] (in) -- node [right=4pt] {} (b);

    \draw[black, thick, fill=gray] (b) circle (0.5cm);

    \node[fill=black,inner sep=0.75pt,circle] at ($(i2)!0.25!(in)$) {};
    \node[fill=black,inner sep=0.75pt,circle] at ($(i2)!0.5!(in)$) {};
    \node[fill=black,inner sep=0.75pt,circle] at ($(i2)!0.75!(in)$) {};
    
\end{feynman}
\end{tikzpicture}
\rightarrow \frac{\mathbf{q}_1 \cdot \mathbf{Q}_{n,m}}{\mathbf{q}_1^2} \dots \frac{\mathbf{q}_m \cdot \mathbf{Q}_{n,m}}{\mathbf{q}_m^2}\;\;
\begin{tikzpicture}[scale=1.0, 
            baseline={([xshift=-5cm,yshift=-\the\dimexpr\fontdimen22\textfont2\relax]
                current bounding box.center)}]
\begin{feynman}
    \vertex (b) ;
    \vertex [above left=of b] (i1) {$\mathbf{q}_{m+1}$};
    \vertex (i2) at (-1.8,+0.5){$\mathbf{q}_{m+2}$};
    \vertex [below left=1 and 1 of b] (in) {${\mathbf{q}_n}$};
    \vertex [right=of b] (o1) {};

    \draw[thick] (b) -- node [below, xshift=10pt] {$\mathbf{Q}_{n,m}$} (o1);
    \draw[thick] (i1) -- node [left=4pt] {} (b);
    \draw[thick] (i2) -- node [right=4pt] {} (b);
    \draw[thick] (in) -- node [right=4pt] {} (b);

    \draw[black, thick, fill=gray] (b) circle (0.5cm);

    \node[fill=black,inner sep=0.75pt,circle] at ($(i2)!0.25!(in)$) {};
    \node[fill=black,inner sep=0.75pt,circle] at ($(i2)!0.5!(in)$) {};
    \node[fill=black,inner sep=0.75pt,circle] at ($(i2)!0.75!(in)$) {};    
\end{feynman}
\end{tikzpicture}\;,
\end{equation}
where $\mathbf{Q}_{i,j}$ denotes the non-linear external leg and $\mathbf{q}_i$ denotes the linear fields.
\vspace{5pt}\\
\underline{\textbf{Next-to-Leading Order (NLO)}}
\vspace{5pt}\\
The next-to-leading order constraints arises when sum of external momenta are taken to be soft. When sum of $m$ external momenta are taken to be soft, the constraints state that the kernel factorises as follows\,\cite{DAmico:2021rdb},
\begin{align}\label{eq:NLO_all_order}
\lim_{\mathbf{Q}_{m,0}\rightarrow 0} K_n \left(\mathbf{q}_1, \ldots, \mathbf{q}_m, \mathbf{q}_{m+1}, \ldots, \mathbf{q}_n \right) &\supset \frac{\mathbf{Q}_{m,0}\cdot\mathbf{Q}_{n,m}}{\mathbf{Q}_{m,0}^2} K_{n-m}\left(\mathbf{q}_{m+1}, \ldots, \mathbf{q}_n ; \eta \right)\nonumber\\
&\times\int d\eta^{\prime}f_{+}(\eta^{\prime})\left(\frac{D_{+}(\eta^{\prime})}{D_{+}\eta}\right)^{m}G_{m}\left(\mathbf{q}_{1}, \ldots, \mathbf{q}_m ; \eta^{\prime} \right)\;.
\end{align}
As for the case of LO, we have a nice diagrammatic representation for NLO constraints. Diagrammatically, Eq.\eqref{eq:NLO_all_order} can be represented as,
\begin{equation}\label{eq:NLO_diag_here}
\begin{aligned}
\lim_{\mathbf{Q}_{m,0} \rightarrow 0}
\begin{tikzpicture}[scale=0.6, 
            baseline={([xshift=-5cm,yshift=-\the\dimexpr\fontdimen22\textfont2\relax]
                current bounding box.center)}]
\begin{feynman}
    \vertex (b) ;
    \vertex (i1) at (-1.2,+1.5) {$\mathbf{q}_1$};
    \vertex (i2) at (-1.8,+0.5){$\mathbf{q}_{2}$};
    \vertex (i3) at (-1.2,-1.5) {${\mathbf{q}_m}$};
    \vertex (o1) at (2.5,0){};
    \vertex (i4) at (4.5,0.5){$\mathbf{q}_{m+1}$};
    \vertex (i5) at (3.6,-1.5){$\mathbf{q}_n$};
    \vertex (c) at (2.5,1.8){};

    \draw[thick] (b) -- node [below, xshift=5pt] {$\mathbf{Q}_{m,0}$} (o1);
    \draw[thick] (i1) -- node [left=4pt] {} (b);
    \draw[thick] (i2) -- node [right=4pt] {} (b);
    \draw[thick] (i3) -- node [right=4pt] {} (b);
    \draw[thick] (i4) -- node [right=4pt] {} (o1);
    \draw[thick] (i5) -- node [right=4pt] {} (o1);
    \draw[thick] (c) -- node [above, xshift=15pt,yshift=5pt] {$\mathbf{Q}_{n,0}$} (o1);

    \draw[black, thick, fill=gray] (b) circle (0.5cm);
    \draw[black, thick, fill=gray] (o1) circle (0.5cm);

    \node[fill=black,inner sep=0.75pt,circle] at ($(i2)!0.25!(i3)$) {};
    \node[fill=black,inner sep=0.75pt,circle] at ($(i2)!0.5!(i3)$) {};
    \node[fill=black,inner sep=0.75pt,circle] at ($(i2)!0.75!(i3)$) {};

    \node[fill=black,inner sep=0.75pt,circle] at ($(i4)!0.25!(i5)$) {};
    \node[fill=black,inner sep=0.75pt,circle] at ($(i4)!0.5!(i5)$) {};
    \node[fill=black,inner sep=0.75pt,circle] at ($(i4)!0.75!(i5)$) {};
\end{feynman}
\end{tikzpicture}
\rightarrow \; \frac{\mathbf{Q}_{m,0}\cdot\mathbf{Q}_{n,m}}{\mathbf{Q}_{m,0}^2}\;\left\{
\begin{tikzpicture}[scale=0.6, 
            baseline={([xshift=-5cm,yshift=-\the\dimexpr\fontdimen22\textfont2\relax]
                current bounding box.center)}]
\begin{feynman}
    \vertex (b) ;
    \vertex (i1) at (-1.2,+1.5) {$\mathbf{q}_{1}$};
    \vertex (i2) at (-1.8,+0.5){$\mathbf{q}_{2}$};
    \vertex (in) at (-1.2,-1.5) {${\mathbf{q}_m}$};
    \vertex (o1) at (1.5,0){};

    \draw[thick] (b) -- node [below, xshift=15pt] {$\mathbf{Q}_{m,0}$} (o1);
    \draw[thick] (i1) -- node [left=4pt] {} (b);
    \draw[thick] (i2) -- node [right=4pt] {} (b);
    \draw[thick] (in) -- node [right=4pt] {} (b);

    \draw[black, thick, fill=gray] (b) circle (0.5cm);

    \node[fill=black,inner sep=0.75pt,circle] at ($(i2)!0.25!(in)$) {};
    \node[fill=black,inner sep=0.75pt,circle] at ($(i2)!0.5!(in)$) {};
    \node[fill=black,inner sep=0.75pt,circle] at ($(i2)!0.75!(in)$) {};
    
\end{feynman}
\end{tikzpicture}
\right\}\left\{
\begin{tikzpicture}[scale=0.6, 
            baseline={([xshift=-5cm,yshift=-\the\dimexpr\fontdimen22\textfont2\relax]
                current bounding box.center)}]
\begin{feynman}
    \vertex (b) ;
    \vertex (i1) at (1.2,1.5) {$\mathbf{q}_{m+1}$};
    \vertex (i2) at (1.8,+0.5){$\mathbf{q}_{m+2}$};
    \vertex (in) at (1.2,-1.5){${\mathbf{q}_n}$};
    \vertex (o1) at (-1.5,0){};

    \draw[thick] (b) -- node [below, xshift=-10pt] {$\mathbf{Q}_{n,m}$} (o1);
    \draw[thick] (i1) -- node [left=4pt] {} (b);
    \draw[thick] (i2) -- node [right=4pt] {} (b);
    \draw[thick] (in) -- node [right=4pt] {} (b);

    \draw[black, thick, fill=gray] (b) circle (0.5cm);

    \node[fill=black,inner sep=0.75pt,circle] at ($(i2)!0.25!(in)$) {};
    \node[fill=black,inner sep=0.75pt,circle] at ($(i2)!0.5!(in)$) {};
    \node[fill=black,inner sep=0.75pt,circle] at ($(i2)!0.75!(in)$) {};
    
\end{feynman}
\end{tikzpicture}
\right\}\;,
\end{aligned}
\end{equation}
where the notations have been given below Eq.\eqref{eq:LO_diag_here}. One can look at NLO constraints as the kinematic limit when different combination of sums of momenta goes soft in contrast with individual momenta going soft which gives LO constraints.
\vspace{5pt}\\
\underline{\textbf{Constraints from "spurious poles"}}
\vspace{5pt}\\
Apart from the usual constraints as given in Eq.\eqref{eq:Lo_const} and \eqref{eq:NLO_all_order}, we also need to impose other constraints which we describe now. When we take the limit as given in Eq.\eqref{eq:NLO_diag_here}, we also get certain momentum structures which are not present on the RHS of Eq.\eqref{eq:NLO_diag_here} but which involve $\mathbf{Q}_{m,0}\rightarrow 0$. For example, in the limit $\mathbf{Q}_{m,0}\rightarrow 0$, terms proportional to
\begin{align}\label{eq:wrong_pol}  \frac{\mathbf{Q}_{m,0}\cdot\mathbf{Q}_{n,m}}{\mathbf{Q}_{n,m}^2}\hspace{0.2cm},\hspace{0.2cm}\frac{(\mathbf{Q}_{m,0}\cdot\mathbf{Q}_{n,m})^2}{\mathbf{Q}_{m,0}^2\mathbf{Q}_{n,m}^2}\;,
\end{align}
are not allowed and should not be present in the kernel. We call these structures ``spurious poles"\footnote{This is just terminology. The structures given in Eq.\eqref{eq:wrong_pol} are not actually poles in the limit $\mathbf{Q}_{m,0}\rightarrow 0$.}. Imposing that the "spurious poles" are absent from the kernel also gives further constraints\,\cite{DAmico:2021rdb}. We will say more about this in Sec.\,\ref{sec:no_wrong_poles}, where we propose a method of constructing the kernels in a manner that these ``spurious poles" are absent by construction. However, for the next few sections, we impose the absence of "spurious pole" as part of the NLO constraints.

Eq.\eqref{eq:Lo_const} and  Eq.\eqref{eq:NLO_all_order} along with imposing the absence of "spurious poles" as given in Eq.\eqref{eq:wrong_pol}, together constitute the leading order (LO) and next-to-leading order (NLO) constraints. These constraints severely restrict the form of perturbative kernels at each order to the extent that one can bootstrap the kernels. 
\vspace{5pt}\\
\underline{\textbf{Matter Kernel: Mass and Momentum constraints}}
\vspace{5pt}\\
In general, tracer overdensity is not a conserved quantity since the number density for tracers does not remain constant with time. However, dark matter overdensity and the total linear momentum of the center of mass is conserved throughout their evolution. Therefore, apart from LO and NLO as given in Eq.\eqref{eq:Lo_const} and \eqref{eq:NLO_all_order}, the kernel for matter satisfies some extra condition. These are called "mass and momentum conservation" and are respectively given as\,\cite{DAmico:2021rdb},
\begin{align}\label{eq:mm_pos}
    \int d^3x\; \delta(\mathbf{x,\eta})=0\;,\quad\quad
     \int d^3x\;\mathbf{x}^i\; \delta(\mathbf{x,\eta})=0\;,
\end{align}
which in momentum space implies,
\begin{align}\label{eq:mass_mom_conserv}
\lim_{\mathbf{Q}_{n,0}\rightarrow 0} F_n(\mathbf{q}_1,...\mathbf{q}_n;\eta)=0\;,\quad
\lim_{\mathbf{Q}_{n,0}\rightarrow 0} \frac{\partial}{\partial\mathbf{q}_1^i}F_n(\mathbf{q}_1,...\mathbf{q}_n;\eta)=0\;.
\end{align}
One should impose Eq.\eqref{eq:mass_mom_conserv} further on $K_m$ to get $F_m$ ($G_m$) and then use the resulting kernel on the R.H.S of Eq.\eqref{eq:NLO_all_order} to put constraints on $K_{n>m}$.  In the next section we will discuss the construction of kernels up to third order and see how these constraints provide a bootstrapping procedure for the kernels.

 \subsection{Kernel for tracers upto third order}\label{ssec:ker_3_const}
In this section, we review the construction of kernels for second and third order and impose EGI constraints to get the tracer kernel up to third order. This was first done in\,\cite{DAmico:2021rdb} and we will be reproducing their results, although working with a different basis. The main purpose of this section is to develop a diagrammatic approach to constructing the kernel as opposed to the analytic method given in\,\cite{DAmico:2021rdb}. We find the diagrammatic approach more convenient when generalizing to higher orders.

Tracer overdensity are scalars and rotationally invariant quantities. Therefore, the most general kernel as given in Eq.\eqref{eq:ker_def} should be written in terms of the most general dimensionless $SO(3)$ scalar constructed out of the momenta. At first order, there is just one momentum and the only dimensionless $SO(3)$ scalar is 1. Therefore, tracer kernel at first order is given as,
\begin{align}\label{eq:K1}
    K_1(\mathbf{q}_1)=a_1\;.
\end{align}
At second order, we have two momenta say $\mathbf{q}_1$ and $\mathbf{q}_2$. The most general dimensionless, rotationally invariant structures that you can build with two momenta is given as\,\cite{DAmico:2021rdb},
\begin{equation}\label{eq:gen_struc_2nd}
    1, \;\;\; \frac{\mathbf{q_1} \cdot \mathbf{q}_2}{\mathbf{q}_1^2}, \;\;\; \frac{\mathbf{q}_1 \cdot \mathbf{q}_2}{\mathbf{q}_2^2}, \;\;\; \frac{(\mathbf{q}_1 \cdot \mathbf{q}_2)^{2}}{ \mathbf{q}_1^2 \mathbf{q}_2^2}\;.
\end{equation}
This can be re-written in another basis as follows,
\begin{align}\label{eq:alter_basis}
    &\quad\quad1\;,
    &\alpha_{+}(\mathbf{q}_1,\mathbf{q}_2) \equiv \frac{1}{2}\left( \frac{\mathbf{q}_1 \cdot \mathbf{q}_2}{\mathbf{q}_1^2} + \frac{\mathbf{q}_1 \cdot \mathbf{q}_2}{\mathbf{q}_2^2}\right)\;,\nonumber\\
    &\beta(\mathbf{q}_1,\mathbf{q}_2) \equiv \frac{(\mathbf{q}_1 \cdot \mathbf{q}_2)^{2}}{ \mathbf{q}_1^2 \mathbf{q}_2^2}\;,
    &\alpha_{-}(\mathbf{q}_1,\mathbf{q}_2) \equiv\frac{1}{2} \left( \frac{\mathbf{q}_1 \cdot \mathbf{q}_2}{\mathbf{q}_1^2} - \frac{\mathbf{q}_1 \cdot \mathbf{q}_2}{\mathbf{q}_2^2}\right)\;.
\end{align}
Therefore, the most general second order kernel has four structures to begin with and is given as,
\begin{align}\label{eq:S2}
     S_2\left( \mathbf{q}_{1}, \mathbf{q}_2 \right) &= \{1, \; \alpha_{+}\left( \mathbf{q}_{1}, \mathbf{q}_2\right), \; \beta(\mathbf{q}_{1},\mathbf{q}_2), \; \alpha_{-}\left( \mathbf{q}_{1}, \mathbf{q}_2 \right)\}\;.
\end{align}
This is what constitutes the kernel for biased tracers at second order. Hence, we can write $K_2$ as,
\begin{align}\label{eq:K2}
     K_2\left( \mathbf{q}_{1}, \mathbf{q}_2 \right)=\sum_{i=1}^{l(S_2)}b_i S_2(i)\;,
\end{align}
where $b_i$ are arbitrary coefficients and $l(X)$ denotes the cardinal number i.e. number of elements in the set $X$. For example here, $l(S_2)=4$ which can be seen from Eq.\eqref{eq:S2}. After imposing symmetrisation and EGI constraints, we get the kernel for biased tracers at second order.

To construct the third order kernel, we can take the set given in Eq.\eqref{eq:S2} and define\,\cite{DAmico:2021rdb},
\begin{align}\label{eq:K3}
K_3(\mathbf{q}_1,\mathbf{q}_2,\mathbf{q}_3)= \left(\sum_i^{l(S_2\otimes S_2)}c_i[S_2(\mathbf{q}_1,\mathbf{q}_2) \otimes S_2(\mathbf{q}_{12},\mathbf{q}_3)](i)\right ) + \text{cyclic permutations}\;,
\end{align}
where $\otimes$ denotes outer product operation and $c_i$ are independent coefficients multiplying different momentum structure. Now we will give a diagrammatic representation of Eq.\eqref{eq:K3} which will be very helpful in construction of higher order kernels.

Note that upto third order, one may not need the diagrammatic approach to construct the kernels as done in\,\cite{DAmico:2021rdb}. However, at higher order, we find the diagrammatic approach to be extremely useful. Let us explain how it works.

We start by defining the basic building block $S_2$ as given in Eq.\eqref{eq:S2} in a diagrammatic way as follows,

\begin{equation}\label{eq:K2_diag}
S_2(\mathbf{q}_1,\mathbf{q}_2)\equiv
\begin{tikzpicture}[scale=1.0, 
            baseline={([xshift=-5cm,yshift=-\the\dimexpr\fontdimen22\textfont2\relax]
                current bounding box.center)},
    ] 
\begin{feynman}
\vertex (b) ;
\vertex [above left=of b] (i1) {$\mathbf{q}_1$};
\vertex [right=of b] (o1) {};
\vertex [below left=1 and 1 of b] (i2) {${\mathbf{q}_2}$};

\draw[thick, postaction={decorate}, decoration={markings, mark=at position 0.5 with {\arrow{>}}}] 
    (b) -- node [below=8pt] {$\mathbf{q}_{12}$} (o1);
\draw[thick, postaction={decorate}, decoration={markings, mark=at position 0.5 with {\arrow{>}}}] 
    (i1) -- (b);
\draw[thick, postaction={decorate}, decoration={markings, mark=at position 0.5 with {\arrow{>}}}] 
    (i2) -- (b);
\end{feynman}

\end{tikzpicture}\;,
\end{equation}
where $\mathbf{q}_{ij}\equiv \mathbf{q}_i+\mathbf{q}_j$ and we have denoted the linear fields with incoming arrows and the non-linear fields with outgoing arrow. Note that the diagram given in Eq.\eqref{eq:K2_diag} represents a set of momentum structure which are given in Eq.\eqref{eq:S2}. Using Eq.\eqref{eq:K2_diag} as building blocks, we can construct higher order kernels. For example, we can join two such vertex as given in Eq.\eqref{eq:K2_diag} and write the higher order vertex as,
\begin{equation}
   \begin{tikzpicture}[scale=0.5, 
    baseline={([xshift=-5cm,yshift=-\the\dimexpr\fontdimen22\textfont2\relax]
        current bounding box.center)},
    every node/.style={font=\small}
] 
\begin{feynman}
\vertex (b) ;
\vertex [above left=of b] (i1) {$\mathbf{q}_1$};
\vertex [right=of b] (o1) {};
\vertex [below left=1 and 1 of b] (i2) {${\mathbf{q}_2}$};

\draw[thick, postaction={decorate}, decoration={markings, mark=at position 0.5 with {\arrow{>}}}] 
    (b) -- node [below=8pt] {$\mathbf{q}_{12}$} (o1); 
\draw[thick, postaction={decorate}, decoration={markings, mark=at position 0.5 with {\arrow{>}}}] 
    (i1) -- (b);
\draw[thick, postaction={decorate}, decoration={markings, mark=at position 0.5 with {\arrow{>}}}] 
    (i2) -- (b);
\end{feynman}

\end{tikzpicture}
\otimes
\begin{tikzpicture}[scale=0.5, 
            baseline={([xshift=-5cm,yshift=-\the\dimexpr\fontdimen22\textfont2\relax]
                current bounding box.center)},
    ] 
\begin{feynman}
\vertex (b) ;
\vertex [above right=of b] (i1) {$\mathbf{q}_{123}$};
\vertex [left=of b] (o1) {};
\vertex [below right=1 and 1 of b] (i2) {${\mathbf{q}_3}$};

\draw[thick, postaction={decorate}, decoration={markings, mark=at position 0.5 with {\arrow{<}}}] 
    (b) -- node [below=8pt] {$\mathbf{q}_{12}$} (o1);
    
\draw[thick, postaction={decorate}, decoration={markings, mark=at position 0.5 with {\arrow{<}}}] 
    (i1) -- (b);
\draw[thick, postaction={decorate}, decoration={markings, mark=at position 0.5 with {\arrow{>}}}] 
    (i2) -- (b);
\end{feynman}

\end{tikzpicture}
=
\begin{tikzpicture}[scale=0.5, 
            baseline={([xshift=-5cm,yshift=-\the\dimexpr\fontdimen22\textfont2\relax]
                current bounding box.center)},
    ] 
\begin{feynman}
\vertex (b) at (0,0) ;
\vertex [above left=of b] (i1) {$\mathbf{q}_1$};
\vertex (o1) at (2,0);
\vertex [below left=1 and 1 of b] (i2) {${\mathbf{q}_2}$};
\vertex [above right=of o1] (v1) {$\mathbf{q}_{123}$};
\vertex [below right=1 and 1 of o1] (v2) {${\mathbf{q}_3}$};

\draw[thick, postaction={decorate}, decoration={markings, mark=at position 0.5 with {\arrow{>}}}] 
    (b) -- node [below=8pt] {$\mathbf{q}_{12}$} (o1);
\draw[thick, postaction={decorate}, decoration={markings, mark=at position 0.5 with {\arrow{>}}}] 
    (i1) -- (b);
\draw[thick, postaction={decorate}, decoration={markings, mark=at position 0.5 with {\arrow{>}}}] 
    (i2) -- (b);
\draw[thick, postaction={decorate}, decoration={markings, mark=at position 0.5 with {\arrow{>}}}] 
    (o1) -- (v1);
\draw[thick, postaction={decorate}, decoration={markings, mark=at position 0.5 with {\arrow{<}}}] 
    (o1) -- (v2);
\end{feynman}

\end{tikzpicture}\;,
\end{equation}
such that the third order kernel $K_3$ can be written as,
\begin{equation} \label{eq:K3_diag}
K_3(\mathbf{q}_1,\mathbf{q}_2,\mathbf{q}_3)\equiv
\begin{tikzpicture}[scale=0.4, 
            baseline={([xshift=-5cm,yshift=-\the\dimexpr\fontdimen22\textfont2\relax]
                current bounding box.center)},
    ] 
\begin{feynman}
\vertex (b) at (0,0) ;
\vertex [above left=of b] (i1) {$\mathbf{q}_1$};
\vertex (o1) at (2,0);
\vertex [below left=1 and 1 of b] (i2) {${\mathbf{q}_2}$};
\vertex [above right=of o1] (v1) {$\mathbf{q}_{123}$};
\vertex [below right=1 and 1 of o1] (v2) {${\mathbf{q}_3}$};

\draw[thick, postaction={decorate}, decoration={markings, mark=at position 0.5 with {\arrow{>}}}] 
    (b) -- node [below=8pt] {$\mathbf{q}_{12}$} (o1);
\draw[thick, postaction={decorate}, decoration={markings, mark=at position 0.5 with {\arrow{>}}}] 
    (i1) -- (b);
\draw[thick, postaction={decorate}, decoration={markings, mark=at position 0.5 with {\arrow{>}}}] 
    (i2) -- (b);
\draw[thick, postaction={decorate}, decoration={markings, mark=at position 0.5 with {\arrow{>}}}] 
    (o1) -- (v1);
\draw[thick, postaction={decorate}, decoration={markings, mark=at position 0.5 with {\arrow{<}}}] 
    (o1) -- (v2);
\end{feynman}
\end{tikzpicture}
+ \text{cyclic permutations}\;.
\end{equation}
Note that Eq.\eqref{eq:K3_diag} has the same interpretation as Eq.\eqref{eq:K3}.

Let us state the results of imposing the EGI constraints on the kernels at second and third order. For details of the calculation, see Appendix.\,\ref{app:third_order}. For the second order kernel given in Eq.\eqref{eq:K2}, imposing symmetrisation under $\mathbf{q}_1, \mathbf{q}_2$ implies $b_4=0$, since it is anti symmetric. This leaves us with three coefficients i.e. $\{b_1,b_2,b_3\}$. Then the LO constraints given in Eq.\eqref{eq:Lo_const} imposes the relation,
\begin{align}
    b_2 = 2 a_1\;.
\end{align}
With this we can write the kernel given in Eq.\eqref{eq:K2} as\,\cite{DAmico:2021rdb},
\begin{align}\label{eq:K2_ind_coeff}
 K_{2}(\mathbf{q}_1,\mathbf{q}_2)= b_1 + 2 a_1 \; \alpha_{+}(\mathbf{q_1,\mathbf{q_2}}) + b_3 \beta(\mathbf{q_1,\mathbf{q_2}})\;.
\end{align}
Note that at second order, there are three independent coefficients out of which, one is from first order and the rest two from second order. It can be checked that there exists an invertible map between the bootstrap operators and the ones obtained in a bias expansion.\footnote{The comparison is done with the basis obtained by using SPT kernel for matter assuming Einstein deSitter (EdS) Universe.} We list the independent set of operators in a bias expansion at second order for reference\footnote{The notation used is $r_{ij}\equiv\partial_i\partial_j\Phi$ and $p_{ij}\equiv\partial_iv_j$. Then the scalars are defined as, $\delta=r_{ii},\theta=p_{ii},r^2=r_{ij}r_{ij}$ and so on.},
\begin{align}
    \{\delta,\delta^2,r^2\}\;.
\end{align}
\textbf{Matter Kernel : Mass and momentum conservation}

The three coefficients given in Eq.\eqref{eq:K2_ind_coeff} fully determine the kernel for tracers at second order. However, in order to get the matter kernel we need to impose mass and momentum conservation given in Eq.\eqref{eq:mass_mom_conserv} further on $K_2$. Doing that we get the following condition,
\begin{align}\label{eq:mass_2order}
    b_3=2a_1-b_1\;.
\end{align}
This gives us the matter kernel at second order as\,\cite{DAmico:2021rdb},
\begin{align}\label{eq:G2_exp}
    G_{2}(\mathbf{q}_1,\mathbf{q}_2)= b_1 + 2 \alpha_{+}(\mathbf{q_1,\mathbf{q_2}}) + (2-b_1) \; \beta(\mathbf{q_1,\mathbf{q_2}})\;.
\end{align}

Here we have assumed that $a_1=1$ which is the case for matter kernels. The kernel for matter given in Eq.\eqref{eq:G2_exp} will be used in to get constraints on higher order tracer kernel.
\vspace{5pt}\\
\underline{\textbf{EGI constraints at third order}}
\vspace{5pt}\\
For the third order, we need to symmetrize the kernel given in Eq.\eqref{eq:K3} under exchange of all momenta. Doing that one finds that there are twelve independent coefficients in $K_3$. 
Imposing all the "leading order" and next-to-leading  order" constraints on $K_3$ as given in Eq.\eqref{eq:K3}, provides us a set of seven independent coefficients at third order\,\cite{DAmico:2021rdb}. We can see from Eq.\eqref{eq:Lo_const} and \eqref{eq:NLO_all_order} that EGI constraints relate the coefficients at a particular order with the coefficients at lower order. Out of the 7 coefficients we have 4 from third order and 2 and 1 from second and first order respectively. This hierarchy holds at higher order as well. The details of the calculation of EGI constraints is given in Appendix.\,\ref{app:third_order}. With 7 independent coefficients we get 7 independent momentum structures at third order for tracers.\footnote{The tracer kernel at third order contains a coefficient $h$ which comes from the matter kernel at second order. It's value depends on the underlying cosmology (Eq.\eqref{eq:def_h}). In order to check for degeneracy, we have fixed the value of $h$ with the corresponding EdS value.}. Bootstrap basis is equivalent to the basis of operators obtained at third order in a bias expansion which are given as,
\begin{align}
    \{\delta,\delta^2,\delta\theta,r^2,rp,\delta^3,r^3\}\;.
\end{align}
\vspace{5pt}\\
\underline{\textbf{Mass and momentum constraints}}
\vspace{5pt}\\
Apart from the EGI constraints, one can now impose mass and momentum conservation as given in Eq.\eqref{eq:mass_mom_conserv}, further on $K_3$ to get the matter kernel $G_3 (F_3)$. This gives only 3 independent coefficients for the matter kernel at third order\,\cite{DAmico:2021rdb}\footnote{We have set $a_1=1$ which is the case for matter kernel at first order.}. The details of the calculation for mass and momentum conservation is given in Appendix.\,\ref{app:third_order}.



\section{Eulerian Bootstrap beyond third order}\label{sec:Eulerian_boot}
In this section, we discuss the bootstrap procedure and impose the LO and NLO constraints to get the tracer kernel at fourth and fifth order. We have found that the bootstrap kernel at fourth and fifth order are consistent with the results obtained with the bias expansion in the sense we describe shortly. Using bootstrap we obtain 17 and 44 independent operators in the tracer kernel at fourth and fifth order respectively. Direct computations give 15 and 29 as the number of independent operators that appear in the bias expansion. We have verified that we can write all 15(29) operators in terms of the bootstrap operators at fourth(fifth) order. Bootstrap basis contains all the time non-local structures which begin to affect tracer kernel starting from fifth order. This tells us that bootstrap captures all the information contained in a bias expansion. However, we can see that there is an apparent discrepancy between the number of independent coefficients as given by bootstrap and direct computations. We elaborate on this issue in Sec.\,\ref{ssec:bias_exp_unfixed_4} and then later in Sec.\,\ref{sec:lag_boot} where we point out the underlying reason for the discrepancy and also provide a remedy for the same. 

Apart from constraints coming from EGI, we impose mass and momentum conservation as given in Eq.\eqref{eq:mass_mom_conserv} to get the matter kernel at fourth and fifth order. Let us start with describing the construction of the kernel at fourth order.
\subsection{Bootstrapping Tracer Kernel at Fourth Order}\label{sec:ker_4_boot}
In this section, we discuss the construction of kernel for tracers at fourth order and impose the EGI constraints as discussed in Sec.\,\ref{ssec:EGI_const}. The leading and next-to-leading order constraints as given in Eq.\eqref{eq:Lo_const} and \eqref{eq:NLO_all_order} impose restrictions on the set of allowed momentum structure in $K_4$. 

Let us now construct the most general kernel at fourth order using the building blocks given in Eq.\eqref{eq:K2_diag}. We have already constructed the kernel up to third order in Eq.\eqref{eq:K3_diag}. For the fourth order one needs to extend the structure at third order as given in Eq.\eqref{eq:K3_diag} using the cubic vertex given in Eq.\eqref{eq:K2_diag}. Diagrammatically, we can see that there are two possible extensions of the diagrams given in Eq.\eqref{eq:K3_diag}. One is when we attach one more cubic vertex at one of the external leg that gives the following diagram,
\begin{equation}\label{eq:K4_diag_1}
   \left( \left(\begin{tikzpicture}[scale=0.62, 
            baseline={([xshift=-5cm,yshift=-\the\dimexpr\fontdimen22\textfont2\relax]
                current bounding box.center)},
    ] 
\begin{feynman}
\vertex (v1) at (0,0) ;
\vertex  (e1) at (-1.5,1.5){$\mathbf{q}_{1}$};
\vertex  (e2) at (-1.5,-1.5){$\mathbf{q}_{2}$};
\vertex  (v2) at (1.5,0);
\vertex  (e3) at (1.5,-1.5){$\mathbf{q}_{3}$};
\vertex  (v3) at (3,0);
\vertex  (e4) at (4.5,1.5){$\mathbf{q}_{1234}$};
\vertex  (e5) at (4.5,-1.5){$\mathbf{q}_{4}$};

\draw[thick, postaction={decorate}, decoration={markings, mark=at position 0.5 with {\arrow{>}}}] 
    (e1) -- (v1);
\draw[thick, postaction={decorate}, decoration={markings, mark=at position 0.5 with {\arrow{>}}}] 
    (e2) -- (v1);
\draw[thick, postaction={decorate}, decoration={markings, mark=at position 0.5 with {\arrow{>}}}] 
    (v1) -- node [above=4pt] {$\mathbf{q}_{12}$} (v2);
\draw[thick, postaction={decorate}, decoration={markings, mark=at position 0.5 with {\arrow{<}}}] 
    (v2) -- (e3);
\draw[thick, postaction={decorate}, decoration={markings, mark=at position 0.5 with {\arrow{>}}}] 
    (v2) -- node [above=4pt] {$\mathbf{q}_{123}$} (v3);
\draw[thick, postaction={decorate}, decoration={markings, mark=at position 0.5 with {\arrow{>}}}] 
    (v3) -- (e4);
\draw[thick, postaction={decorate}, decoration={markings, mark=at position 0.5 with {\arrow{<}}}] 
    (v3) -- (e5);
\end{feynman}

\end{tikzpicture}+\text{cyc. perm. of}\{\mathbf{q}_1,\mathbf{q}_2,\mathbf{q}_3\}\right)+\text{cyc. perm. of}\{\mathbf{q}_1,\mathbf{q}_2,\mathbf{q}_3,\mathbf{q}_4\}\right)\;.
\end{equation}
The other way is when we attach the cubic vertex at the internal line which gives the following diagram,
\begin{equation}\label{eq:K4_diag_2}
    \begin{tikzpicture}[scale=0.7, 
            baseline={([xshift=-5cm,yshift=-\the\dimexpr\fontdimen22\textfont2\relax]
                current bounding box.center)},
    ] 
\begin{feynman}
\vertex (v1) at (0,0) ;
\vertex  (e1) at (-1.5,1.5){$\mathbf{q}_{1}$};
\vertex  (e2) at (-1.5,-1.5){$\mathbf{q}_{2}$};
\vertex  (v2) at (1.5,0);
\vertex  (e3) at (1.5,-1.5){$\mathbf{q}_{1234}$};
\vertex  (v3) at (3,0);
\vertex  (e4) at (4.5,1.5){$\mathbf{q}_{4}$};
\vertex  (e5) at (4.5,-1.5){$\mathbf{q}_{3}$};

\draw[thick, postaction={decorate}, decoration={markings, mark=at position 0.5 with {\arrow{>}}}] 
    (e1) -- (v1);
\draw[thick, postaction={decorate}, decoration={markings, mark=at position 0.5 with {\arrow{>}}}] 
    (e2) -- (v1);
\draw[thick, postaction={decorate}, decoration={markings, mark=at position 0.5 with {\arrow{>}}}] 
    (v1) -- node [above=4pt] {$\mathbf{q}_{12}$} (v2);
\draw[thick, postaction={decorate}, decoration={markings, mark=at position 0.5 with {\arrow{>}}}] 
    (v2) -- (e3);
\draw[thick, postaction={decorate}, decoration={markings, mark=at position 0.5 with {\arrow{<}}}] 
    (v2) -- node [above=4pt] {$\mathbf{q}_{34}$} (v3);
\draw[thick, postaction={decorate}, decoration={markings, mark=at position 0.5 with {\arrow{<}}}] 
    (v3) -- (e4);
\draw[thick, postaction={decorate}, decoration={markings, mark=at position 0.5 with {\arrow{<}}}] 
    (v3) -- (e5);
\end{feynman}

\end{tikzpicture}+\text{t and u channel}\;,
\end{equation}
where $\mathbf{q}_{1234}$ denotes the external leg.
Therefore, we can write the momentum structures at fourth order as,
\begin{align}\label{eq:K4_analytic}  T_4(\mathbf{q}_1,\mathbf{q}_2,\mathbf{q}_3,\mathbf{q}_4)= \{\{   S_2(\mathbf{q}_1,\mathbf{q}_2) \otimes S_2(\mathbf{q}_{12},\mathbf{q}_3) +\text{cyclic perm.}\}&\otimes S_2(\mathbf{q}_{123},\mathbf{q}_{4}) + \text{cyclic perm}\} \nonumber\\
    &\cup \nonumber \\
    S_2(\mathbf{q}_1,\mathbf{q}_2) \otimes S_2(\mathbf{q}_3,\mathbf{q}_4) \otimes S_2(\mathbf{q}_{12},\mathbf{q}_{34}) &+ (2 \leftrightarrow 3) + (2 \leftrightarrow 4)\;,
\end{align}
where the first line of Eq.\eqref{eq:K4_analytic} denotes Eq.\eqref{eq:K4_diag_1} while the second line denotes Eq.\eqref{eq:K4_diag_2}. Note that $T_4$ is a set and the diagrams as given in Eq.\eqref{eq:K4_diag_1} and \eqref{eq:K4_diag_2} represent the set of momentum structures given in the first and second line of Eq.\eqref{eq:K4_analytic}.

This implies that the most general kernel for biased tracers at fourth order is given as,
\begin{align}\label{eq:ker4}
K_4(\mathbf{q}_1,\mathbf{q}_2,\mathbf{q}_3,\mathbf{q}_4)=\sum_i^{l(T_4)}d_i [T_4](i)\;,
\end{align}
where $d_i$ are independent coefficients and $l(T_4)$ is the cardinal number of the set $T_4$. After imposing symmetrisation on Eq.\eqref{eq:ker4}, we find that there are 66 independent $d_i$ coefficients. These $d_i$'s get related to each other further after imposing the the EGI constraints. In the next section, we will discuss the EGI constraints that has to be satisfied by $K_4$ according to Eq.\eqref{eq:Lo_const} and \eqref{eq:NLO_all_order} and then impose them to obtain the kernel for biased tracers at fourth order.

\subsubsection{Constraints from EGI}\label{ssec:K4_constraints}
In this section, we will explicitly state the constraints that $K_4$ as given in Eq.\eqref{eq:ker4} has to satisfy in accordance with Eq.\eqref{eq:Lo_const} and Eq.\eqref{eq:NLO_all_order}. We will see that the diagrammatic approach developed in Sec.\,\ref{ssec:ker_3_const} for third order will also be helpful in understanding and imposing the constraints at fourth order.

Let us start by listing out all the EGI constraint coming from leading order as given in Eq.\eqref{eq:Lo_const}. For the case of $K_4$ we can take up to three external momenta to be soft which provides us with three sets of constraints. From Eq.\eqref{eq:K4_diag_1} and \eqref{eq:K4_diag_2}, we can see that taking one momenta soft on $K_4$ will be as follows,
\begin{equation}\label{eq:K4_lo_diag_1mom}
    \lim_{\mathbf{q}_1\rightarrow 0}\begin{tikzpicture}[scale=0.58, 
            baseline={([xshift=-5cm,yshift=-\the\dimexpr\fontdimen22\textfont2\relax]
                current bounding box.center)},
    ] 
\begin{feynman}
\vertex (v1) at (0,0) ;
\vertex  (e1) at (-1.5,1.5){$\mathbf{q}_{1}$};
\vertex  (e2) at (-1.5,-1.5){$\mathbf{q}_{2}$};
\vertex  (v2) at (1.5,0);
\vertex  (e3) at (1.5,-1.5){$\mathbf{q}_{3}$};
\vertex  (v3) at (3,0);
\vertex  (e4) at (4.5,1.5){$\mathbf{q}_{1234}$};
\vertex  (e5) at (4.5,-1.5){$\mathbf{q}_{4}$};

\draw[thick, postaction={decorate}, decoration={markings, mark=at position 0.5 with {\arrow{>}}}] 
    (e1) -- (v1);
\draw[thick, postaction={decorate}, decoration={markings, mark=at position 0.5 with {\arrow{>}}}] 
    (e2) -- (v1);
\draw[thick, postaction={decorate}, decoration={markings, mark=at position 0.5 with {\arrow{>}}}] 
    (v1) -- node [above=4pt] {$\mathbf{q}_{12}$} (v2);
\draw[thick, postaction={decorate}, decoration={markings, mark=at position 0.5 with {\arrow{<}}}] 
    (v2) -- (e3);
\draw[thick, postaction={decorate}, decoration={markings, mark=at position 0.5 with {\arrow{>}}}] 
    (v2) -- node [above=4pt] {$\mathbf{q}_{123}$} (v3);
\draw[thick, postaction={decorate}, decoration={markings, mark=at position 0.5 with {\arrow{>}}}] 
    (v3) -- (e4);
\draw[thick, postaction={decorate}, decoration={markings, mark=at position 0.5 with {\arrow{<}}}] 
    (v3) -- (e5);
\end{feynman}

\end{tikzpicture}
=\frac{\mathbf{q}_1\cdot\mathbf{q}_{234}}{\mathbf{q}_1^2}
\left\{\begin{tikzpicture}[scale=0.6, 
            baseline={([xshift=-5cm,yshift=-\the\dimexpr\fontdimen22\textfont2\relax]
                current bounding box.center)},
    ] 
\begin{feynman}
\vertex (v1) at (0,0) ;
\vertex  (e1) at (-1.5,1.5){$\mathbf{q}_{2}$};
\vertex  (e2) at (-1.5,-1.5){$\mathbf{q}_{3}$};
\vertex  (v3) at (3,0);
\vertex  (e4) at (4.5,1.5){$\mathbf{q}_{234}$};
\vertex  (e5) at (4.5,-1.5){$\mathbf{q}_{4}$};

\draw[thick, postaction={decorate}, decoration={markings, mark=at position 0.5 with {\arrow{>}}}] 
    (e1) -- (v1);
\draw[thick, postaction={decorate}, decoration={markings, mark=at position 0.5 with {\arrow{>}}}] 
    (e2) -- (v1);
\draw[thick, postaction={decorate}, decoration={markings, mark=at position 0.5 with {\arrow{>}}}] 
    (v1) -- node [above=4pt] {$\mathbf{q}_{23}$} (v3);
\draw[thick, postaction={decorate}, decoration={markings, mark=at position 0.5 with {\arrow{>}}}] 
    (v3) -- (e4);
\draw[thick, postaction={decorate}, decoration={markings, mark=at position 0.5 with {\arrow{<}}}] 
    (v3) -- (e5);
\end{feynman}

\end{tikzpicture}\right\}+\mathcal{O}(q_1^0)\;,
\end{equation}
which is identical to the following constraint\,\cite{DAmico:2021rdb},
\begin{align}\label{eq:lo_4th_1}
  \lim_{\mathbf{q}_1\rightarrow 0}  K_4(\mathbf{q}_1,\mathbf{q}_2,\mathbf{q}_3,\mathbf{q}_4)=\frac{\mathbf{q}_1\cdot(\mathbf{q}_2+\mathbf{q}_3+\mathbf{q}_4)}{\mathbf{q}_1^2}K_3(\mathbf{q}_2,\mathbf{q}_3,\mathbf{q}_4)+\mathcal{O}(q_1^0)\;.
\end{align}

Similarly we can take two and three momenta to be soft in $K_4$ as given in Eq.\eqref{eq:K4_diag_1} and \eqref{eq:K4_diag_2}. This is same as  taking further soft limits on Eq.\eqref{eq:K4_lo_diag_1mom}  which gives us the following ,
\begin{equation}
     \lim_{\mathbf{q}_1,\mathbf{q}_2\rightarrow 0}\begin{tikzpicture}[scale=0.58, 
            baseline={([xshift=-5cm,yshift=-\the\dimexpr\fontdimen22\textfont2\relax]
                current bounding box.center)},
    ] 
\begin{feynman}
\vertex (v1) at (0,0) ;
\vertex  (e1) at (-1.5,1.5){$\mathbf{q}_{1}$};
\vertex  (e2) at (-1.5,-1.5){$\mathbf{q}_{2}$};
\vertex  (v2) at (1.5,0);
\vertex  (e3) at (1.5,-1.5){$\mathbf{q}_{3}$};
\vertex  (v3) at (3,0);
\vertex  (e4) at (4.5,1.5){$\mathbf{q}_{1234}$};
\vertex  (e5) at (4.5,-1.5){$\mathbf{q}_{4}$};

\draw[thick, postaction={decorate}, decoration={markings, mark=at position 0.5 with {\arrow{>}}}] 
    (e1) -- (v1);
\draw[thick, postaction={decorate}, decoration={markings, mark=at position 0.5 with {\arrow{>}}}] 
    (e2) -- (v1);
\draw[thick, postaction={decorate}, decoration={markings, mark=at position 0.5 with {\arrow{>}}}] 
    (v1) -- node [above=4pt] {$\mathbf{q}_{12}$} (v2);
\draw[thick, postaction={decorate}, decoration={markings, mark=at position 0.5 with {\arrow{<}}}] 
    (v2) -- (e3);
\draw[thick, postaction={decorate}, decoration={markings, mark=at position 0.5 with {\arrow{>}}}] 
    (v2) -- node [above=4pt] {$\mathbf{q}_{123}$} (v3);
\draw[thick, postaction={decorate}, decoration={markings, mark=at position 0.5 with {\arrow{>}}}] 
    (v3) -- (e4);
\draw[thick, postaction={decorate}, decoration={markings, mark=at position 0.5 with {\arrow{<}}}] 
    (v3) -- (e5);
\end{feynman}

\end{tikzpicture}
=\frac{\mathbf{q}_1\cdot(\mathbf{q}_3+\mathbf{q}_4)}{\mathbf{q}_1^2}\frac{\mathbf{q}_2\cdot(\mathbf{q}_3+\mathbf{q}_4)}{\mathbf{q}_2^2}
\begin{tikzpicture}[scale=0.6, 
            baseline={([xshift=-5cm,yshift=-\the\dimexpr\fontdimen22\textfont2\relax]
                current bounding box.center)},
    ] 
\begin{feynman}
\vertex (v1) at (0,0) ;
\vertex  (e1) at (-1.5,1.5){$\mathbf{q}_{3}$};
\vertex  (e2) at (-1.5,-1.5){$\mathbf{q}_{4}$};
\vertex  (v2) at (2,0) {$\mathbf{q}_{34}$};

\draw[thick, postaction={decorate}, decoration={markings, mark=at position 0.5 with {\arrow{>}}}] 
    (e1) -- (v1);
\draw[thick, postaction={decorate}, decoration={markings, mark=at position 0.5 with {\arrow{>}}}] 
    (e2) -- (v1);
\draw[thick, postaction={decorate}, decoration={markings, mark=at position 0.5 with {\arrow{>}}}] 
    (v1) -- (v2);
\end{feynman}

\end{tikzpicture}\nonumber\;,
\end{equation}
\begin{equation}\label{eq:K4_lo_diag_23mom}
     \lim_{\mathbf{q}_1,\mathbf{q}_2,\mathbf{q}_3\rightarrow 0}\begin{tikzpicture}[scale=0.58, 
            baseline={([xshift=-5cm,yshift=-\the\dimexpr\fontdimen22\textfont2\relax]
                current bounding box.center)},
    ] 
\begin{feynman}
\vertex (v1) at (0,0) ;
\vertex  (e1) at (-1.5,1.5){$\mathbf{q}_{1}$};
\vertex  (e2) at (-1.5,-1.5){$\mathbf{q}_{2}$};
\vertex  (v2) at (1.5,0);
\vertex  (e3) at (1.5,-1.5){$\mathbf{q}_{3}$};
\vertex  (v3) at (3,0);
\vertex  (e4) at (4.5,1.5){$\mathbf{q}_{1234}$};
\vertex  (e5) at (4.5,-1.5){$\mathbf{q}_{4}$};

\draw[thick, postaction={decorate}, decoration={markings, mark=at position 0.5 with {\arrow{>}}}] 
    (e1) -- (v1);
\draw[thick, postaction={decorate}, decoration={markings, mark=at position 0.5 with {\arrow{>}}}] 
    (e2) -- (v1);
\draw[thick, postaction={decorate}, decoration={markings, mark=at position 0.5 with {\arrow{>}}}] 
    (v1) -- node [above=4pt] {$\mathbf{q}_{12}$} (v2);
\draw[thick, postaction={decorate}, decoration={markings, mark=at position 0.5 with {\arrow{<}}}] 
    (v2) -- (e3);
\draw[thick, postaction={decorate}, decoration={markings, mark=at position 0.5 with {\arrow{>}}}] 
    (v2) -- node [above=4pt] {$\mathbf{q}_{123}$} (v3);
\draw[thick, postaction={decorate}, decoration={markings, mark=at position 0.5 with {\arrow{>}}}] 
    (v3) -- (e4);
\draw[thick, postaction={decorate}, decoration={markings, mark=at position 0.5 with {\arrow{<}}}] 
    (v3) -- (e5);
\end{feynman}

\end{tikzpicture}
=\frac{\mathbf{q}_1\cdot\mathbf{q}_4}{\mathbf{q}_1^2}\frac{\mathbf{q}_2\cdot\mathbf{q}_4}{\mathbf{q}_2^2}\frac{\mathbf{q}_3\cdot\mathbf{q}_4}{\mathbf{q}_3^2}
\begin{tikzpicture}[scale=0.6, 
            baseline={([xshift=-5cm,yshift=-\the\dimexpr\fontdimen22\textfont2\relax]
                current bounding box.center)},
    ] 
\begin{feynman}
\vertex (v1) at (0,0) {$\mathbf{q}_4$} ;
\vertex (v2) at (0,-3) {$\mathbf{q}_{4}$};

\draw[thick, postaction={decorate}, decoration={markings, mark=at position 0.5 with {\arrow{<}}}] 
    (v1) -- (v2);
\end{feynman}

\end{tikzpicture}+\mathcal{O}(q^{-2})\;,
\end{equation}
which has the following analytic interpretation,
\begin{align}\label{eq:lo_4th_2}
&\lim_{\mathbf{q}_1,\mathbf{q}_2\rightarrow 0}K_4(\mathbf{q}_1,\mathbf{q}_2,\mathbf{q}_3,\mathbf{q}_4)=\frac{\mathbf{q}_1\cdot(\mathbf{q}_3+\mathbf{q}_4)}{\mathbf{q}_1^2}\frac{\mathbf{q}_2\cdot(\mathbf{q}_3+\mathbf{q}_4)}{\mathbf{q}_2^2}K_2(\mathbf{q}_3,\mathbf{q}_4)+\mathcal{O}(q^{-1})\;,\nonumber\\
&\lim_{\mathbf{q}_1,\mathbf{q}_2,\mathbf{q}_3\rightarrow 0}K_4(\mathbf{q}_1,\mathbf{q}_2,\mathbf{q}_3,\mathbf{q}_4)=\frac{\mathbf{q}_1\cdot(\mathbf{q}_4)}{\mathbf{q}_1^2}\frac{\mathbf{q}_2\cdot(\mathbf{q}_4)}{\mathbf{q}_2^2}\frac{\mathbf{q}_3\cdot(\mathbf{q}_4)}{\mathbf{q}_3^2}K_1(\mathbf{q}_4)+\mathcal{O}(q^{-2})\;.
\end{align}
Now we move on to next-to-leading order (NLO) and next-to-next-to leading order (NNLO) constraints which are a consequence of Eq.\eqref{eq:NLO_all_order}. These constraints are a straight forward generalisation of Eq.\eqref{eq:K3_NLO_diag} and are given as,

\begin{equation}\label{eq:K4_nlo_nnlo_diag_2}
  \lim_{\mathbf{q}_{12}\rightarrow 0}\begin{tikzpicture}[scale=0.58, 
            baseline={([xshift=-5cm,yshift=-\the\dimexpr\fontdimen22\textfont2\relax]
                current bounding box.center)},
    ] 
\begin{feynman}
\vertex (v1) at (0,0) ;
\vertex  (e1) at (-1.5,1.5){$\mathbf{q}_{1}$};
\vertex  (e2) at (-1.5,-1.5){$\mathbf{q}_{2}$};
\vertex  (v2) at (1.5,0);
\vertex  (e3) at (1.5,-1.5){$\mathbf{q}_{3}$};
\vertex  (v3) at (3,0);
\vertex  (e4) at (4.5,1.5){$\mathbf{q}_{1234}$};
\vertex  (e5) at (4.5,-1.5){$\mathbf{q}_{4}$};

\draw[thick, postaction={decorate}, decoration={markings, mark=at position 0.5 with {\arrow{>}}}] 
    (e1) -- (v1);
\draw[thick, postaction={decorate}, decoration={markings, mark=at position 0.5 with {\arrow{>}}}] 
    (e2) -- (v1);
\draw[thick, postaction={decorate}, decoration={markings, mark=at position 0.5 with {\arrow{>}}}] 
    (v1) -- node [above=4pt] {$\mathbf{q}_{12}$} (v2);
\draw[thick, postaction={decorate}, decoration={markings, mark=at position 0.5 with {\arrow{<}}}] 
    (v2) -- (e3);
\draw[thick, postaction={decorate}, decoration={markings, mark=at position 0.5 with {\arrow{>}}}] 
    (v2) -- node [above=4pt] {$\mathbf{q}_{123}$} (v3);
\draw[thick, postaction={decorate}, decoration={markings, mark=at position 0.5 with {\arrow{>}}}] 
    (v3) -- (e4);
\draw[thick, postaction={decorate}, decoration={markings, mark=at position 0.5 with {\arrow{<}}}] 
    (v3) -- (e5);
\end{feynman}

\end{tikzpicture}
\supset\frac{\mathbf{q}_{12}\cdot(\mathbf{q}_3+\mathbf{q}_4)}{\mathbf{q}_{12}^2}
\left\{\begin{tikzpicture}[scale=0.55, 
            baseline={([xshift=-5cm,yshift=-\the\dimexpr\fontdimen22\textfont2\relax]
                current bounding box.center)},
    ] 
\begin{feynman}
\vertex (v1) at (0,0) ;
\vertex  (e1) at (-1.5,1.5){$\mathbf{q}_{3}$};
\vertex  (e2) at (-1.5,-1.5){$\mathbf{q}_{4}$};
\vertex  (v2) at (2,0) {$\mathbf{q}_{34}$};

\draw[thick, postaction={decorate}, decoration={markings, mark=at position 0.5 with {\arrow{>}}}] 
    (e1) -- (v1);
\draw[thick, postaction={decorate}, decoration={markings, mark=at position 0.5 with {\arrow{>}}}] 
    (e2) -- (v1);
\draw[thick, postaction={decorate}, decoration={markings, mark=at position 0.5 with {\arrow{>}}}] 
    (v1) -- (v2);
\end{feynman}

\end{tikzpicture}\right\}\left\{\begin{tikzpicture}[scale=0.55, 
            baseline={([xshift=-5cm,yshift=-\the\dimexpr\fontdimen22\textfont2\relax]
                current bounding box.center)},
    ] 
\begin{feynman}
\vertex (v1) at (0,0) ;
\vertex  (e1) at (-1.5,1.5){$\mathbf{q}_{1}$};
\vertex  (e2) at (-1.5,-1.5){$\mathbf{q}_{2}$};
\vertex  (v2) at (2,0) {$\mathbf{q}_{12}$};

\draw[thick, postaction={decorate}, decoration={markings, mark=at position 0.5 with {\arrow{>}}}] 
    (e1) -- (v1);
\draw[thick, postaction={decorate}, decoration={markings, mark=at position 0.5 with {\arrow{>}}}] 
    (e2) -- (v1);
\draw[thick, postaction={decorate}, decoration={markings, mark=at position 0.5 with {\arrow{>}}}] 
    (v1) -- (v2);
\end{feynman}

\end{tikzpicture}\right\},
\end{equation}

\begin{equation}\label{eq:K4_nlo_nnlo_diag}
  \lim_{\mathbf{q}_{123}\rightarrow 0}\begin{tikzpicture}[scale=0.58, 
            baseline={([xshift=-5cm,yshift=-\the\dimexpr\fontdimen22\textfont2\relax]
                current bounding box.center)},
    ] 
\begin{feynman}
\vertex (v1) at (0,0) ;
\vertex  (e1) at (-1.5,1.5){$\mathbf{q}_{1}$};
\vertex  (e2) at (-1.5,-1.5){$\mathbf{q}_{2}$};
\vertex  (v2) at (1.5,0);
\vertex  (e3) at (1.5,-1.5){$\mathbf{q}_{3}$};
\vertex  (v3) at (3,0);
\vertex  (e4) at (4.5,1.5){$\mathbf{q}_{1234}$};
\vertex  (e5) at (4.5,-1.5){$\mathbf{q}_{4}$};

\draw[thick, postaction={decorate}, decoration={markings, mark=at position 0.5 with {\arrow{>}}}] 
    (e1) -- (v1);
\draw[thick, postaction={decorate}, decoration={markings, mark=at position 0.5 with {\arrow{>}}}] 
    (e2) -- (v1);
\draw[thick, postaction={decorate}, decoration={markings, mark=at position 0.5 with {\arrow{>}}}] 
    (v1) -- node [above=4pt] {$\mathbf{q}_{12}$} (v2);
\draw[thick, postaction={decorate}, decoration={markings, mark=at position 0.5 with {\arrow{<}}}] 
    (v2) -- (e3);
\draw[thick, postaction={decorate}, decoration={markings, mark=at position 0.5 with {\arrow{>}}}] 
    (v2) -- node [above=4pt] {$\mathbf{q}_{123}$} (v3);
\draw[thick, postaction={decorate}, decoration={markings, mark=at position 0.5 with {\arrow{>}}}] 
    (v3) -- (e4);
\draw[thick, postaction={decorate}, decoration={markings, mark=at position 0.5 with {\arrow{<}}}] 
    (v3) -- (e5);
\end{feynman}

\end{tikzpicture}
\supset\frac{\mathbf{q}_{123}\cdot\mathbf{q}_4}{\mathbf{q}_{123}^2}
\left\{\begin{tikzpicture}[scale=0.55, 
            baseline={([xshift=-5cm,yshift=-\the\dimexpr\fontdimen22\textfont2\relax]
                current bounding box.center)},
    ] 
\begin{feynman}
\vertex (v1) at (0,0) ;
\vertex  (e1) at (-1.5,1.5){$\mathbf{q}_{1}$};
\vertex  (e2) at (-1.5,-1.5){$\mathbf{q}_{2}$};
\vertex  (v3) at (3,0);
\vertex  (e4) at (4.5,1.5){$\mathbf{q}_{123}$};
\vertex  (e5) at (4.5,-1.5){$\mathbf{q}_{3}$};

\draw[thick, postaction={decorate}, decoration={markings, mark=at position 0.5 with {\arrow{>}}}] 
    (e1) -- (v1);
\draw[thick, postaction={decorate}, decoration={markings, mark=at position 0.5 with {\arrow{>}}}] 
    (e2) -- (v1);
\draw[thick, postaction={decorate}, decoration={markings, mark=at position 0.5 with {\arrow{>}}}] 
    (v1) -- node[above, yshift=2pt] {$\mathbf{q}_{12}$} (v3);
\draw[thick, postaction={decorate}, decoration={markings, mark=at position 0.5 with {\arrow{>}}}] 
    (v3) -- (e4);
\draw[thick, postaction={decorate}, decoration={markings, mark=at position 0.5 with {\arrow{<}}}] 
    (v3) -- (e5);
\end{feynman}

\end{tikzpicture}\right\}\left\{\begin{tikzpicture}[scale=0.55, 
            baseline={([xshift=-5cm,yshift=-\the\dimexpr\fontdimen22\textfont2\relax]
                current bounding box.center)},
    ] 
\begin{feynman}
\vertex (v1) at (0,0) {$\mathbf{q}_4$} ;
\vertex  (v2) at (0,-3) {$\mathbf{q}_{4}$};

\draw[thick, postaction={decorate}, decoration={markings, mark=at position 0.5 with {\arrow{<}}}] 
    (v1) -- (v2);
\end{feynman}

\end{tikzpicture}\right\}\;.
\end{equation}
Analytically, the NLO and NNLO constraints state that\,\cite{DAmico:2021rdb},
\begin{align}\label{eq:NLO_4th}
&\lim_{\mathbf{q}_{12}\rightarrow 0}  K_4(\mathbf{q}_1,\mathbf{q}_2,\mathbf{q}_3,\mathbf{q}_4)\supset \frac{\mathbf{q}_{12}\cdot(\mathbf{q}_3+\mathbf{q}_4)}{\mathbf{q}_{12}^2}K_2(\mathbf{q}_3,\mathbf{q}_4)\int^\eta d\eta'f_{+}(\eta^{\prime})\left(\frac{D_{+}(\eta^{\prime})}{D_{+}(\eta)}\right)^{2}G_2(\mathbf{q}_1,\mathbf{q}_2,\eta')\nonumber\\
&\lim_{\mathbf{q}_{123}\rightarrow 0}  K_4(\mathbf{q}_1,\mathbf{q}_2,\mathbf{q}_3,\mathbf{q}_4)\supset \frac{\mathbf{q}_{123}\cdot\mathbf{q}_4}{\mathbf{q}_{123}^2}K_1(\mathbf{q}_4)\int^\eta d\eta'f_{+}(\eta^{\prime})\left(\frac{D_{+}(\eta^{\prime})}{D_{+}(\eta)}\right)^{3}G_3(\mathbf{q}_1,\mathbf{q}_2,\mathbf{q}_3,\eta')\;.
\end{align}
From Eq.\eqref{eq:K4_nlo_nnlo_diag_2}, we can see that after taking the limit $\mathbf{q}_{12}\rightarrow 0$, we can further take $\mathbf{q}_3\rightarrow 0$ and that gives us the following,
\begin{equation}\label{eq:K4_d2d1del1_diag}
\lim_{\mathbf{q}_{12},\mathbf{q}_3\rightarrow 0}\begin{tikzpicture}[scale=0.58, 
            baseline={([xshift=-5cm,yshift=-\the\dimexpr\fontdimen22\textfont2\relax]
                current bounding box.center)},
    ] 
\begin{feynman}
\vertex (v1) at (0,0) ;
\vertex  (e1) at (-1.5,1.5){$\mathbf{q}_{1}$};
\vertex  (e2) at (-1.5,-1.5){$\mathbf{q}_{2}$};
\vertex  (v2) at (1.5,0);
\vertex  (e3) at (1.5,-1.5){$\mathbf{q}_{3}$};
\vertex  (v3) at (3,0);
\vertex  (e4) at (4.5,1.5){$\mathbf{q}_{1234}$};
\vertex  (e5) at (4.5,-1.5){$\mathbf{q}_{4}$};

\draw[thick, postaction={decorate}, decoration={markings, mark=at position 0.5 with {\arrow{>}}}] 
    (e1) -- (v1);
\draw[thick, postaction={decorate}, decoration={markings, mark=at position 0.5 with {\arrow{>}}}] 
    (e2) -- (v1);
\draw[thick, postaction={decorate}, decoration={markings, mark=at position 0.5 with {\arrow{>}}}] 
    (v1) -- node [above=4pt] {$\mathbf{q}_{12}$} (v2);
\draw[thick, postaction={decorate}, decoration={markings, mark=at position 0.5 with {\arrow{<}}}] 
    (v2) -- (e3);
\draw[thick, postaction={decorate}, decoration={markings, mark=at position 0.5 with {\arrow{>}}}] 
    (v2) -- node [above=4pt] {$\mathbf{q}_{123}$} (v3);
\draw[thick, postaction={decorate}, decoration={markings, mark=at position 0.5 with {\arrow{>}}}] 
    (v3) -- (e4);
\draw[thick, postaction={decorate}, decoration={markings, mark=at position 0.5 with {\arrow{<}}}] 
    (v3) -- (e5);
\end{feynman}

\end{tikzpicture}
\supset\frac{\mathbf{q}_{12}\cdot\mathbf{q}_4}{\mathbf{q}_{12}^2}\frac{\mathbf{q}_{3}\cdot\mathbf{q}_4}{\mathbf{q}_{3}^2}
\left\{\begin{tikzpicture}[scale=0.55, 
            baseline={([xshift=-5cm,yshift=-\the\dimexpr\fontdimen22\textfont2\relax]
                current bounding box.center)},
    ] 
\begin{feynman}
\vertex (v1) at (0,0) ;
\vertex  (e1) at (-1.5,1.5){$\mathbf{q}_{1}$};
\vertex  (e2) at (-1.5,-1.5){$\mathbf{q}_{2}$};
\vertex  (v2) at (2,0) ;

\draw[thick, postaction={decorate}, decoration={markings, mark=at position 0.5 with {\arrow{>}}}] 
    (e1) -- (v1);
\draw[thick, postaction={decorate}, decoration={markings, mark=at position 0.5 with {\arrow{>}}}] 
    (e2) -- (v1);
\draw[thick, postaction={decorate}, decoration={markings, mark=at position 0.5 with {\arrow{>}}}] 
    (v1) -- node[above, yshift=2pt] {$\mathbf{q}_{12}$} (v2);
\end{feynman}

\end{tikzpicture}\right\}\left\{\begin{tikzpicture}[scale=0.55, 
            baseline={([xshift=-5cm,yshift=-\the\dimexpr\fontdimen22\textfont2\relax]
                current bounding box.center)},
    ] 
\begin{feynman}
\vertex (v1) at (0,0) {$\mathbf{q}_3$} ;
\vertex  (v2) at (0,-3) {$\mathbf{q}_{3}$};

\draw[thick, postaction={decorate}, decoration={markings, mark=at position 0.5 with {\arrow{<}}}] 
    (v1) -- (v2);
\end{feynman}

\end{tikzpicture}
\begin{tikzpicture}[scale=0.55, 
            baseline={([xshift=-5cm,yshift=-\the\dimexpr\fontdimen22\textfont2\relax]
                current bounding box.center)},
    ] 
\begin{feynman}
\vertex (v1) at (0,0) {$\mathbf{q}_{4}$} ;
\vertex  (v2) at (0,-3) {$\mathbf{q}_{4}$};

\draw[thick, postaction={decorate}, decoration={markings, mark=at position 0.5 with {\arrow{<}}}] 
    (v1) -- (v2);
\end{feynman}

\end{tikzpicture}
\right\}\;.
\end{equation}
Therefore, we see that there exists one more kinematic limit which is allowed at fourth order which is different from the ones given in Eq.\eqref{eq:K4_nlo_nnlo_diag}. Hence, we see that the analytical constraint for this kinematic limit is given as,
\begin{align}\label{eq:K4_nlo_d2d1del1} \lim_{\mathbf{q}_{12},\mathbf{q}_3\rightarrow 0}  K_4(\mathbf{q}_1,\mathbf{q}_2,\mathbf{q}_3,\mathbf{q}_4)\supset \frac{\mathbf{q}_{12}\cdot\mathbf{q}_4}{\mathbf{q}_{12}^2}\frac{\mathbf{q}_{3}\cdot\mathbf{q}_4}{\mathbf{q}_{3}^2}K_1(\mathbf{q}_4)\int^\eta d\eta'f_{+}(\eta^{\prime})\left(\frac{D_{+}(\eta^{\prime})}{D_{+}(\eta)}\right)^{2}G_2(\mathbf{q}_1,\mathbf{q}_2,\eta')\;.
\end{align}
We have stated all possible kinematic limits that can be taken on the kernel at fourth order and the consistency condition associated with the kinematic limits. The constraint relations that $K_4$ kernel, as given in Eq.\eqref{eq:ker4}, has to satisfy are given in Eqns.\eqref{eq:lo_4th_1}, \eqref{eq:lo_4th_2}, \eqref{eq:NLO_4th} and \eqref{eq:K4_nlo_d2d1del1}. The LO and NLO constraints relate the coefficients, given in Eq.\eqref{eq:ker4}, among themselves and to coefficients of lower order. This gives us the independent coefficient at fourth order and their associated momentum structure for biased tracers at fourth order. Note that here NLO constraints also include all the constraints coming from imposing the absence of "spurious poles". For details of the calculation, see Appendix.\,\ref{app:fourth_order}. It is common in the literature to call the momentum structure as "operators". Therefore, we will use momentum structures and operators interchangeably throughout the text.

Solving all the LO, NLO, and NNLO constraints together provides us fifty five constraints, which gives a set of 18 coefficients at fourth order.
Each coefficient listed has a momentum structure associated with it. We found that 1 of the momentum structures is linearly dependent with the other 17. Hence, we found that there are 17 independent set of momentum structures at fourth order.\footnote{Note that the tracer kernel at fourth order contains coefficients from matter kernel at third order. These are $s_1, s_2, s_3$ and $s_4$ defined in Eq.\eqref{eq:def_s1_s2_etc}. For the purpose of defining operators and checking degeneracy, we have fixed the values of these coefficients with their corresponding EdS values.}
\vspace{5pt}\\
\underline{\textbf{Discrepancy with direct computation}}
\vspace{5pt}\\
By direct computations, we know that the basis of operators at fourth order contains 15 operators\,\cite{DAmico:2022osl}. We have verified that all the 15 basis operators can be written in terms of the 17 bootstrap operators. This may imply an apparent discrepancy between the boostrap approach and direct computations. The reason for the discrepancy and it's resolution will be discussed later when we discuss Lagrangian bootstrap for tracers in Sec.\,\ref{sec:lag_boot}. We will say more about this in Sec.\,\ref{ssec:bias_exp_unfixed_4} when we discuss the bias expansion for unfixed cosmology.

Now, we impose mass and momentum conservation, given in Eq.\eqref{eq:mass_mom_conserv}, further on the fourth order kernel. This leaves us with 8 independent coefficients,
which is in agreement with the recently proposed Lagrangian bootstrap techniques for matter\,\cite{Marinucci:2024add}, where it has been pointed out that among the eight, the coefficients coming from second, third and fourth order are one, two and five respectively. We have also obtained the same hierarchy among the coefficients. In the next section, we are going to explore the bias expansion for unfixed cosmology which provides some justification for the number of independent coefficients to be 17 instead of 15 which we get in a bias expansion with fixed cosmology.
\subsubsection{Bias expansion with unfixed cosmology}\label{ssec:bias_exp_unfixed_4}
We have seen in Sec.\,\ref{sec:ker_4_boot} that by imposing EGI, we get 17 independent coefficients in the tracer kernel at fourth order. However, in a bias expansion, we get 15 independent coefficients at fourth order\,\cite{DAmico:2022osl}. The difference is that in the bias expansion we use the SPT kernel for matter after solving the equations of motion for a fixed cosmology, usually for EdS Universe\,\cite{DAmico:2022osl,Donath:2023sav}. Hence, the matter kernel in any bias expansion is completely fixed. However, the bootstrap approach is blind to the underlying cosmology. As a result, the matter kernels themselves are not fully fixed as shown in Eq.\eqref{eq:G2_exp}. This is also the reason for the bootstrap approach being useful in probing the cosmology. It is interesting to ask the independent operators appearing in the bias expansion when matter kernel are not fully fixed. Therefore, in this section, we are going to write the bias expansion and find the number of independent operators (coefficients) at fourth order but for unfixed cosmology. That means for writing kernel of operators in the bias expansion, we are going to use the matter kernel obtained from bootstrap in Secs.\,\ref{ssec:ker_3_const} and \ref{sec:ker_4_boot}.

So we start by writing all possible scalars
made out of the tidal tensor ($\partial_i\partial_j \Phi$) and velocity gradient ($\partial_i v_j$). We get the following list of 27 scalars,
\begin{align}\label{eq:bias_exp_scalars_4}
    &\delta,\theta,\nonumber\\[3pt]
    &\delta^2,\delta\theta,\theta^2,r^2,rp,p^2,\nonumber\\[3pt]
    &\delta^3,\delta^2\theta,\delta\theta^2,\theta^3,r^3,r^2p,rp^2,p^3,r^2\delta,r^2\theta,rp\delta,rp\theta,p^2\delta,p^2\theta,\nonumber\\[3pt]
    &\delta^4,r^4,r^2\delta^2,r^2r^2,r^3\delta\;.
\end{align}

While checking for degeneracy among operators appearing in the bias expansion, one usually writes the kernels of each operator in momentum space and remove linearly dependent operators. To check for linear dependence, one uses the SPT kernels which are completely fixed with no unfixed coefficients. However, in Eq.\eqref{eq:bias_exp_scalars_4}, instead of the SPT kernels, we are going to write the kernels for operators using bootstrap kernel which has some unfixed coefficients. We state our findings below,
\begin{align}\label{eq:bias_unfixed_results}
\text{Eq.\eqref{eq:bias_exp_scalars_4}}+\text{SPT kernels}+\text{degeneracy}+\text{EGI}&=15\;,\nonumber\\
\text{Eq.\eqref{eq:bias_exp_scalars_4}}+\text{Bootstrap kernels}+\text{degeneracy}+\text{EGI}&=17\;.
\end{align}
where bootstrap kernel refers to the unfixed matter kernel. For example, the second order matter kernel $G_2$ as given in Eq.\eqref{eq:G2_exp} contains one unfixed coefficients. The explicit form of kernels at higher order are given in the Mathematica file attached with the manuscript.
We have found that there exists an invertible map between the operators appearing in the kernel for bias expansion with unfixed cosmology and the set we obtain in Sec.\,\ref{sec:ker_4_boot} using the bootstrap technique. This implies that if we write the bias expansion for unfixed cosmology, then at fourth order we get 17 independent coefficients and not 15 that is obtained for fixed cosmology.

\subsection{Bootstrapping Tracers at Fifth Order}\label{sec:fifth_order}
In this section, we discuss the bootstrapping procedure for biased tracers at fifth order. First we discuss how to write the most general form of the tracer kernel at order five. Then we impose EGI constraints and obtain the fifth order tracer kernel. We will also comment on the relation between the bootstrap kernel and "time non-locality" which manifests itself at fifth order as pointed out in\,\cite{Donath:2023sav}.
\subsubsection{Kernel and EGI constraints}\label{ssec:k5_EGI_const}
Now, we write down the most general kernel for tracer at fifth order, using the procedure as we did for constructing kernel at fourth order in Sec.\,\ref{sec:ker_4_boot}. Basically, we extend the structures given in Eq.\eqref{eq:K4_diag_1} and \eqref{eq:K4_diag_2} by the cubic vertex given in Eq.\eqref{eq:K2_diag}. There is a straight forward way to do that, by attaching the cubic vertex to the external legs of Eq.\eqref{eq:K4_diag_1} and \eqref{eq:K4_diag_2}. That gives us the following two sectors for the kernel at fifth order,
\begin{equation}
 \left(\left( \left(\begin{tikzpicture}[scale=0.65, 
            baseline={([xshift=-5cm,yshift=-\the\dimexpr\fontdimen22\textfont2\relax]
                current bounding box.center)},
    ] 
\begin{feynman}
\vertex (v1) at (0,0) ;
\vertex  (e1) at (-1.5,1.5){$\mathbf{q}_{1}$};
\vertex  (e2) at (-1.5,-1.5){$\mathbf{q}_{2}$};
\vertex  (v2) at (1.5,0);
\vertex  (e3) at (1.5,-1.5){$\mathbf{q}_{3}$};
\vertex  (v3) at (4.5,0);
\vertex  (e4) at (6,1.5){$\mathbf{q}_{12345}$};
\vertex  (e5) at (6,-1.5){$\mathbf{q}_{5}$};
\vertex  (v4) at (3,0);
\vertex  (e6) at (3,-1.5){$\mathbf{q}_{4}$};

\draw[thick, postaction={decorate}, decoration={markings, mark=at position 0.5 with {\arrow{>}}}] 
    (e1) -- (v1);
\draw[thick, postaction={decorate}, decoration={markings, mark=at position 0.5 with {\arrow{>}}}] 
    (e2) -- (v1);
\draw[thick, postaction={decorate}, decoration={markings, mark=at position 0.5 with {\arrow{>}}}] 
    (v1) -- node [above, yshift=4pt] {$\mathbf{q}_{12}$} (v2);
\draw[thick, postaction={decorate}, decoration={markings, mark=at position 0.5 with {\arrow{<}}}] 
    (v2) -- (e3);
\draw[thick, postaction={decorate}, decoration={markings, mark=at position 0.5 with {\arrow{>}}}] 
    (v2) -- node [above, yshift=4pt] {$\mathbf{q}_{123}$} (v4);
\draw[thick, postaction={decorate}, decoration={markings, mark=at position 0.5 with {\arrow{>}}}] 
    (e6) -- (v4);
\draw[thick, postaction={decorate}, decoration={markings, mark=at position 0.5 with {\arrow{>}}}] 
    (v4) -- node [above, yshift=4pt] {$\mathbf{q}_{1234}$} (v3);
\draw[thick, postaction={decorate}, decoration={markings, mark=at position 0.5 with {\arrow{>}}}] 
    (v3) -- (e4);
\draw[thick, postaction={decorate}, decoration={markings, mark=at position 0.5 with {\arrow{<}}}] 
    (v3) -- (e5);
\end{feynman}

\end{tikzpicture}+\text{cyc.}(\mathbf{q}_1,\mathbf{q}_2,\mathbf{q}_3)\right)+\text{cyc.}(\mathbf{q}_1,\mathbf{q}_2,\mathbf{q}_3,\mathbf{q}_4)\right)\right.\nonumber
\end{equation}

\begin{equation}
    +\text{cyc.}(\mathbf{q}_1,\mathbf{q}_2,\mathbf{q}_3,\mathbf{q}_4,\mathbf{q}_5)\Bigg)\;,\nonumber
\end{equation}

\begin{equation}\label{eq:K5_sec_1_2_diag}
    \left(\left( \begin{tikzpicture}[scale=0.6, 
            baseline={([xshift=-5cm,yshift=-\the\dimexpr\fontdimen22\textfont2\relax]
                current bounding box.center)},
    ] 
\begin{feynman}
\vertex (v1) at (0,0) ;
\vertex  (e1) at (-1.5,1.5){$\mathbf{q}_{1}$};
\vertex  (e2) at (-1.5,-1.5){$\mathbf{q}_{2}$};
\vertex  (v2) at (1.5,0);
\vertex  (v4) at (1.5,-1.5);
\vertex  (e6) at (3,-2.5){$\mathbf{q}_{5}$};
\vertex  (e7) at (0,-2.5){$\mathbf{q}_{12345}$};
\vertex  (v3) at (3,0);
\vertex  (e4) at (4.5,1.5){$\mathbf{q}_{4}$};
\vertex  (e5) at (4.5,-1.5){$\mathbf{q}_{3}$};

\draw[thick, postaction={decorate}, decoration={markings, mark=at position 0.5 with {\arrow{>}}}] 
    (e1) -- (v1);
\draw[thick, postaction={decorate}, decoration={markings, mark=at position 0.5 with {\arrow{>}}}] 
    (e2) -- (v1);
\draw[thick, postaction={decorate}, decoration={markings, mark=at position 0.5 with {\arrow{>}}}] 
    (v1) -- node [above, yshift=4pt] {$\mathbf{q}_{12}$} (v2);
\draw[thick, postaction={decorate}, decoration={markings, mark=at position 0.5 with {\arrow{>}}}] 
    (v2) -- (v4);
\draw[thick, postaction={decorate}, decoration={markings, mark=at position 0.5 with {\arrow{<}}}] 
    (v2) -- node [above, yshift=4pt] {$\mathbf{q}_{34}$} (v3);
\draw[thick, postaction={decorate}, decoration={markings, mark=at position 0.5 with {\arrow{<}}}] 
    (v3) -- (e4);
\draw[thick, postaction={decorate}, decoration={markings, mark=at position 0.5 with {\arrow{<}}}] 
    (v3) -- (e5);
\draw[thick, postaction={decorate}, decoration={markings, mark=at position 0.5 with {\arrow{<}}}] 
    (v4) -- (e6);
\draw[thick, postaction={decorate}, decoration={markings, mark=at position 0.5 with {\arrow{>}}}] 
    (v4) -- (e7);
\end{feynman}

\end{tikzpicture}+(\mathbf{q}_2\leftrightarrow\mathbf{q}_3)+(\mathbf{q}_2\leftrightarrow\mathbf{q}_4)\right)+\text{cyc.}(\mathbf{q}_1,\mathbf{q}_2,\mathbf{q}_3,\mathbf{q}_4,\mathbf{q}_5)\right)\;.
\end{equation}
However, there is a third sector which can be obtained by attaching  a cubic vertex as given in Eq.\eqref{eq:K2_diag} to an internal leg of Eq.\eqref{eq:K4_diag_2}. It can be written as follows,
\begin{equation}\label{eq:K5_sec3_diag}
    \begin{tikzpicture}[scale=0.65, 
            baseline={([xshift=-5cm,yshift=-\the\dimexpr\fontdimen22\textfont2\relax]
                current bounding box.center)},
    ] 
\begin{feynman}
\vertex (v1) at (0,0) ;
\vertex  (e1) at (-1.5,1.5){$\mathbf{q}_{1}$};
\vertex  (e2) at (-1.5,-1.5){$\mathbf{q}_{2}$};
\vertex  (v2) at (1.5,0);
\vertex  (e3) at (1.5,-1.5){$\mathbf{q}_{3}$};
\vertex  (v3) at (4.5,0);
\vertex  (e4) at (6,1.5){$\mathbf{q}_{12345}$};
\vertex  (e5) at (6,-1.5){$\mathbf{q}_{5}$};
\vertex  (v4) at (3,0);
\vertex  (e6) at (3,-1.5){$\mathbf{q}_{4}$};

\draw[thick, postaction={decorate}, decoration={markings, mark=at position 0.5 with {\arrow{>}}}] 
    (e1) -- (v1);
\draw[thick, postaction={decorate}, decoration={markings, mark=at position 0.5 with {\arrow{>}}}] 
    (e2) -- (v1);
\draw[thick, postaction={decorate}, decoration={markings, mark=at position 0.5 with {\arrow{>}}}] 
    (v1) -- node [above, yshift=4pt] {$\mathbf{q}_{12}$} (v2);
\draw[thick, postaction={decorate}, decoration={markings, mark=at position 0.5 with {\arrow{<}}}] 
    (v2) -- (e3);
\draw[thick, postaction={decorate}, decoration={markings, mark=at position 0.5 with {\arrow{>}}}] 
    (v2) -- node [above, yshift=4pt] {$\mathbf{q}_{123}$} (v4);
\draw[thick, postaction={decorate}, decoration={markings, mark=at position 0.5 with {\arrow{>}}}] 
    (e6) -- (v4);
\draw[thick, postaction={decorate}, decoration={markings, mark=at position 0.5 with {\arrow{>}}}] 
    (v4) -- node [above, yshift=4pt] {$\mathbf{q}_{1234}$} (v3);
\draw[thick, postaction={decorate}, decoration={markings, mark=at position 0.5 with {\arrow{>}}}] 
    (v3) -- (e4);
\draw[thick, postaction={decorate}, decoration={markings, mark=at position 0.5 with {\arrow{<}}}] 
    (v3) -- (e5);
\end{feynman}

\end{tikzpicture}+ \text{(9 more channels)}
\end{equation}
From Eq.\eqref{eq:K5_sec_1_2_diag} and \eqref{eq:K5_sec3_diag}, we can see that the momentum structures that constitutes the kernel are given by the following,
\begin{align}\label{eq:ker5_1}
    T_5(\mathbf{q}_1,\mathbf{q}_2,\mathbf{q}_3,\mathbf{q}_4,\mathbf{q}_5)&= \{\{T_{4}\left(\ \mathbf{q}_{1},\mathbf{q}_{2},\mathbf{q}_{3},\mathbf{q}_{4} \right)\}
    \otimes S_2(\mathbf{q}_{1234},\mathbf{q}_{5}) + \text{cyclic perm.}\} \; + \nonumber\\[4pt]
    &\{ S_2(\mathbf{q}_1,\mathbf{q}_2) \otimes T_{3}\left(\mathbf{q}_3,\mathbf{q}_4,\mathbf{q}_5\right) \otimes{S_{2}\left( \mathbf{q}_{12},\mathbf{q}_{345} \right)}\;+\;\text{all channels}\}
\end{align}
where $T_4$ is given in Eq.\eqref{eq:ker4}. Note that $T_5$ as given in Eq.\eqref{eq:ker5_1} denotes a set of momentum structures. In the second line of Eq.\eqref{eq:ker5_1} we 
have to include all 10 channels which are given as,
\begin{equation}\label{eq:K5_channels}
    \begin{aligned}
        (1,2,3,4,5)\;,\;(1,3,2,4,5)\;,\;(1,4,3,2,5)\;,\;(1,5,3,4,2)\;,\;(2,3,1,4,5)\;,\\
        (2,4,3,1,5)\;,\;(2,5,3,4,1)\;,\;(3,4,1,2,5)\;,\;(3,5,1,2,4)\;,\;(4,5,1,2,3)\;.
    \end{aligned}
\end{equation}
The two sectors of $T_4$ corresponds to two diagrams in Eq.\eqref{eq:K5_sec_1_2_diag} as given in Eq.\eqref{eq:ker4}. Therefore, the first line of Eq.\eqref{eq:ker5} is the analogue of Eq.\eqref{eq:K5_sec_1_2_diag}. While the second line of Eq.\eqref{eq:ker5} is the analogue of Eq.\eqref{eq:K5_sec3_diag}. Now that we have the momentum structures, we can write the fifth order kernel as,
\begin{align}\label{eq:ker5}
K_5(\mathbf{q}_1,\mathbf{q}_2,\mathbf{q}_3,\mathbf{q}_4,\mathbf{q}_5)=\sum_{i=1}^{l(T_5)}e_i [T_5](i)\;,
\end{align}
where $l(T_5)=394$. 
Now, we impose all the EGI constraints on Eq.\eqref{eq:ker5} which comes as  a consequence of Eqs.\eqref{eq:Lo_const} and \eqref{eq:NLO_all_order}. These constraints correspond to taking external and internal momenta to be soft, respectively.

Taking external momenta soft gives us four sets of constraints as there are four momenta that can be taken to be soft at fifth order. These constitute the leading order (LO) constraints at fifth order. These constraints relate the soft limits of Eq.\eqref{eq:ker5} with lower order kernels which relates the $e_i$ coefficients to the coefficients of lower order kernel. 

Next, we have the next-to-leading order (NLO), NNLO and NNNLO constraints which comes from taking internal momenta to be soft. Taking the internal momenta soft breaks the diagram into two subdiagrams as was shown for the case of fourth order in Eq.\eqref{eq:K4_nlo_nnlo_diag}. For example, on taking $\mathbf{q}_{12}\rightarrow 0$ on Eq.\eqref{eq:ker5}, the following relation holds,
\begin{align}\label{eq:K5_nlo}
\lim_{\mathbf{q}_{12}\rightarrow 0}K_5(\mathbf{q}_1,\mathbf{q}_2,\mathbf{q}_3,\mathbf{q}_4,\mathbf{q}_5)&\supset \frac{\mathbf{q}_{12}\cdot\mathbf{q}_{345}}{\mathbf{q}_{12}^2}K_3(\mathbf{q}_{3},\mathbf{q}_{4},\mathbf{q}_{5})\int^\eta d\eta'f_{+}(\eta^{\prime})\left(\frac{D_{+}(\eta^{\prime})}{D_{+}(\eta)}\right)^{2}G_2(\mathbf{q}_1,\mathbf{q}_2,\eta')\nonumber\\
&\equiv i (\mathbf{k}\cdot\mathbf{d}^{(2)})\delta_{\mathbf{k}}^{(3)}\;.
\end{align}
From Eq.\eqref{eq:K5_nlo}, we can see that we can associate the limit $\mathbf{q}_{12}\rightarrow 0$ with the term $i (\mathbf{k}\cdot\mathbf{d}^{(2)})\delta_{\mathbf{k}}^{(3)}$ which may arise from Eq.\eqref{eq:delta_mom_space_trans}. Similarly, we have other beyond leading order constraints similar in structure as Eq.\eqref{eq:K5_nlo} but for different limits. The NLO, NNLO and NNNLO contributions arise from the following terms,
\begin{align}\label{eq:K5_nlo_nnlo_nnnlo}
(i\mathbf{k}\cdot\mathbf{d}^{(2)})\delta_{\mathbf{k}}^{(3)},\hspace{0.5cm}  (i\mathbf{k}\cdot\mathbf{d}^{(3)})\delta_{\mathbf{k}}^{(2)},\hspace{0.5cm} (i\mathbf{k}\cdot\mathbf{d}^{(4)})\delta_{\mathbf{k}}^{(1)}\;,
\end{align}
which corresponds to taking $\mathbf{q}_{12}\rightarrow 0$, $\mathbf{q}_{123}\rightarrow 0$ and $\mathbf{q}_{1234}\rightarrow 0$ respectively on Eq.\eqref{eq:ker5}. We can take further limits on the first and second term in Eq.\eqref{eq:K5_nlo_nnlo_nnnlo}. These can be denoted as follows,
\begin{align}\label{eq:K5_beyond_nlo}
(i\mathbf{k}\cdot\mathbf{d}^{(2)})(i\mathbf{k}\cdot\mathbf{d}^{(2)})\delta_{\mathbf{k}}^{(1)},\hspace{0.5cm} &(i\mathbf{k}\cdot\mathbf{d}^{(2)})(i\mathbf{k}\cdot\mathbf{d}^{(1)})\delta_{\mathbf{k}}^{(2)},\hspace{0.5cm} (i\mathbf{k}\cdot\mathbf{d}^{(2)})(i\mathbf{k}\cdot\mathbf{d}^{(1)})(i\mathbf{k}\cdot\mathbf{d}^{(1)})\delta_{\mathbf{k}}^{(1)}\nonumber\\
& (i\mathbf{k}\cdot\mathbf{d}^{(3)})(i\mathbf{k}\cdot\mathbf{d}^{(1)})\delta_{\mathbf{k}}^{(1)}\;,
\end{align}
where the first line in Eq.\eqref{eq:K5_beyond_nlo} comes from taking further limits on the first term in Eq.\eqref{eq:K5_nlo} and the second line in Eq.\eqref{eq:K5_beyond_nlo} comes from taking one more momenta soft on the second term in Eq.\eqref{eq:K5_nlo}. We impose these constraints on the full kernel at fifth order given in Eq.\eqref{eq:ker5}. Note that these NLO constraints also include constraints coming from imposing the absence of "spurious poles". We must state that finding all the "spurious poles" in the various limits is, in general, a very tediuos task. Therefore, in Sec.\,\ref{sec:no_wrong_poles}, we propose a new way to construct the kernels in which there are no "spurious poles" by construction.

All in all, imposing all the LO, NLO and NNLO constraints give us the kernel for biased tracers at fifth order. We have found that there are 44 independent coefficients left after imposing all the EGI constraints. The explicit calculations and expressions involved are not very illuminating. Hence, we have not included them here. The calculations are given in the Mathematica file provided with the manuscript.

Recently, it has been pointed out that the basis of operators for biased tracers at fifth order contains 29 independent coefficients (operators)\,\cite{Donath:2023sav}. As shown in\,\cite{Donath:2023sav}, the basis of operators at fifth order contains certain "time non-local" structures which cannot be written in terms of local operators made of dark matter fields. We have shown that all 29 operators can be written in terms of the 44 operators obtained from bootstrap. This implies that bootstrap approach is able to capture the "time non-local" structures. However, at fifth order as well, we see that bootstrap method gives more number of independent coefficients as compared to direct computations. This discrepancy is of the same nature as we found at fourth order in Sec.\,\ref{sec:ker_4_boot} and we have discussed the resolution in Appendix.\,\ref{app:5_order}. 
\vspace{5pt}\\
\underline{\textbf{Matter kernel}}
\vspace{5pt}\\
Now we impose mass and momentum constraints as given in Eq.\eqref{eq:mass_mom_conserv} further on the fifth order kernel. This leaves us with 21 free coefficients, see Appendix.\,\ref{app:5_order} for details. Note that this is a disagreement with the Lagrangian bootstrap for matter\,\cite{Marinucci:2024add}, where it has been pointed out that the matter kernel at fifth order has 23 independent coefficients. However, upon imposing mass and momentum constraints, we have found that 2 of the coefficients provided in\,\cite{Marinucci:2024add} are related to other coefficients which brings the number down to 21. This agrees with our result for matter kernel at fifth order.
\section{Eulerian Bootstrap recipe with no ``spurious poles"}\label{sec:no_wrong_poles}
As stated before, one of the major challenges in implementing the Eulerian bootstrap is to identify all possible "spurious poles" in the NLO constraints and put them to zero. The number of such structure grows very quickly as we move to higher orders. This is one of the major obstacles in implementing Eulerian bootstrap to higher orders. Therefore, in this section, we provide an alternative method for the construction of the kernel such that it does not contain any "spurious poles". 

Let us restate the type of structures we are referring to. For example, at third order we have one such structure given as,
\begin{align}\label{eq:wrong_3rd_new}
    \frac{(\mathbf{q}_{12}\cdot\mathbf{q}_3)^2}{\mathbf{q}_{12}^2\mathbf{q}_{3}^2}\;,
\end{align}
which does not occur on the R.H.S of NLO constraints as given in Eq.\eqref{eq:NLO_all_order} and must be removed.
Identifying all such poles is in general tedious and difficult. This holds true especially beyond third order, when the number of NLO limits increase rapidly. This motivated us to construct the kernel in a slightly different manner, such that all these ``spurious" momentum structures are absent by construction. This is an advantageous method, as it reduces the number of momentum structures in the kernel significantly even before imposing EGI constraints. This also reduces the computation cost and enhances the utility of Eulerian bootstrap beyond third order. We describe the procedure below in a bit more detail.
\subsection{Construction of the kernel}
Let us define two sets as follows ,
\begin{align}\label{eq:S2_prime}
    S_2'\left( \mathbf{q}_{1}, \mathbf{q}_2 \right) &= \{{  \alpha_{+}\left( \mathbf{q}_{1}, \mathbf{q}_2\right), \; \alpha_{-}\left( \mathbf{q}_{1}, \mathbf{q}_2 \right), \; \beta(\mathbf{q}_{1},\mathbf{q}_2)}\}\;,\nonumber\\[5pt]
    T_2'\left( \mathbf{q}_{1}, \mathbf{q}_2 \right) &=S_2'\left( \mathbf{q}_{1}, \mathbf{q}_2 \right)\cup\{1\}\;.
\end{align}
Note that $ T_2'\left( \mathbf{q}_{1}, \mathbf{q}_2 \right)$ is same as $S_2\left( \mathbf{q}_{1}, \mathbf{q}_2 \right)$ as given in Eq.\eqref{eq:S2}, but we define it separately to avoid confusion. Given $T_2'\left( \mathbf{q}_{1}, \mathbf{q}_2 \right)$ we can now define,
\begin{align}\label{eq:K2_prime}
    K_2'\left( \mathbf{q}_{1}, \mathbf{q}_2 \right)=\sum_{i=1}^{l(T_2')}b_i'T_2'(i)\;.
\end{align}
The explicit form being,
\begin{align}
    K_2^{\prime}(\mathbf{q}_1,\mathbf{q}_2)=b_1+b_2\alpha(\mathbf{q}_1,\mathbf{q}_2)+b_3 \beta(\mathbf{q}_1,\mathbf{q}_2)\;.
\end{align}
It should be noted that this is exactly the same as $K_2(\mathbf{q}_1,\mathbf{q}_2)$ given in Eq.\eqref{eq:K2_ind_coeff}. This is precisely due to the lack of any NLO constraints and hence any "spurious structures" at second order. We expect to see a difference starting at third order.

Let us define another set $M_2\left( \mathbf{q}_{1}, \mathbf{q}_2 \right)$ as the set of all momentum structures that we get after imposing symmetrization and mass and momentum constraints, given in Eq.\eqref{eq:mass_mom_conserv}, on $K_2'\left( \mathbf{q}_{1}, \mathbf{q}_2\right)$ as given in Eq.\eqref{eq:K2_prime}. Now we have all the building blocks necessary to construct higher order kernels. At third order, we define,
\begin{align}\label{eq:T3_prime}
T_3'(\mathbf{q}_1,\mathbf{q}_2,\mathbf{q}_3)=    M_2(\mathbf{q}_1,\mathbf{q}_2) \otimes S_2'(\mathbf{q}_{12},\mathbf{q}_3)\cup T_2'(\mathbf{q}_1,\mathbf{q}_2)\;,
\end{align}
which gives the kernel at third order as,
\begin{align}\label{eq:K3_prime}
     K_3'\left( \mathbf{q}_{1}, \mathbf{q}_2, \mathbf{q}_3 \right)=\sum_{i=1}^{l(T_3')}c_i'T_3'(i)+\text{cyc.}(\mathbf{q}_1,\mathbf{q}_2,\mathbf{q}_3)\;,
\end{align}
where $l(T_3')=9$, which is starkly different from Eq.\eqref{eq:K3} having 12 coefficients. One of the missing "spurious structure" in this new definition, is of the form given in Eq.\eqref{eq:wrong_3rd_new}. For the interested reader, the explicit forms of the kernels have been provided in the attached Mathematica file.

At fourth order we need $M_3\left( \mathbf{q}_{1}, \mathbf{q}_2, \mathbf{q}_3 \right)$ which we define as the set of all momentum structures that we get after imposing symmetrization and mass and momentum constraints, given in Eq.\eqref{eq:mass_mom_conserv}, on $K_3'\left( \mathbf{q}_{1}, \mathbf{q}_2, \mathbf{q}_3\right)$ as given in Eq.\eqref{eq:K3_prime}. This way we can define the fourth order kernel as,
\begin{align}\label{eq:T4_K4_prime}
T_4'(\mathbf{q}_1,\mathbf{q}_2,\mathbf{q}_3,\mathbf{q}_4)= \{M_3\left( \mathbf{q}_{1}, \mathbf{q}_2, \mathbf{q}_3 \right)\otimes S_2'(\mathbf{q}_{123},\mathbf{q}_{4})+&T_3'\left( \mathbf{q}_{1}, \mathbf{q}_2, \mathbf{q}_3 \right)\}+\text{cyc.}(\mathbf{q}_1,\mathbf{q}_2,\mathbf{q}_3,\mathbf{q}_4) \nonumber\\
    &\cup \nonumber \\
    \{M_2(\mathbf{q}_1,\mathbf{q}_2) \otimes M_2(\mathbf{q}_3,\mathbf{q}_4) \otimes S_2'(\mathbf{q}_{12},\mathbf{q}_{34})&\cup T_2'(\mathbf{q}_1,\mathbf{q}_2) \otimes T_2'(\mathbf{q}_3,\mathbf{q}_4)\} \nonumber\\
    +& (2 \leftrightarrow 3) + (2 \leftrightarrow 4)\;,\nonumber\\
     K_4'\left( \mathbf{q}_{1}, \mathbf{q}_2, \mathbf{q}_3, \mathbf{q}_4  \right)=\sum_{i=1}^{l(T_4')}c_i'T_4'(i)\;,
\end{align}
and similarly at fifth order,
\begin{align}\label{eq:T5_K5_prime}
T_5'(\mathbf{q}_1,\mathbf{q}_2,\mathbf{q}_3,\mathbf{q}_4,\mathbf{q}_5)= \{\{M_4\left( \mathbf{q}_{1}, \mathbf{q}_2, \mathbf{q}_3, \mathbf{q}_4 \right) \}\otimes S_2'(\mathbf{q}_{1234},\mathbf{q}_{5})+&T_4'\left( \mathbf{q}_{1}, \mathbf{q}_2, \mathbf{q}_3, \mathbf{q}_4 \right)\}+\text{cyc. perm} \nonumber\\
    &\cup \nonumber \\
    \{M_2(\mathbf{q}_1,\mathbf{q}_2) \otimes M_3(\mathbf{q}_3,\mathbf{q}_4,\mathbf{q}_5) \otimes S_2'(\mathbf{q}_{12},\mathbf{q}_{345})&\cup T_2'(\mathbf{q}_1,\mathbf{q}_2) \otimes T_3'(\mathbf{q}_3,\mathbf{q}_4,\mathbf{q}_5)\} \nonumber\\
    +& \text{all channels in Eq.\eqref{eq:K5_channels}}\;,\nonumber\\
     K_5'\left( \mathbf{q}_{1}, \mathbf{q}_2, \mathbf{q}_3, \mathbf{q}_4, \mathbf{q}_5  \right)=\sum_{i=1}^{l(T_5')}c_i'T_5'(i)\;.
\end{align}
We can perform this procedure recursively to get higher-order kernels. In the next section, we compare the results obtained with the method given here and the conventional approach as given in Sec.\,\ref{sec:Eulerian_boot}.
\subsection{Comparison with the conventional approach}
We have verified the absence of spurious structures in this construction upto fifth order, and believe it to hold true for higher orders. After imposing EGI, we have explicitly verified that this way of defining the kernels gives the same results as we have obtained in Secs.\,\ref{ssec:ker_3_const}, \ref{sec:ker_4_boot} and \ref{sec:fifth_order}. This construction streamlines the implementation of the Eulerian bootstrap because now we only need to impose the LO and NLO constraints as given in Eq.\eqref{eq:Lo_const} and Eq.\eqref{eq:NLO_all_order}. All the "spurious pole" constraints are identically satisfied by the kernels defined in the manner given in Eq.\eqref{eq:T4_K4_prime} and Eq.\eqref{eq:T5_K5_prime}. This method also extends the practical application of Eulerian bootstrap at higher orders. 

Recently, EGI has been employed in Lagrangian space to bootstrap matter displacement field. Exploiting the fact that tracers are made of dark matter field, we have generalised it to bootstrap tracer overdensity. In the next section, we describe this method of Lagrangian bootstrap for tracers in detail. We have found that the results obtained using Lagrangian boostrap agree with the Eulerian results obtained in Sec.\,\ref{sec:Eulerian_boot} for the kernel at fourth and fifth order. 

\section{Lagrangian Bootstrap for tracers}\label{sec:lag_boot}
In this section, we discuss a bootstrap method for tracers in Lagrangian space, generalizing the method recently developed for dark matter\,\cite{Marinucci:2024add}. Before moving on to tracers, let us quickly state the bootstrap approach adopted for dark matter overdensity in Lagrangian space.
The Eulerian coordinate $\mathbf{x}$ is related to the Lagrangian coordinate $\mathbf{q}$ as follows\,\cite{Mirbabayi:2014zca},
\begin{align}\label{eq:lag_coor_trans}
    \mathbf{x}(\mathbf{q},\eta)=\mathbf{q}+\boldsymbol{\psi}(\mathbf{q},\eta)
\end{align}
where $\boldsymbol{\psi}(\mathbf{q},\eta)$ is the displacement field such that $\boldsymbol{\psi}(\mathbf{q},0)=0$. The objective of the Lagrangian bootstrap for dark matter is to obtain the most general form of the displacement field that is consistent with constraints from EGI and conservation of mass and momentum. Matter overdensity can then be obtained in terms of the displacement field as follows\,\cite{Marinucci:2024add},
\begin{align}\label{eq:matter_disp_rel}
    \delta=\frac{1}{\mathcal{J}}-1
\end{align}
where $\mathcal{J}=\abs{\frac{\partial x_i}{\partial q_j}}$ is the Jacobian of the coordinate transformation given in Eq.\eqref{eq:lag_coor_trans} and contains information about the displacement field. 

\begin{figure}
    \centering
    \includegraphics[scale=0.4]{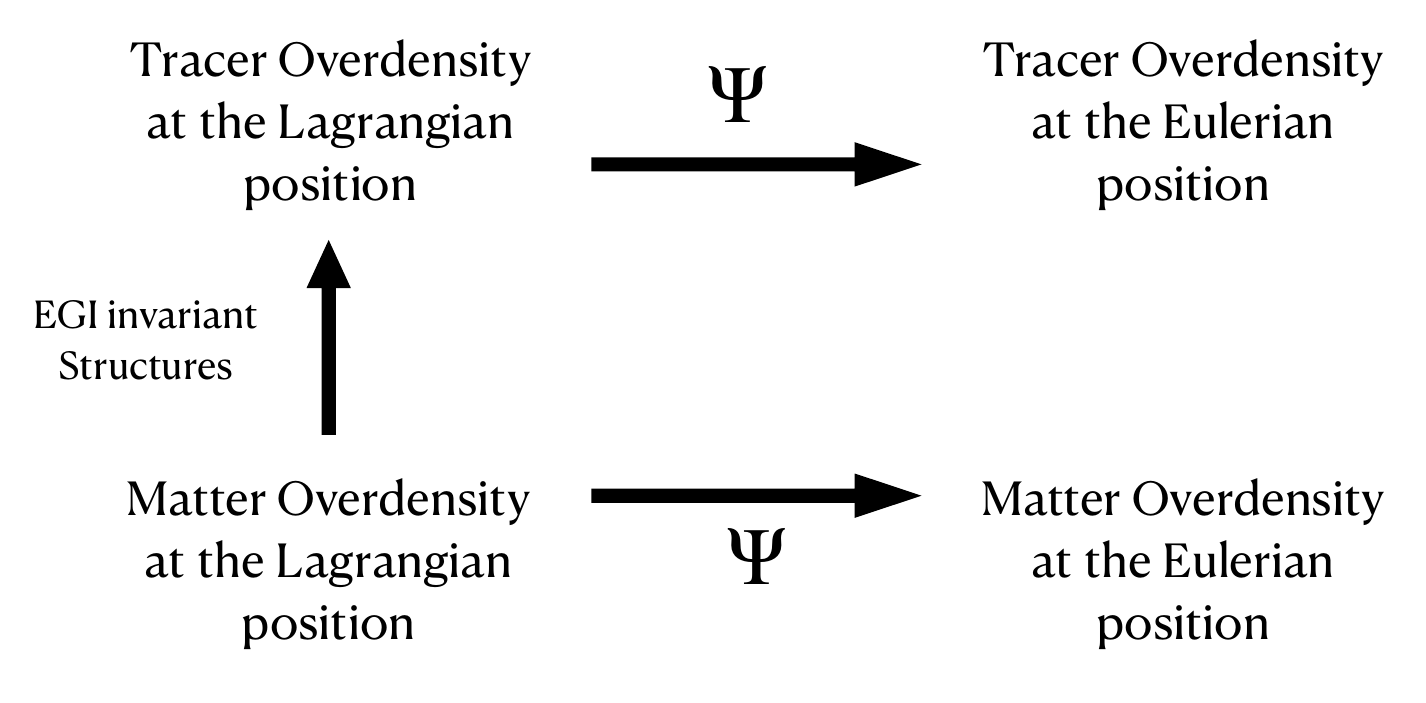}
    \caption{Schematic diagram showing the relationship between Lagrangian and Eulerian space for tracer and matter overdensity through the displacement field. We have written the tracer overdensity in Lagrangian space (top left) as function of EGI invariant structures made out of dark matter fields (bottom left). Using the displacement field for matter we displace the underlying dark matter fields and get the tracer overdensity in the Eulerian space (top right).}
    \label{fig:diag_lag_euler}
\end{figure}

For tracers, since number density is not a conserved quantity, displacement field is not well defined. However, one can reasonably assume that tracer overdensity is a function
of dark matter overdensity. The basic idea relating tracer and matter overdensities at the Lagrangian and Eulerian position is nicely depicted in Fig.\,\ref{fig:diag_lag_euler} which we explain now. For the purpose of bootstrap, we define the
overdensity for tracers at the Lagrangian position ($\mathbf{q}$) in terms of all the EGI invariant scalars made
of dark matter fields at that point. To get the same at the Eulerian position ($\mathbf{x}$), we simply displace the underlying matter fields using $\boldsymbol{\psi}$. One notable difference between the bootstrap approach and writing a bias expansion is the verticle arrow shown on the left side of Fig.\,\ref{fig:diag_lag_euler}. In the conventional approach, that arrow corresponds to writing locally observable scalars while in the bootstrap we allow for more general EGI invariant scalars.

Let us mathematically elaborate the bootstrap procedure mentioned in the last paragraph. For that, we have to start by defining the relation between Eulerian and Lagrangian overdensities for matter. Non perturbatively, the Eulerian and the Lagrangian expansion are equal and therefore we can write\,\cite{Mirbabayi:2014zca},
\begin{align}\label{eq:lag_vs_euler_matter}
    \delta^{\text{E}}(\mathbf{x},\eta)=\delta^{\text{L}}(\mathbf{q},\eta)=\delta^{\text{L}}(\mathbf{x}-\boldsymbol{\psi}(\mathbf{q},\eta),\eta)\;,
\end{align}
where $\delta^{\text{E}}$ and $ \delta^{\text{L}}$ have different IR properties in momentum space. When expanded in Eulerian coordinates, matter overdensity contains IR divergent terms\,\cite{Mirbabayi:2014zca}. However, the same quantity in Lagrangian space does not contain any IR divergent terms. One might think that Eq.\eqref{eq:lag_vs_euler_matter} is inconsistent since the L.H.S has IR divergent terms while the R.H.S does not. That is not true, since one can use Eq.\eqref{eq:lag_coor_trans} recursively in Eq.\eqref{eq:lag_vs_euler_matter} and Taylor expand around $\mathbf{x}$. This brings back the IR divergent terms through the displacement field.  
With the assumption that tracers are made of dark matter fields, a relation like Eq.\eqref{eq:lag_vs_euler_matter} holds for them as well allowing us to write the tracer overdensity in Eulerian space as,
\begin{align}\label{eq:lag_vs_euler_tracer}
\delta^{\text{E}}_t(\mathbf{x},\eta)=\delta^{\text{L}}_t(\mathbf{q},\eta)=\delta^{\text{L}}_t(\mathbf{x}-\boldsymbol{\psi}(\mathbf{q},\eta),\eta)\;.
\end{align}
where the subscript 't' denotes that $\delta_t$ is the tracer overdensity. Now, we can use Eq.\eqref{eq:lag_coor_trans} in Eq.\eqref{eq:lag_vs_euler_tracer} recursively and expand around $\mathbf{x}$ to get the following,
\begin{align}\label{eq:tracer_lag_expan} \delta_{\text{t}}^{\text{E}}=\delta_{\text{t}}^{\text{L}}-\partial_i\delta_{\text{t}}^{\text{L}}\psi^i+\partial_i\delta_{\text{t}}^{\text{L}}\partial_j\psi^i\psi^j+\frac{1}{2}\partial_i\partial_j\delta_{\text{t}}^{\text{L}}\psi^i\psi^j+..
\end{align}
where we have only kept terms relevant up to third order. From Eq.\eqref{eq:tracer_lag_expan}, we can see that $\delta_{\text{t}}^{\text{E}}$ depends on two quantities: the matter displacement field ($\boldsymbol{\psi}$) and the Lagrangian tracer overdensity $\delta_{\text{t}}^{\text{L}}$. The form of $\boldsymbol{\psi}$ is already known\,\cite{Marinucci:2024add} and we can use it directly in Eq.\eqref{eq:tracer_lag_expan}, whereas for tracer overdensity $\delta_{\text{t}}^{\text{L}}$, we write down all possible EGI invariant scalars made from dark matter fields. A systematic procedure to do so has already been devised in\,\cite{Marinucci:2024add} in the context of matter. One striking difference is that for matter one requires mass and momentum conservation to hold whereas for tracers we can relax this condition.

Let us explain this relaxation condition with the help of an example. At fourth order, one of the scalars considered in the context of matter has the following form,
\begin{align}\label{eq:lag_sca_4}
     \phi_{ii}^{(4)}&=\frac{1}{3!}\epsilon^{ikm}\epsilon^{jln}\phi^{(2)}_{,ij}\phi^{(1)}_{,kl}\phi^{(1)}_{,mn}
\end{align}
where $\phi^{(n)}_{,ij}\equiv\partial_i\partial_j\phi^{(n)}$, $\phi^{(1)}_{,ii}=-\delta^{(1)}$ and $\phi^{(2)}_{,ii}= 1/2(\phi^{(1)}_{,ii}\phi^{(1)}_{,ll}-\phi^{(1)}_{,il}\phi^{(1)}_{,il})$. The products of Levi-Civita symbols in Eq.\eqref{eq:lag_sca_4} are present to ensure mass and momentum conservation, which also implies that we cannot have products of more than 3 fields. So for example, a scalar like $\phi^{(1)}_{,ij}\phi^{(1)}_{,jk}\phi^{(1)}_{,kl}\phi^{(1)}_{,li}$ is not allowed. Even a scalar like $\phi^{(2)}_{,ij}\phi^{(1)}_{,jk}\phi^{(1)}_{,ki}$ cannot appear with independent coefficient but only in the linear combination given in Eq.\eqref{eq:lag_sca_4}. 

\vspace{5pt}
For tracers, we drop these requirements coming from mass and momentum conservation and generalize the construction in the following manner,
\begin{itemize}
    \item Consider all possible contractions with independent coefficients and not just the linear combination given in Eq.\eqref{eq:lag_sca_4}.
    \item Allow for scalars with products of more than 3 fields in the contraction.
\end{itemize}
These are the two criteria we follow to write scalars that constitute $\delta_{\text{t}}^{\text{L}}$ at each order. Combining $\delta_{\text{t}}^{\text{L}}$ with the displacement field $\boldsymbol{\psi}$ in the manner given in Eq.\eqref{eq:tracer_lag_expan} gives us $\delta_{\text{t}}^{\text{E}}$ which we expect to be equivalent to the overdensity that we obtain using the Eulerian bootstrap methods in Sec.\,\ref{sec:Eulerian_boot}. We have verified this to hold true up to fifth order by obtaining an invertible map between the independent operators occuring in both\footnote{The explicit mapping is done for the EdS cosmology for matter kernel.}.

\vspace{4pt}

Few comments on the advantages of Lagrangian bootstrap are in order. 
\begin{itemize}
    \item Firstly, it can be checked that writing the Eulerian tracer overdensity as given in Eq.\eqref{eq:tracer_lag_expan}, automatically satisfies all the EGI constraints discussed in Sec.\,\ref{sec:eulerian_boot}. This includes all the LO and NLO constraints.
    \item  Secondly, one can check that there are no "spurious poles" in the Eulerian overdensity as given in Eq.\eqref{eq:tracer_lag_expan}. Therefore, we find that Eq.\eqref{eq:tracer_lag_expan} is a nice way to arrange the terms in Eulerian overdensity such that EGI is explicit.
\end{itemize}

Let us state that both the Eulerian and Lagrangian approaches give 17 and 44 independent coefficients at fourth and fifth order respectively for tracers. As discussed earlier in Sec.\,\ref{sec:Eulerian_boot}, these numbers differ from what we get from a bias expansion. In the next section, we discuss the reason for the discrepancy and provide a resolution for the same.
\subsection{Beyond EGI: Matching with the bias expansion}\label{ssec:lag_cosmo_fix}
In this section, we provide a resolution to the apparent discrepancy in the number of independent coefficients obtained from bootstrap approach (Sec.\,\ref{sec:Eulerian_boot} and \ref{sec:lag_boot}) and the one we get from the usual bias expansion as mentioned in the last paragraph. This discrepancy occurs for the first time at fourth order, so let us start from there.

Expanding Eq.\eqref{eq:tracer_lag_expan} up to fourth order, we find that tracer overdensity in Eulerian space contains 17 independent coefficients. This does not agree with the number obtained in the bias expansion which is 15\,\cite{DAmico:2022ukl}. Out of the 17 coefficients obtained through bootstrap in accordance with Eq.\eqref{eq:tracer_lag_expan}, 10 come from the Lagrangian overdensity $\delta_{\text{t}}^{(4),\text{L}}$ whereas 7 come from lower orders through the displacement field. Note that we have exhausted all the constraints coming from EGI, hence the number 17 cannot be further reduced.

The origin of this discrepancy stems from a fundamental difference in operator construction between the bootstrap approach and conventional bias expansion when writing $\delta_{\text{t}}^{\text{L}}$. Let us elaborate with an example. In the Lagrangian overdensity $\delta_{\text{t}}^{(4),\text{L}}$, out of 10 there are three (EGI invariant) scalars which have following form\,\cite{Marinucci:2024add}\footnote{Here $\varphi^{(3)}_{1,ij}$, $\varphi^{(3)}_{2,ij}$ are given as\,\cite{Marinucci:2024add},
\begin{align}\label{eq:lag_matter_scalar_3}
    \phi_{1,ii}^{(3)}&=\frac{1}{3!}\epsilon^{ikm}\epsilon^{jln}\phi^{(1)}_{,ij}\phi^{(1)}_{,kl}\phi^{(1)}_{,mn}\nonumber\\
     \phi_{2,ii}^{(3)}&=\frac{1}{2!}\epsilon^{ijk}\epsilon^{lmk}\phi^{(2)}_{,il}\phi^{(1)}_{,jm}
\end{align}
and 
\begin{equation}\label{eq:v_exp_lag}
v^{(3)i}_{,kk}=\epsilon^{iln}\varphi^{(2)}_{,lj}\varphi^{(1)}_{,jn}\;.
\end{equation}},
\begin{align}\label{eq:lag_scalars_4}
\varphi^{(3)}_{1,ij}\varphi^{(1)}_{ij}\;,\;\varphi^{(3)}_{2,ij}\varphi^{(1)}_{ij}\;,\;\varphi^{(1)}_{lj}\left(\epsilon^{jmn}v^{(3)n}_{,lm}+\epsilon^{lmn}v^{(3)n}_{,jm}\right)\;.
\end{align}
In the bootstrap approach, we assign independent coefficients (say $\alpha_1,\alpha_2$ and $\alpha_3$) to all the 3 structures given in Eq.\eqref{eq:lag_scalars_4} when writing $\delta_{\text{t}}^{\text{L},(4)}$. However, in a bias expansion\,\cite{Mirbabayi:2014zca}, such terms appear in a particular linear combination as a single scalar ($\psi_{i,j}^{(3)}\psi_{i,j}^{(1)}$) to which we assign a single bias parameter (say $\alpha$). For Einstein deSitter cosmology, we have the following,
\begin{align}\label{eq:lag_sca_4_combined}
   \psi_{i,j}^{(3)}\psi_{i,j}^{(1)}= -\frac{1}{3}\varphi^{(3)}_{1,ij}\varphi^{(1)}_{ij}+\frac{10}{21}\varphi^{(3)}_{2,ij}\varphi^{(1)}_{ij}-\frac{1}{7}\varphi^{(1)}_{lj}\left(\epsilon^{jmn}v^{(3)n}_{,lm}+\epsilon^{lmn}v^{(3)n}_{,jm}\right)\;,
\end{align}
where the explicit values of the coefficients given in Eq.\eqref{eq:lag_sca_4_combined} depend on the cosmology. This implies the following linear relations on $\alpha_i$'s,
\begin{align}\label{eq:ratio}
    \frac{\alpha_1}{\alpha_2}=-\frac{7}{10}\hspace{0.5cm},\hspace{0.5cm}  \frac{\alpha_1}{\alpha_3}=\frac{7}{3}\;.
\end{align}
With the relations given in Eq.\eqref{eq:ratio}, the independent number of coefficients reduces from 17 to 15. This resolves the discrepancy which we have verified by obtaining an invertible map between bootstrap basis and the basis we get from bias expansion. Since the relations given in Eq.\eqref{eq:ratio} depend on cosmology, one should not think of them as constraint relations. In fact, these relations are not induced by some symmetry principle, such as EGI, but by imposing that bias expansion is the most general description for defining tracer overdensity. For observational purposes, we can still assume the 3 parameters to be independent and measure them separately from data. Then the ratios given in Eq.\eqref{eq:ratio} may serve as probes to determine the underlying cosmology.

At fifth order, we expand Eq.\eqref{eq:tracer_lag_expan} up to fifth order which gives 44 independent coefficients for tracer overdensity in Eulerian space, see Appendix.\,\ref{app:5_order}. Note that this already agrees with the result from Eulerian bootstrap discussed in Sec.\,\ref{sec:fifth_order}. But in order to match with the bias expansion, we can further reduce this number by forming linear combinations similar to what we did at fourth order. After combination, we obtain 29 independent operators which are related with the time non-local basis obtained from bias expansion\,\cite{Donath:2023sav}. We do not state here the relations among momentum structures at fifth order for brevity. The details of which structures
need to be combined and in what manner is provided in the Appendix.\,\ref{app:5_order}. 
\vspace{5pt} \\
\textbf{\underline{Mass and momentum conservation}} \vspace{5pt} \\
Imposing mass and momentum conservation further on the tracer kernel leaves us with 8 and 21 independent coefficients at fourth and fifth order respectively. This agrees with what we got from Eulerian bootstrap in Sec.\,\ref{sec:Eulerian_boot}. However, at fifth order, it still is in disagreement with the Lagrangian bootstrap for matter\,\cite{Marinucci:2024add}. As we have already stated in Sec.\,\ref{sec:fifth_order}, the matter overdensity as obtained from Lagrangian bootstrap in\,\cite{Marinucci:2024add} does not satisfy momentum conservation. Instead, upon imposing momentum conservation 2 of the 23 coefficients, as given in\,\cite{Marinucci:2024add}, get related with the other coefficients which gives us 21 independent coefficients at fifth order. This agrees with our result obtained in this work. The details of the calculation is provided in the Mathematica file attached with the manuscript, also see Appendix.\,\ref{app:5_order}.
\section{Conclusion}\label{sec:conc}
Understanding the distribution of dark matter across the Universe is a fundamental objective in cosmology. However, directly observing dark matter remains highly challenging, as it can only be probed at large scales through gravitational lensing effects. Consequently, one has to rely on observables derived from baryonic matter, which serves as tracers for the underlying dark matter distribution. Given the complexity of modeling the dynamics of tracers, a common approach is to write a bias expansion, which expresses the overdensity of tracers in terms of the underlying dark matter fields. While bias expansion make no assumptions about tracer dynamics, they remain cosmology-dependent, requiring a fixed dark matter model. However, one can employ bootstrap methods, where symmetries play the central role, to get the tracer overdensity that is completely model independent. Such methods for obtaining Eulerian tracer overdensity were developed in\,\cite{DAmico:2021rdb} where Extended Galilean symmetry was used to bootstrap tracer and matter overdensity up to third order.

We use the techniques of Eulerian bootstrap to derive tracer and matter overdensity at fourth and fifth order which is discussed in Sec.\,\ref{sec:ker_4_boot} and Sec.\,\ref{sec:fifth_order} respectively. This is relevant for analyzing higher point or higher loop statistics of tracers and matter overdensities in a model independent way. We summarize the key findings of our work below,
\begin{itemize}
    \item At fourth order, the tracer overdensity field is fully determined by EGI up to 17 independent coefficients which is discussed in Sec.\,\ref{sec:ker_4_boot}. This count differs from the 15 independent coefficients obtained in a standard bias expansion, a discrepancy whose origin and resolution we have addressed in Sec.\,\ref{ssec:lag_cosmo_fix}. One interesting observation is that out of 17, there are 3 coefficients whose ratios admit universal values independent of the properties of specific tracer. We have shown this explicitly in Eq.\eqref{eq:ratio}. These ratios only depend on the underlying dynamics of dark matter and in principle, may serve as direct probes of the cosmology purely through the observation of tracer field.

    \item At fifth order, implementing all the EGI constraints leaves us with 44 independent coefficients as shown in Sec.\,\ref{sec:fifth_order}. Again, this count is different from what we obtain in a bias expansion where the number is 29. Interestingly, at fifth order, the bias expansion contains time non-local operators\,\cite{Donath:2023sav}. We have found that all the 29 operators are writable in terms of the 44 obtained using bootstrap implying that time non-local effects are captured by the bootstrap approach.

    \item Going beyond fourth order, we found that implementing EGI in Eulerian space becomes tedious. The primary difficulty stems from the need to systematically identify and eliminate all "spurious poles" in the calculation. Therefore, we propose a new way of constructing the kernels such that there are no "spurious poles" present. This makes the calculation more efficient and still gives the correct results.

    \item There also exists Lagrangian bootstrap method which was originally developed for matter\,\cite{Marinucci:2024add}. We have generalized it for tracers in Sec.\ref{sec:lag_boot} and found agreement with the results obtained through Eulerian method. From the Lagrangian perspective, we have argued how certain momentum structures can be combined to give 15 and 29 unfixed coefficients at fourth and fifth order, respectively. This makes the bootstrap basis in direct agreement with the basis obtained from brute force computation of bias expansion at fourth and fifth order.
\end{itemize} 

We have also obtained matter overdensity by imposing mass and momentum conservation, as given in Eq.\eqref{eq:mass_mom_conserv}, further on tracer overdensities which gives 8 and 21 unfixed coefficients at fourth and fifth order respectively. This agrees with the Lagrangian results obtained in\,\cite{Marinucci:2024add}. Except that at fifth order, \cite{Marinucci:2024add} reports 23 unfixed coefficients for matter overdensity. However, when momentum conservation is implemented properly, the coefficients reduce to 21, see Appendix.\,\ref{app:5_order} for details. Below we list some directions which we believe is worthy for future exploration: 
\begin{itemize}
    \item The bootstrap approach should be extended to include higher derivative operators which are not part of the current study. 
   
    \item The bootstrap basis that we obtain at fifth order contains time non-local operators. This means that EGI constraints alone gives us a basis which spans a larger space that what is spanned by local operators. However, it would be interesting to find out what further conditions must be imposed on the kernels so as to get the local basis.

    \item It would also be interesting to think how one can employ bootstrap methods for tracers in Lagrangian space independently, without referring to matter.

    \item The actual observations from various surveys are done in redshift space. Therefore, it is important to generalize the bootstrap methods for redshift space for a model independent analysis of data.
    
\end{itemize}
\acknowledgments
AA acknowledges the support from a Senior Research Fellowship, granted by the Human
Resource Development Group, Council of Scientific and Industrial Research, Government
of India. AB acknowledges support from Science and Engineering Research Board (SERB)
India via the Startup Research Grant SRG/2023/000378. AA would like to thank M. Marinucci for useful email exchange. The authors would also like to acknowledge their debt to the people of
India for their steady support of research in basic sciences.
\appendix
\section{Details of Third order calculations in Eulerian space}\label{app:third_order}

This Appendix details the calculation for bootstrapping the tracer and matter kernel at third order. We will be reproducing the calculation given in\;\cite{DAmico:2021rdb}, although using a different basis, which is given in \eqref{eq:alter_basis}.
\subsection{Setup of LO and NLO constraints}
Having constructed the kernel for the second and third order as given in Eq.\eqref{eq:K2} and \eqref{eq:K3}, we can now state the EGI constraints which are to be satisfied by $K_2$ and $K_3$. Imposing the constraints given in Eq.\eqref{eq:Lo_const} and Eq.\eqref{eq:NLO_all_order}, will relate certain coefficients among each other or with the coefficients at lower order. 

For the third order, there are two constraints from Eq.\eqref{eq:Lo_const}, arising from taking one and two momenta soft, respectively. They are given as,
\begin{align}\label{eq:lo_3rd}
  &\lim_{\mathbf{q}_3\rightarrow 0} K_3(\mathbf{q}_1,\mathbf{q}_2,\mathbf{q}_3)=\frac{\mathbf{q}_3\cdot(\mathbf{q}_2+\mathbf{q}_1)}{\mathbf{q}_3^2}K_2(\mathbf{q}_2,\mathbf{q}_1)+\mathcal{O}(q^0)\nonumber\\
  &\lim_{\mathbf{q}_3\rightarrow 0,\mathbf{q}_2\rightarrow 0}  K_3(\mathbf{q}_1,\mathbf{q}_2,\mathbf{q}_3)=\frac{\mathbf{q}_3\cdot\mathbf{q}_1}{\mathbf{q}_3^2}\frac{\mathbf{q}_2\cdot\mathbf{q}_1}{\mathbf{q}_2^2}K_1(\mathbf{q}_1)+\mathcal{O}(q^{-1})\;,
\end{align}
with the corresponding diagrammatic representation,
\begin{equation}\label{eq:K3_LO}
\lim_{\mathbf{q}_3\rightarrow 0}
\begin{tikzpicture}[scale=0.5, 
            baseline={([xshift=-5cm,yshift=-\the\dimexpr\fontdimen22\textfont2\relax]
                current bounding box.center)},
    ] 
\begin{feynman}
\vertex (b) at (0,0) ;
\vertex [above left=of b] (i1) {$\mathbf{q}_1$};
\vertex (o1) at (2,0);
\vertex [below left=1 and 1 of b] (i2) {${\mathbf{q}_2}$};
\vertex [above right=of o1] (v1) {$\mathbf{q}_{123}$};
\vertex [below right=1 and 1 of o1] (v2) {${\mathbf{q}_3}$};

\draw[thick, postaction={decorate}, decoration={markings, mark=at position 0.5 with {\arrow{>}}}] 
    (b) -- node [above, yshift=2pt] {$\mathbf{q}_{12}$} (o1);
\draw[thick, postaction={decorate}, decoration={markings, mark=at position 0.5 with {\arrow{>}}}] 
    (i1) -- (b);
\draw[thick, postaction={decorate}, decoration={markings, mark=at position 0.5 with {\arrow{>}}}] 
    (i2) -- (b);
\draw[thick, postaction={decorate}, decoration={markings, mark=at position 0.5 with {\arrow{>}}}] 
    (o1) -- (v1);
\draw[thick, postaction={decorate}, decoration={markings, mark=at position 0.5 with {\arrow{<}}}] 
    (o1) -- (v2);
\end{feynman}
\end{tikzpicture}
=\frac{\mathbf{q}_3\cdot(\mathbf{q}_2+\mathbf{q}_1)}{\mathbf{q}_3^2}
\left\{\begin{tikzpicture}[scale=1.0, 
            baseline={([xshift=-5cm,yshift=-\the\dimexpr\fontdimen22\textfont2\relax]
                current bounding box.center)},
    ] 
\begin{feynman}
\vertex (b) ;
\vertex [above left=of b] (i1) {$\mathbf{q}_1$};
\vertex [right=of b] (o1) {};
\vertex [below left=1 and 1 of b] (i2) {${\mathbf{q}_2}$};

\draw[thick, postaction={decorate}, decoration={markings, mark=at position 0.5 with {\arrow{>}}}] 
    (b) -- node [above, yshift=2pt] {$\mathbf{q}_{12}$} (o1);
\draw[thick, postaction={decorate}, decoration={markings, mark=at position 0.5 with {\arrow{>}}}] 
    (i1) -- (b);
\draw[thick, postaction={decorate}, decoration={markings, mark=at position 0.5 with {\arrow{>}}}] 
    (i2) -- (b);
\end{feynman}
\end{tikzpicture}\right\}
\end{equation}

\begin{equation}\label{eq:K3_LO_2mom}
\lim_{\mathbf{q}_3\rightarrow 0,\mathbf{q}_2\rightarrow 0}
\begin{tikzpicture}[scale=0.5, 
            baseline={([xshift=-5cm,yshift=-\the\dimexpr\fontdimen22\textfont2\relax]
                current bounding box.center)},
    ] 
\begin{feynman}
\vertex (b) at (0,0) ;
\vertex [above left=of b] (i1) {$\mathbf{q}_1$};
\vertex (o1) at (2,0);
\vertex [below left=1 and 1 of b] (i2) {${\mathbf{q}_2}$};
\vertex [above right=of o1] (v1) {$\mathbf{q}_{123}$};
\vertex [below right=1 and 1 of o1] (v2) {${\mathbf{q}_3}$};

\draw[thick, postaction={decorate}, decoration={markings, mark=at position 0.5 with {\arrow{>}}}] 
    (b) -- node [above, yshift=2pt] {$\mathbf{q}_{12}$} (o1);
\draw[thick, postaction={decorate}, decoration={markings, mark=at position 0.5 with {\arrow{>}}}] 
    (i1) -- (b);
\draw[thick, postaction={decorate}, decoration={markings, mark=at position 0.5 with {\arrow{>}}}] 
    (i2) -- (b);
\draw[thick, postaction={decorate}, decoration={markings, mark=at position 0.5 with {\arrow{>}}}] 
    (o1) -- (v1);
\draw[thick, postaction={decorate}, decoration={markings, mark=at position 0.5 with {\arrow{<}}}] 
    (o1) -- (v2);
\end{feynman}
\end{tikzpicture}
=\frac{\mathbf{q}_3\cdot\mathbf{q}_1}{\mathbf{q}_3^2}\frac{\mathbf{q}_2\cdot\mathbf{q}_1}{\mathbf{q}_2^2}
\begin{tikzpicture}[scale=1.0, 
            baseline={([xshift=-5cm,yshift=-\the\dimexpr\fontdimen22\textfont2\relax]
                current bounding box.center)},
    ] 
\begin{feynman}
\vertex (b) at (0,0) {$\mathbf{q}_1$} ;
\vertex (i1) at (0,2) {$\mathbf{q}_1$};

\draw[thick, postaction={decorate}, decoration={markings, mark=at position 0.5 with {\arrow{<}}}] 
    (i1) -- (b);
\end{feynman}
\end{tikzpicture}+\mathcal{O}(q^{-1})\;,
\end{equation}
Apart from the LO constraints given in Eq.\eqref{eq:lo_3rd}, EGI also imposes NLO constraints coming from Eq.\eqref{eq:NLO_all_order}. The NLO constraint at third order is given as,
\begin{align}\label{eq:NLO_3rd}
\lim_{\mathbf{q}_{12}\rightarrow 0}  K_3(\mathbf{q}_1,\mathbf{q}_2,\mathbf{q}_3)\supset \frac{\mathbf{q}_{12}\cdot\mathbf{q}_3}{\mathbf{q}_{12}^2}K_1(\mathbf{q}_3)\int^\eta d\eta'f_{+}(\eta^{\prime})\left(\frac{D_{+}(\eta^{\prime})}{D_{+}\eta}\right)^{2}G_2(\mathbf{q}_1,\mathbf{q}_2,\eta')\;.
\end{align}
The L.H.S of Eq.\eqref{eq:NLO_3rd} has the following diagrammatic representation,
\begin{equation}\label{eq:K3_NLO_diag}
\lim_{\mathbf{q}_{12}\rightarrow 0}
\begin{tikzpicture}[scale=0.5, 
            baseline={([xshift=-5cm,yshift=-\the\dimexpr\fontdimen22\textfont2\relax]
                current bounding box.center)},
    ] 
\begin{feynman}
\vertex (b) at (0,0) ;
\vertex [above left=of b] (i1) {$\mathbf{q}_1$};
\vertex (o1) at (2,0);
\vertex [below left=1 and 1 of b] (i2) {${\mathbf{q}_2}$};
\vertex [above right=of o1] (v1) {$\mathbf{q}_{123}$};
\vertex [below right=1 and 1 of o1] (v2) {${\mathbf{q}_3}$};

\draw[thick, postaction={decorate}, decoration={markings, mark=at position 0.5 with {\arrow{>}}}] 
    (b) -- node [above, yshift=2pt] {$\mathbf{q}_{12}$} (o1);
\draw[thick, postaction={decorate}, decoration={markings, mark=at position 0.5 with {\arrow{>}}}] 
    (i1) -- (b);
\draw[thick, postaction={decorate}, decoration={markings, mark=at position 0.5 with {\arrow{>}}}] 
    (i2) -- (b);
\draw[thick, postaction={decorate}, decoration={markings, mark=at position 0.5 with {\arrow{>}}}] 
    (o1) -- (v1);
\draw[thick, postaction={decorate}, decoration={markings, mark=at position 0.5 with {\arrow{<}}}] 
    (o1) -- (v2);
\end{feynman}
\end{tikzpicture}
\supset\frac{\mathbf{q}_3\cdot(\mathbf{q}_{12})}{\mathbf{q}_{12}^2}
\left\{\begin{tikzpicture}[scale=1.0, 
            baseline={([xshift=-5cm,yshift=-\the\dimexpr\fontdimen22\textfont2\relax]
                current bounding box.center)},
    ] 
\begin{feynman}
\vertex (b) ;
\vertex [above left=of b] (i1) {$\mathbf{q}_1$};
\vertex [right=of b] (o1) {};
\vertex [below left=1 and 1 of b] (i2) {${\mathbf{q}_2}$};

\draw[thick, postaction={decorate}, decoration={markings, mark=at position 0.5 with {\arrow{>}}}] 
    (b) -- node [above, yshift=2pt] {$\mathbf{q}_{12}$} (o1);
\draw[thick, postaction={decorate}, decoration={markings, mark=at position 0.5 with {\arrow{>}}}] 
    (i1) -- (b);
\draw[thick, postaction={decorate}, decoration={markings, mark=at position 0.5 with {\arrow{>}}}] 
    (i2) -- (b);
\end{feynman}
\end{tikzpicture}\right\}
\begin{tikzpicture}[scale=1.0, 
            baseline={([xshift=-5cm,yshift=-\the\dimexpr\fontdimen22\textfont2\relax]
                current bounding box.center)},
    ] 
\begin{feynman}
\vertex (b) {$\mathbf{q}_3$} ;
\vertex [above=of b] (i1) {$\mathbf{q}_3$};

\draw[thick, postaction={decorate}, decoration={markings, mark=at position 0.5 with {\arrow{<}}}] 
    (i1) -- (b);
\end{feynman}
\end{tikzpicture}\;,
\end{equation}
which has to be compared with the R.H.S of Eq.\eqref{eq:NLO_3rd} to get constraints on $K_3$. Note that only one diagram from the three given in Eq.\eqref{eq:K3_diag} participates in Eq.\eqref{eq:K3_NLO_diag}. This is the case since only the diagram with $\mathbf{q}_{12}$ as an internal is considered under this limit. The other two diagrams participate on taking $\mathbf{q}_{13}$, $\mathbf{q}_{23}$ to be soft. But the symmetry of $K_3(\mathbf{q}_1,\mathbf{q}_2,\mathbf{q}_3)$ under exchange of momenta ensures that one gets the same constraints irrespective of the choice of the momenta taken to be soft.

We see from Eq.\eqref{eq:K3_NLO_diag}, that taking $\mathbf{q}_{12}\rightarrow 0$, which is an internal line in the diagram, breaks the diagram into two sub-diagrams. This is similar to taking the collinear limit of scattering amplitudes in gauge theories, where similar decomposition occurs when internal particles go on-shell.

Note that in the limit $\mathbf{q}_{12}$, terms proportional to,
\begin{align}\label{eq:wrong_mom}  \frac{\mathbf{q}_{12}\cdot\mathbf{q}_3}{\mathbf{q}_{3}^2}\hspace{0.2cm},\hspace{0.2cm}\frac{(\mathbf{q}_{12}\cdot\mathbf{q}_3)^2}{\mathbf{q}_{12}^2\mathbf{q}_{3}^2}\;,
\end{align}
are not allowed and constitute the ``spurious" kind of momentum structures that should not be present. Imposing that the spurious structures are absent from the kernel gives further constraints\,\cite{DAmico:2021rdb}.

\subsection{LO and NLO}

In this section we give a qualitative idea of the procedure we use to obtain constraints. We refer the reader to the Mathematica file for exact calculations and momentum structures. At third order we have three types of constraint relations. Two arising from leading order (LO) given in Eq.\eqref{eq:lo_3rd} and one from next-to-leading order (NLO) given in Eq.\eqref{eq:NLO_3rd}. Taking $\mathbf{q}_1$ soft on $K_3$, we expect \eqref{eq:lo_3rd} to be satisfied. The LHS in this limit gives two types of pole structures,
\begin{align}
    \frac{\mathbf{q}_1 \cdot \mathbf{q}_{2}}{\mathbf{q}_1^2} \;\textbf{M}^{\text{LO}}_1\left(\mathbf{q}_2,\mathbf{q}_3\right)\;, \frac{\mathbf{q}_1 \cdot \mathbf{q}_{3}}{\mathbf{q}_1^2} \;\textbf{M}^{\text{LO}}_2\left(\mathbf{q}_2,\mathbf{q}_3\right),
\end{align}
Symmetrizing $\text{M}_1^{\text{LO}}$ and $\text{M}_2^{\text{LO}}$ gives one constraint which also makes them equal and of the form,
\begin{equation}\label{eq:3rd_lo_sym}
    \begin{aligned}
        \textbf{M}^{\text{LO}} (\mathbf{q}_2,\mathbf{q}_3) = \frac{1}{2}\left(c_2 + c_4 - c_7\right) + c_5 \; \alpha_{+}(\mathbf{q}_2,\mathbf{q}_3) \; +
        \frac{1}{2} \left( c_{11} + c_6 -c_9 \right) \beta(\mathbf{q}_2,\mathbf{q}_3)\,.
    \end{aligned}
\end{equation}
Comparing \eqref{eq:3rd_lo_sym} with the RHS of \eqref{eq:lo_3rd} gives three more constraints. Next, imposing and solving two momenta soft gives one more constraint, which we found to be redundant to one momenta soft.

\paragraph{}
Next we impose NLO constraints as given in \eqref{eq:NLO_3rd}. The LHS in the $\mathbf{q}_{12} \rightarrow 0$ limit gives three types of structures,
\begin{align}
    \frac{\mathbf{q}_{12} \cdot \mathbf{q}_{3}}{\mathbf{q}_{12}^2} \;\textbf{N}^{\text{NLO}}_1\left(\mathbf{q}_1,\mathbf{q}_2\right)\;,\; \frac{\mathbf{q}_{12} \cdot \mathbf{q}_{3}}{\mathbf{q}_3^2} \;\textbf{N}^{\text{NLO}}_2\left(\mathbf{q}_1,\mathbf{q}_2\right)\;,\; \frac{\left(\mathbf{q}_{12} \cdot \mathbf{q}_{3}\right)^2}{\mathbf{q}_{12}^2 \mathbf{q}_3^2} \;\textbf{N}^{\text{NLO}}_3\left(\mathbf{q}_1,\mathbf{q}_2\right)\;,
\end{align}
Comparing $\textbf{N}^{\text{NLO}}_1$ with the R.H.S of Eq.\eqref{eq:NLO_3rd} gives three constraints,
\begin{align}\label{eq:K3_nlo1_const}
    c_{7} = -c_{4} + 2 a_{1} h\;,\quad c_{8} = 2 a_{1} - c_{5}\;,\quad c_{9} = 2 a_{1} - c_{6} - 2 a_{1} h
\end{align}
where,
\begin{align}\label{eq:def_h}
    & h (\eta) = \int^{\eta} d \eta^{\prime} f_{+}(\eta^{\prime})\left[\frac{D_{+}(\eta^{\prime})}{D_{+}(\eta)}\right]^{2} b_{1}(\eta^{\prime})\;,
\end{align}
and two more constraints from putting $\textbf{N}^{\text{NLO}}_2$ and $\textbf{N}^{\text{NLO}}_3$ to zero which are the spurious poles at third order. \vspace{5pt} \\
\underline{\textbf{Tracer kernel}} \vspace{5pt}\\
Solving LO and NLO together gives us eight independent constraints on the kernel $K_3$, leaving a set of 7 independent coefficients in $K_3$, of which 4 are from $3^{\text{rd}}$ order, 2 from $2^{\text{nd}}$ order and 1 from the $1^{\text{st}}$ order. \vspace{5pt} \\
\underline{\textbf{Matter kernel}}
\vspace{5pt}\\
Further, imposing mass and momentum conservation \eqref{eq:mass_mom_conserv} with the limit $\mathbf{Q}_{3,0}\rightarrow 0$ gives three constraints, leaving 3 independent coefficients for the matter kernel, where 2 are from $3^{\text{rd}}$ order and 1 from $2^{\text{nd}}$ order.\footnote{$a_1 = 1$ for matter.}

\section{Details of fourth order calculations in Eulerian space}\label{app:fourth_order}
The following sections qualitatively list the calculations to obtain the $4^\text{th}$ order tracer and matter kernel. We provided the diagrammatic motivation for these limits in section \eqref{sec:ker_4_boot}.

\subsection{Leading Order Constraints}\label{ssec:LO _4}

We now impose 4th order leading order constraints given in Eq.\eqref{eq:lo_4th_1} and \eqref{eq:lo_4th_2}. Taking $\mathbf{q}_1$ soft we expect the following consistency relation to be satisfied,
\begin{align}\label{eq:lo_4_1}
    \lim_{\mathbf{q}_1\rightarrow 0}  K_4(\mathbf{q}_1,\mathbf{q}_2,\mathbf{q}_3,\mathbf{q}_4)=\frac{\mathbf{q}_1\cdot(\mathbf{q}_2+\mathbf{q}_3+\mathbf{q}_4)}{\mathbf{q}_1^2}K_3(\mathbf{q}_2,\mathbf{q}_3,\mathbf{q}_4)+\mathcal{O}(q^0)\;.
\end{align}
The LHS in this limit gives three kinds of pole structures in $\mathbf{q}_1$,
\begin{align}\label{eq:LO4_F1}
    \frac{\mathbf{q}_1 \cdot \mathbf{q}_{2}}{\mathbf{q}_1^2} \;\textbf{F}^{\text{LO}}_1\left(\mathbf{q}_2,\mathbf{q}_3,\mathbf{q}_4\right)\;, \quad \frac{\mathbf{q}_1 \cdot \mathbf{q}_{3}}{\mathbf{q}_1^2} \;\textbf{F}^{\text{LO}}_2\left(\mathbf{q}_2,\mathbf{q}_3,\mathbf{q}_4\right)\;, \quad \frac{\mathbf{q}_1 \cdot \mathbf{q}_{4}}{\mathbf{q}_1^2} \;\textbf{F}^{\text{LO}}_3\left(\mathbf{q}_2,\mathbf{q}_3,\mathbf{q}_4\right)\;.
\end{align}
Since the RHS of \eqref{eq:lo_4_1} requires a structure symmetric in $\{\mathbf{q}_2,\mathbf{q}_3,\mathbf{q}_4\}$, we impose the same on $\textbf{F}^{\text{LO}}_1, \textbf{F}^{\text{LO}}_2, \textbf{F}^{\text{LO}}_3$ individually and obtain 14 symmetrization constraints. Imposing the symmetrization constraints on the L.H.S of Eq.\eqref{eq:lo_4_1}, we get,
\begin{align}\label{eq:lo_4th_sym_lhs}
     \frac{\mathbf{q}_1 \cdot \left(\mathbf{q}_{2}+\mathbf{q}_{3}+\mathbf{q}_{4}\right)}{\mathbf{q}_1^2} \;\textbf{F}^{\text{LO}}_{\text{sym}}\left(\mathbf{q}_2,\mathbf{q}_3,\mathbf{q}_4\right)\;.
\end{align}
Comparing Eq.\eqref{eq:lo_4th_sym_lhs}
with the RHS in Eq.\eqref{eq:lo_4_1} gives us 10 more constraints.

\paragraph{}
Next we take two momenta soft and impose the condition given in eq.\eqref{eq:lo_4th_2}. Here we rewrite the constraint relation for convenience as,
\begin{align}\label{eq:lo_4_2}
    \lim_{\mathbf{q}_1,\mathbf{q}_2\rightarrow 0}K_4(\mathbf{q}_1,\mathbf{q}_2,\mathbf{q}_3,\mathbf{q}_4)=\frac{\mathbf{q}_1\cdot(\mathbf{q}_3+\mathbf{q}_4)}{\mathbf{q}_1^2}\frac{\mathbf{q}_2\cdot(\mathbf{q}_3+\mathbf{q}_4)}{\mathbf{q}_2^2}K_2(\mathbf{q}_3,\mathbf{q}_4)+\mathcal{O}(q^{-1})\;.
\end{align}
The L.H.S of Eq.\eqref{eq:lo_4_2} gives the following four different pole structures,
\begin{align}\label{eq:glo_4_structure}
    &\frac{\mathbf{q}_1 \cdot \mathbf{q}_{3}}{\mathbf{q}_1^2}\;\frac{\mathbf{q}_2 \cdot \mathbf{q}_{3}}{\mathbf{q}_2^2} \;\textbf{G}^{\text{LO}}_1\left(\mathbf{q}_3,\mathbf{q}_4\right)\;, \;\; \frac{\mathbf{q}_1 \cdot \mathbf{q}_{3}}{\mathbf{q}_1^2}\;\frac{\mathbf{q}_2 \cdot \mathbf{q}_{4}}{\mathbf{q}_2^2} \;\textbf{G}^{\text{LO}}_2\left(\mathbf{q}_3,\mathbf{q}_4\right)\;,\nonumber \\[4pt]
    &\frac{\mathbf{q}_1 \cdot \mathbf{q}_{4}}{\mathbf{q}_1^2}\;\frac{\mathbf{q}_2 \cdot \mathbf{q}_{3}}{\mathbf{q}_2^2} \;\textbf{G}^{\text{LO}}_3\left(\mathbf{q}_3,\mathbf{q}_4\right)\;, \quad \frac{\mathbf{q}_1 \cdot \mathbf{q}_{4}}{\mathbf{q}_1^2}\;\frac{\mathbf{q}_2 \cdot \mathbf{q}_{4}}{\mathbf{q}_2^2} \;\textbf{G}^{\text{LO}}_4\left(\mathbf{q}_3,\mathbf{q}_4\right)\;,
\end{align}
where each $\textbf{G}^{\text{LO}}_i$'s are function of $\{\mathbf{q}_3,\mathbf{q}_4\}$ satisfying the following relations,
\begin{align}\label{eq:Glo_rel}
\textbf{G}^{\text{LO}}_1(\mathbf{q}_3,\mathbf{q}_4)&=\textbf{G}^{\text{LO}}_4(\mathbf{q}_4,\mathbf{q}_3)\nonumber\\
\textbf{G}^{\text{LO}}_2(\mathbf{q}_3,\mathbf{q}_4)&=\textbf{G}^{\text{LO}}_3(\mathbf{q}_4,\mathbf{q}_3)\;.
\end{align}
This is expected, from the structure of Eq.\eqref{eq:glo_4_structure}. Symmetrizing each of the $\textbf{G}^{\text{LO}}_i$'s under $\{\mathbf{q}_3,\mathbf{q}_4\}$ and imposing that all $\textbf{G}^{\text{LO}}_i$'s are equal gives us 4 constraints, which
when imposed on Eq.\eqref{eq:glo_4_structure} gives,
\begin{align}\label{eq:lo_4th_2mom_sym}
     \frac{\mathbf{q}_1 \cdot \left(\mathbf{q}_{3}+\mathbf{q}_{4}\right)}{\mathbf{q}_1^2} \; \frac{\mathbf{q}_2 \cdot \left(\mathbf{q}_{3}+\mathbf{q}_{4}\right)}{\mathbf{q}_2^2} \;\textbf{G}^{\text{LO}}_{\text{sym}}\left(\mathbf{q}_3,\mathbf{q}_4\right)\;,
\end{align}
where,
\begin{equation}
    \begin{aligned}
        \mathbf{G}_{\text{sym}}^{\text{LO}}\left(\mathbf{q}_3,\mathbf{q}_4\right) =\; & \frac{1}{2}\left(d_{14} + d_{16} - d_{19} - d_{26} - d_{28} + d_{31} + \frac{d_{49}}{2}\right) + \\
        & \frac{1}{2} \left( d_{18} - d_{21} + d_{23} - d_{30} + d_{33} - d_{35} + d_{41} - d_{44} \right) \, \beta (\mathbf{q}_3, \mathbf{q}_4) \; + \\
        &\frac{3}{4} \left( 2 d_{17} - d_{20} - d_{29} \right) \, \alpha_{+} (\mathbf{q}_3, \mathbf{q}_4)\;.
    \end{aligned}
\end{equation}
This has $3$ momentum structures and is seen to be comparable to $\mathbf{K}_2$. Comparing Eq.\eqref{eq:lo_4th_2mom_sym} with the RHS of Eq.\eqref{eq:lo_4_2}, gives us three more constraints.

\paragraph{}
Similarly, after taking three momenta soft we have the following relation,
\begin{align}\label{eq:lo_4_3}
\lim_{\mathbf{q}_1,\mathbf{q}_2,\mathbf{q}_3\rightarrow 0}K_4(\mathbf{q}_1,\mathbf{q}_2,\mathbf{q}_3,\mathbf{q}_4)=\frac{\mathbf{q}_1\cdot\mathbf{q}_4}{\mathbf{q}_1^2}\frac{\mathbf{q}_2\cdot\mathbf{q}_4}{\mathbf{q}_2^2}\frac{\mathbf{q}_3\cdot\mathbf{q}_4}{\mathbf{q}_3^2}K_1(\mathbf{q}_4)+\mathcal{O}(q^{-2})\;.
\end{align}
\paragraph{}
The L.H.S of Eq.\eqref{eq:lo_4_3} take the following form,
\begin{align}
    \frac{\mathbf{q}_1\cdot \mathbf{q}_4}{\mathbf{q}_1^2}\;\frac{\mathbf{q}_2\cdot\mathbf{q}_4}{\mathbf{q}_2^2}\;\frac{\mathbf{q}_3\cdot \mathbf{q}_4}{\mathbf{q}_3^2}\;\left( \frac{3 d_{17}}{4} - \frac{3 d_{20}}{4} - \frac{3 d_{29}}{4} + \frac{3 d_{32}}{4} \right)+\mathcal{O}(q^{-2})\;,
\end{align}
\paragraph{}
which upon comparing with the RHS of Eq.\eqref{eq:lo_4_3} gives
\begin{align}\label{eq:const_lo43_1}
    d_{29} = - \frac{4 a_1}{3} + d_{17} - d_{20} + d_{32}\;.
\end{align}

Solving one momenta, two momenta and three momenta soft together constitute the "leading order" constraints. We have observed that the two momenta soft and three momenta soft constraints
are redundant with those from one momenta soft, once we impose the constraints coming from second and third order. In the next section, we will impose the NLO constraints on $K_4$ and finally obtain the independent set of operators for tracers at fourth order.

\subsection{NLO and NNLO constraints}\label{ssec:NLO_4}
In Sec. (\ref{ssec:LO _4}), we have studied one kinematic limit of the kernel $K_4$, i.e. when one or a subset of external momenta go soft. In this section, we will study another kinematic limit, which takes an internal momenta to be soft. These limits are otherwise termed as next-to-leading order (NLO) and next-to-next-to leading order (NNLO)\,\cite{DAmico:2021rdb}. We will discuss what EGI says about such soft limits and derive constraints on the coefficients of the fourth order kernel as given in Eq.\eqref{eq:ker4}. 

\paragraph{}
Let us start with the limit $\mathbf{q}_{12}\rightarrow 0$. Only three diagrams from Eq.\eqref{eq:K4_diag_1} and \eqref{eq:K4_diag_2} participate in this limit as they contain $\mathbf{q}_{12}$ as an internal momenta. These are,
\begin{equation}
   \begin{tikzpicture}[scale=0.58, 
            baseline={([xshift=-5cm,yshift=-\the\dimexpr\fontdimen22\textfont2\relax]
                current bounding box.center)},
    ] 
\begin{feynman}
\vertex (v1) at (0,0) ;
\vertex  (e1) at (-1.5,1.5){$\mathbf{q}_{1}$};
\vertex  (e2) at (-1.5,-1.5){$\mathbf{q}_{2}$};
\vertex  (v2) at (1.5,0);
\vertex  (e3) at (1.5,-1.5){$\mathbf{q}_{3}$};
\vertex  (v3) at (3,0);
\vertex  (e4) at (4.5,1.5){$\mathbf{q}_{1234}$};
\vertex  (e5) at (4.5,-1.5){$\mathbf{q}_{4}$};

\draw[thick, postaction={decorate}, decoration={markings, mark=at position 0.5 with {\arrow{>}}}] 
    (e1) -- (v1);
\draw[thick, postaction={decorate}, decoration={markings, mark=at position 0.5 with {\arrow{>}}}] 
    (e2) -- (v1);
\draw[thick, postaction={decorate}, decoration={markings, mark=at position 0.5 with {\arrow{>}}}] 
    (v1) -- node [above=4pt] {$\mathbf{q}_{12}$} (v2);
\draw[thick, postaction={decorate}, decoration={markings, mark=at position 0.5 with {\arrow{<}}}] 
    (v2) -- (e3);
\draw[thick, postaction={decorate}, decoration={markings, mark=at position 0.5 with {\arrow{>}}}] 
    (v2) -- node [above=4pt] {$\mathbf{q}_{123}$} (v3);
\draw[thick, postaction={decorate}, decoration={markings, mark=at position 0.5 with {\arrow{>}}}] 
    (v3) -- (e4);
\draw[thick, postaction={decorate}, decoration={markings, mark=at position 0.5 with {\arrow{<}}}] 
    (v3) -- (e5);
\end{feynman}

\end{tikzpicture}\hspace{1cm},\hspace{1cm}
\begin{tikzpicture}[scale=0.58, 
            baseline={([xshift=-5cm,yshift=-\the\dimexpr\fontdimen22\textfont2\relax]
                current bounding box.center)},
    ] 
\begin{feynman}
\vertex (v1) at (0,0) ;
\vertex  (e1) at (-1.5,1.5){$\mathbf{q}_{1}$};
\vertex  (e2) at (-1.5,-1.5){$\mathbf{q}_{2}$};
\vertex  (v2) at (1.5,0);
\vertex  (e3) at (1.5,-1.5){$\mathbf{q}_{4}$};
\vertex  (v3) at (3,0);
\vertex  (e4) at (4.5,1.5){$\mathbf{q}_{1234}$};
\vertex  (e5) at (4.5,-1.5){$\mathbf{q}_{3}$};

\draw[thick, postaction={decorate}, decoration={markings, mark=at position 0.5 with {\arrow{>}}}] 
    (e1) -- (v1);
\draw[thick, postaction={decorate}, decoration={markings, mark=at position 0.5 with {\arrow{>}}}] 
    (e2) -- (v1);
\draw[thick, postaction={decorate}, decoration={markings, mark=at position 0.5 with {\arrow{>}}}] 
    (v1) -- node[above, yshift=2pt] {$\mathbf{q}_{12}$} (v2);
\draw[thick, postaction={decorate}, decoration={markings, mark=at position 0.5 with {\arrow{<}}}] 
    (v2) -- (e3);
\draw[thick, postaction={decorate}, decoration={markings, mark=at position 0.5 with {\arrow{>}}}] 
    (v2) -- node[above, yshift=2pt] {$\mathbf{q}_{124}$} (v3);
\draw[thick, postaction={decorate}, decoration={markings, mark=at position 0.5 with {\arrow{>}}}] 
    (v3) -- (e4);
\draw[thick, postaction={decorate}, decoration={markings, mark=at position 0.5 with {\arrow{<}}}] 
    (v3) -- (e5);
\end{feynman}

\end{tikzpicture}\nonumber
\end{equation}

\begin{equation}\label{eq:K4_nlo_q12_diag}
    \begin{tikzpicture}[scale=0.7, 
            baseline={([xshift=-5cm,yshift=-\the\dimexpr\fontdimen22\textfont2\relax]
                current bounding box.center)},
    ] 
\begin{feynman}
\vertex (v1) at (0,0) ;
\vertex  (e1) at (-1.5,1.5){$\mathbf{q}_{1}$};
\vertex  (e2) at (-1.5,-1.5){$\mathbf{q}_{2}$};
\vertex  (v2) at (1.5,0);
\vertex  (e3) at (1.5,-1.5){$\mathbf{q}_{1234}$};
\vertex  (v3) at (3,0);
\vertex  (e4) at (4.5,1.5){$\mathbf{q}_{4}$};
\vertex  (e5) at (4.5,-1.5){$\mathbf{q}_{3}$};

\draw[thick, postaction={decorate}, decoration={markings, mark=at position 0.5 with {\arrow{>}}}] 
    (e1) -- (v1);
\draw[thick, postaction={decorate}, decoration={markings, mark=at position 0.5 with {\arrow{>}}}] 
    (e2) -- (v1);
\draw[thick, postaction={decorate}, decoration={markings, mark=at position 0.5 with {\arrow{>}}}] 
    (v1) -- node [above=4pt] {$\mathbf{q}_{12}$} (v2);
\draw[thick, postaction={decorate}, decoration={markings, mark=at position 0.5 with {\arrow{>}}}] 
    (v2) -- (e3);
\draw[thick, postaction={decorate}, decoration={markings, mark=at position 0.5 with {\arrow{<}}}] 
    (v2) -- node [above=4pt] {$\mathbf{q}_{34}$} (v3);
\draw[thick, postaction={decorate}, decoration={markings, mark=at position 0.5 with {\arrow{<}}}] 
    (v3) -- (e4);
\draw[thick, postaction={decorate}, decoration={markings, mark=at position 0.5 with {\arrow{<}}}] 
    (v3) -- (e5);
\end{feynman}

\end{tikzpicture}
\;.
\end{equation}
EGI states that under the limit $\mathbf{q}_{12}\rightarrow 0$, the kernel should behave as (Eq.\eqref{eq:NLO_4th}),
\begin{align}\label{eq:nlo4_1}
    \lim_{\mathbf{q}_{12}\rightarrow 0}  K_4(\mathbf{q}_1,\mathbf{q}_2,\mathbf{q}_3,\mathbf{q}_4)\supset \frac{\mathbf{q}_{12}\cdot(\mathbf{q}_3+\mathbf{q}_4)}{\mathbf{q}_{12}^2}K_2(\mathbf{q}_3,\mathbf{q}_4)\int^\eta d\eta'f_{+}(\eta^{\prime})\left(\frac{D_{+}(\eta^{\prime})}{D_{+}(\eta)}\right)^{2}G_2(\mathbf{q}_1,\mathbf{q}_2,\eta')\;.
\end{align}
\paragraph{}
The terms in the L.H.S of Eq.\eqref{eq:nlo4_1} which have $\mathbf{q}_{12}^2$ in the denominator, under the limit $\mathbf{q}_{12}\rightarrow 0$, take the following form,
\begin{align}\label{eq:nlo41}
    \frac{\mathbf{q}_{12} \cdot \mathbf{q}_{3}}{\mathbf{q}_{12}^2} \;\mathbf{F}^{\text{NLO}}_{1}\left(\mathbf{q}_1,\mathbf{q}_2,\mathbf{q}_3,\mathbf{q}_4\right)\quad+ \quad \frac{\mathbf{q}_{12} \cdot \mathbf{q}_{4}}{\mathbf{q}_{12}^2} \;\mathbf{F}^{\text{NLO}}_{2}\left(\mathbf{q}_1,\mathbf{q}_2,\mathbf{q}_3,\mathbf{q}_4\right)\;,
\end{align}
where $\mathbf{F}^{\text{NLO}}_{i}$ are functions of $\{\mathbf{q}_1,\mathbf{q}_2,\mathbf{q}_3,\mathbf{q}_4\}$ which are symmetric under $\mathbf{q}_1\leftrightarrow \mathbf{q}_2$ and satisfy,
\begin{align}
\mathbf{F}^{\text{NLO}}_{1}\left(\mathbf{q}_1,\mathbf{q}_2,\mathbf{q}_3,\mathbf{q}_4\right)=\mathbf{F}^{\text{NLO}}_{2}\left(\mathbf{q}_1,\mathbf{q}_2,\mathbf{q}_4,\mathbf{q}_3\right)\;.
\end{align}
Symmetrizing $\mathbf{F}^{\text{NLO}}_{i}$ under $\{\mathbf{q}_3,\mathbf{q}_4\}$ gives us three constraints. Imposing them on the L.H.S of Eq.\eqref{eq:nlo4_1} we get,
\begin{align}\label{eq:K4_nlo_sym_final}
    \frac{\mathbf{q}_{12} \cdot \left(\mathbf{q}_{3}+\mathbf{q}_{4}\right)}{\mathbf{q}_{12}^2} \;\mathbf{F}^{\text{NLO}}_{\text{sym}}\left(\mathbf{q}_1,\mathbf{q}_2,\mathbf{q}_3,\mathbf{q}_4\right)\;,
\end{align}
where,
\begin{equation}
    \begin{aligned}
        \mathbf{F}_{\text{sym}}^{\text{NLO}}&\left(\mathbf{q}_1,\mathbf{q}_2,\mathbf{q}_3,\mathbf{q}_4\right) =  \frac{1}{2}(d_4 + d_7) + \frac{1}{2}\alpha_{+}(\mathbf{q}_3, \mathbf{q}_4) \left( d_{16} + d_{19} + \left( d_{18} + d_{21} \right) \beta(\mathbf{q}_1, \mathbf{q}_2) \right) + \\
        & \frac{1}{2} \beta(\mathbf{q}_3, \mathbf{q}_4) \left( d_{40} + d_{43} \right) + \frac{1}{2} \beta(\mathbf{q}_1, \mathbf{q}_2) \left( d_6 + d_9 + \left( d_{42} + d_{45} \right) \beta(\mathbf{q}_3, \mathbf{q}_4) \right) +\\
        & \frac{1}{2} \alpha_{+}(\mathbf{q}_1, \mathbf{q}_2) \left( d_5 + d_8 + \left( d_{17} + d_{20} \right) \alpha_{+}(\mathbf{q}_3, \mathbf{q}_4) + \left( d_{41} + d_{44} \right) \beta(\mathbf{q}_3, \mathbf{q}_4) \right).
    \end{aligned}
\end{equation}
This expression is seen to have $9$ momentum structures, analogous to the most general $\mathbf{K}_2 \cross \mathbf{K}_2$ construction. This is anticipated, as it is compared to $\mathbf{K}_2(\mathbf{q}_3,\mathbf{q}_4) \cross \mathbf{G}_2(\mathbf{q}_1,\mathbf{q}_2)$ in the RHS of \eqref{eq:nlo4_1}.
Doing so gives us 9 more constraints. 
\paragraph{}
Under the $\mathbf{q}_{12}\rightarrow 0$ limit, the L.H.S of Eq.\eqref{eq:nlo4_1} also contains terms which look like,
\begin{align}\label{eq:K4_nlo_wrong}
    \frac{\mathbf{q}_{12} \cdot \mathbf{q}_{34}}{\mathbf{q}_{34}^2} \;\mathbf{F}^{\text{NLO}}_{3}\left(\mathbf{q}_1,\mathbf{q}_2,\mathbf{q}_3,\mathbf{q}_4\right)\;, \quad \frac{\left(\mathbf{q}_{12} \cdot \mathbf{q}_{34}\right)^2}{\mathbf{q}_{12}^2\;\mathbf{q}_{34}^2} \;\mathbf{F}^{\text{NLO}}_{4}\left(\mathbf{q}_1,\mathbf{q}_2,\mathbf{q}_3,\mathbf{q}_4\right)\;.
\end{align}
These are the "spurious poles" each of which is separately put to zero as dictated by the RHS, giving us 6 more constraints.
These constitute all the constraints coming from taking $\mathbf{q}_{12}\rightarrow 0$. 

\paragraph{}
Now we move on to study another kinematic limit i.e. $\mathbf{q}_{123}\rightarrow 0$ on $K_4$. For consistency with EGI we need $K_4$ to behave as (Eq.\eqref{eq:NLO_4th}),
\begin{align}\label{eq:nlo4_2}
    \lim_{\mathbf{q}_{123}\rightarrow 0}  K_4(\mathbf{q}_1,\mathbf{q}_2,\mathbf{q}_3,\mathbf{q}_4)\supset \frac{\mathbf{q}_{123}\cdot\mathbf{q}_4}{\mathbf{q}_{123}^2}K_1(\mathbf{q}_4)\int^\eta d\eta'd\eta'f_{+}(\eta^{\prime})\left(\frac{D_{+}(\eta^{\prime})}{D_{+}(\eta)}\right)^{3}G_3(\mathbf{q}_1,\mathbf{q}_2,\mathbf{q}_3,\eta')\;,
\end{align}
Taking the soft limit $\mathbf{q}_{123}\rightarrow 0$ on $K_4$ and looking at terms which have $\mathbf{q}_{123}^2$ in the denominator we get,
\begin{align}\label{eq:K4_nnlo_lim}
    \frac{\mathbf{q}_{123} \cdot \mathbf{q}_4}{\mathbf{q}_{123}^2} \textbf{G}^{\text{NNLO}}\left(\mathbf{q}_1,\mathbf{q}_2,\mathbf{q}_3,\mathbf{q}_4\right)\;,
\end{align}
where $G^{\text{NNLO}}$ has 12 momentum structures, comparable to the most generally defined $\mathbf{K}_3$. Comparing Eq.\eqref{eq:K4_nnlo_lim} with the RHS of Eq.\eqref{eq:nlo4_2} gives 11 constraints, which include quantities like $s_1,s_2,s_3,s_4$. These are defined as,
\begin{align}\label{eq:def_s1_s2_etc}
    & s_1 = \int^{\eta} d \eta^{\prime} f_{+}(\eta^{\prime})\left[\frac{D_{+}(\eta^{\prime})}{D_{+}(\eta)}\right]^{3} h(\eta^{\prime})\;,\;\;\;s_{2} = \int^{\eta} d \eta^{\prime} f_{+}(\eta^{\prime})\left[\frac{D_{+}(\eta^{\prime})}{D_{+}(\eta)}\right]^{3} b_{1}(\eta^{\prime})\nonumber\\[4pt]
    & s_{3} = \int^{\eta} d \eta^{\prime} f_{+}(\eta^{\prime})\left[\frac{D_{+}(\eta^{\prime})}{D_{+}(\eta)}\right]^{3} c_{6}(\eta^{\prime})\;,\;\;\;s_{4} = \int^{\eta} d \eta^{\prime} f_{+}(\eta^{\prime})\left[\frac{D_{+}(\eta^{\prime})}{D_{+}(\eta)}\right]^{3} c_{12}(\eta^{\prime})\;,
\end{align}
which come from the time integral in Eq.\eqref{eq:nlo4_2}. Under the $\mathbf{q}_{123}\rightarrow 0$ limit, the L.H.S of Eq.\eqref{eq:nlo4_2} also contains "spurious" terms which look like,
\begin{align}\label{eq:K4_nnlo_wrong}
    \frac{\mathbf{q}_{123} \cdot \mathbf{q}_{4}}{\mathbf{q}_{4}^2} \;\textbf{G}^{\text{NNLO}}_1\left(\mathbf{q}_1,\mathbf{q}_2,\mathbf{q}_3,\mathbf{q}_4\right)\;, \quad \frac{\left(\mathbf{q}_{123} \cdot \mathbf{q}_{4}\right)^2}{\mathbf{q}_{123}^2\;\mathbf{q}_{4}^2} \;\textbf{G}^{\text{NNLO}}_2\left(\mathbf{q}_1,\mathbf{q}_2,\mathbf{q}_3,\mathbf{q}_4\right)\;.
\end{align}
Imposing that such structures vanish in the limit $\mathbf{q}_{123}$ gives us 8 more constraints.

\paragraph{}
We have exhausted all the limits where an internal momenta goes soft. However, there exists another kinematic limit as stated in Eq.\eqref{eq:K4_nlo_d2d1del1}. This limit includes taking one internal momenta soft followed by taking one external leg soft. For the kernel to be consistent with EGI, we require that under the limit,
\begin{align}\label{eq:nlo4_d2d1del1} \lim_{\mathbf{q}_{12},\mathbf{q}_3\rightarrow 0}  K_4(\mathbf{q}_1,\mathbf{q}_2,\mathbf{q}_3,\mathbf{q}_4)\supset \frac{\mathbf{q}_{12}\cdot\mathbf{q}_4}{\mathbf{q}_{12}^2}\frac{\mathbf{q}_{3}\cdot\mathbf{q}_4}{\mathbf{q}_{3}^2}K_1(\mathbf{q}_4)\int^\eta d\eta'f_{+}(\eta^{\prime})\left(\frac{D_{+}(\eta^{\prime})}{D_{+}(\eta)}\right)^{2}G_2(\mathbf{q}_1,\mathbf{q}_2,\eta')\;.
\end{align}
The term on the L.H.S of Eq.\eqref{eq:nlo4_d2d1del1} contains a structure similar to the R.H.S. It is of the following form,
\begin{align}\label{k4_d2d1_corrrect}
    \frac{\mathbf{q}_{12}\cdot\mathbf{q}_4}{\mathbf{q}_{12}^2}\frac{\mathbf{q}_{3}\cdot\mathbf{q}_4}{\mathbf{q}_{3}^2} \textbf{H}^{\text{NLO}}\left(\mathbf{q}_1,\mathbf{q}_2,\mathbf{q}_3,\mathbf{q}_4\right)\;,
\end{align}
Comparing \textbf{H}$^{\text{NLO}}$ to the RHS of Eq.\eqref{eq:nlo4_d2d1del1} gives three constraints. Under the same limit, $\mathbf{q}_{12}\rightarrow 0,\mathbf{q}_{3}\rightarrow 0$, we also get structures of the following form,
\begin{equation}\label{eq:nlo4_d2d1del1_wrong}
    \begin{aligned}
    &\frac{\mathbf{q}_{12} \cdot \mathbf{q}_{4}}{\mathbf{q}_{12}^2}\;\frac{\mathbf{q}_{3} \cdot \mathbf{q}_{4}}{\mathbf{q}_{4}^2} \;\textbf{B}_1,\quad \frac{\mathbf{q}_{12} \cdot \mathbf{q}_{4}}{\mathbf{q}_{4}^2}\;\frac{\mathbf{q}_{3} \cdot \mathbf{q}_{4}}{\mathbf{q}_{3}^2} \;\textbf{B}_2,\quad \frac{\mathbf{q}_{12} \cdot \mathbf{q}_{4}}{\mathbf{q}_{4}^2}\;\frac{\mathbf{q}_{3} \cdot \mathbf{q}_{4}}{\mathbf{q}_{4}^2} \;\textbf{B}_3,\\[4pt]
    &\frac{\left(\mathbf{q}_{12} \cdot \mathbf{q}_{4}\right)^2}{\mathbf{q}_{12}^2\;\mathbf{q}_{4}^2}\;\frac{\mathbf{q}_{3} \cdot \mathbf{q}_{4}}{\mathbf{q}_{4}^2} \;\textbf{B}_4,\quad\frac{\left(\mathbf{q}_{12} \cdot \mathbf{q}_{4}\right)^2}{\mathbf{q}_{12}^2\;\mathbf{q}_{4}^2}\;\frac{\mathbf{q}_{3} \cdot \mathbf{q}_{4}}{\mathbf{q}_{3}^2} \;\textbf{B}_5,\quad \frac{\mathbf{q}_{12} \cdot \mathbf{q}_{4}}{\mathbf{q}_{12}^2}\;\frac{\left(\mathbf{q}_{3} \cdot \mathbf{q}_{4}\right)^2}{\mathbf{q}_{3}^2\;\mathbf{q}_{4}^2} \;\textbf{B}_6,\\[4pt]
    &\frac{\mathbf{q}_{12} \cdot \mathbf{q}_{4}}{\mathbf{q}_{4}^2}\;\frac{\left(\mathbf{q}_{3} \cdot \mathbf{q}_{4}\right)^2}{\mathbf{q}_{3}^2\;\mathbf{q}_{4}^2} \;\textbf{B}_7,\quad \frac{\left(\mathbf{q}_{12} \cdot \mathbf{q}_{4}\right)^2}{\mathbf{q}_{12}^2\;\mathbf{q}_{4}^2}\;\frac{\left(\mathbf{q}_{3} \cdot \mathbf{q}_{4}\right)^2}{\mathbf{q}_{3}^2\;\mathbf{q}_{4}^2} \;\textbf{B}_8,\\[4pt]
    &\frac{\mathbf{q}_{12} \cdot \mathbf{q}_{4}}{\mathbf{q}_{12}^2}\;\textbf{B}_9,\;\quad\quad \frac{\mathbf{q}_{12} \cdot \mathbf{q}_{4}}{\mathbf{q}_{3}^2}\;\textbf{B}_{10},\;\quad\quad  \frac{\left(\mathbf{q}_{12} \cdot \mathbf{q}_{4}\right)}{\mathbf{q}_{12}^2 \mathbf{q}_{4}^2}\;\textbf{B}_{11}\;.
    \end{aligned}
\end{equation}
Since none of the momentum structure given in Eq.\eqref{eq:nlo4_d2d1del1_wrong} resemble the R.H.S of Eq.\eqref{eq:nlo4_d2d1del1}, these constitute the "spurious pole" and we put all of them to zero that gives us 16 more constraints.

We now have all the constraints obtained by taking an internal momenta soft. Together they provide us with constraints beyond the leading order. Putting these together we obtain the set of 17 independent coefficients at 4th order.

In the following subsection, we review the constraints from mass and momentum conservation. Imposing these extra constraints on the tracer kernel leads to the kernel for matter.


\subsection{Mass and Momentum Conservation}

We impose mass and momentum conservation given in Eq.\eqref{eq:mass_mom_conserv}, on the $4^\text{th}$ order kernel for tracers. This amounts to imposing,
\begin{align}
\lim_{\mathbf{q}_{1234}\rightarrow 0} K_4(\mathbf{q}_1,...\mathbf{q}_4;\eta)&=0\;,\nonumber\\
\lim_{\mathbf{q}_{1234}\rightarrow 0} \frac{\partial}{\partial\mathbf{q}_1^i}K_4(\mathbf{q}_1,...\mathbf{q}_4;\eta)&=0\;.
\end{align}
on the tracer kernel given in the previous order.
Doing so gives us 8 constraints, which when imposed on the $K_4$ kernel gives the $F_4$ and $G_4$ kernel, with 8 independent coefficients.\footnote{Here we have taken $a_1 = 1$ which is true for matter}.

\section{Fifth order calculations}\label{app:5_order}
In this Apppendix, we outline the calculation of expanding Eq.\eqref{eq:tracer_lag_expan} up to fifth order which gives the Eulerian overdensity for tracers. Then we discuss about the momentum structures that need to be combined in order to be directly compared with the bias expansion. This is analogous to the combination at fourth order as given in Eq.\eqref{eq:lag_sca_4_combined}. Finally, we discuss the discrepancy between the independent coefficients obtained in this work and in\,\cite{Marinucci:2024add} and also providing the resolution for the same.

The tracer
overdensity at the Eulerian position as given in Eq.\eqref{eq:tracer_lag_expan} can be expanded up to fifth order as follows,
\begin{align} \label{eq:tracer_lag_expan_order_5}\delta_{\text{t}}^{\text{E}}=\delta_{\text{t}}^{\text{L}}-\;&\partial_i\delta_{\text{t}}^{\text{L}}\left(\psi^i-\partial_j\psi^i\psi^j+\partial_j\psi^i\partial_k\psi^j\psi^k+\frac{1}{2}\partial_{j}\partial_{k}\psi^i\psi^j\psi^k-\partial_j\psi^i\partial_k\psi^j\partial_l\psi^k\psi^l \right.\nonumber\\
&\left.\quad\quad\quad - \frac{1}{2}\partial_j\psi^i\partial_k\partial_l\psi^j\psi^k\psi^l-\partial_j\partial_k\psi^i\psi^j\partial_l\psi^k\psi^l-\frac{1}{6}\partial_l\partial_j\partial_k\psi^i\psi^j\psi^k\psi^l\right)+\nonumber\\
&\frac{1}{2}\partial_i\partial_j\delta_{\text{t}}^{\text{L}}\left(\psi^i\psi^j-2\psi^k\partial_k\psi^i\psi^j+\psi^i\partial_k\psi^j\partial_l\psi^k\psi^l +\right.\nonumber\\
&\left. \quad \quad \quad \quad \partial_k\psi^i\psi^k\partial_l\psi^j\psi^l+ \partial_k \psi^i\partial_l\psi^k\psi^l\psi^j+\frac{1}{2}\psi^i\partial_k\partial_l\psi^j\psi^k\psi^l\right)-\nonumber\\
&\frac{1}{6}\partial_i\partial_j\partial_k\delta_{\text{t}}^{\text{L}}\left(\psi^i\psi^j\psi^k-3\psi^i\psi^j\partial_l\psi^k\psi^l\right)+\frac{1}{24}\partial_i\partial_j\partial_k\partial_l\delta_{\text{t}}^{\text{L}}\psi^i\psi^j\psi^k\psi^l\;.
\end{align}

The Lagrangian overdensity gives 27 independent coefficients corresponding to fifth order, while the rest of the expansion provides 17 independent coefficients from the previous orders. Therefore in total, we have 44 independent coefficients for the fifth order tracer kernel. We know that the bias expansion at fifth order gives rise to 29 operators. To obtain them, it is possible to combine some of the coefficients in Lagrangian overdensity in the following manner,

\begin{align}\label{lag_scalars_5}
    \mathcal{O}_1=&-\frac{1}{3}\varphi^{(3)}_{1,ij}\varphi^{(1)}_{,jk}\varphi^{(1)}_{,ki}+\frac{10}{21}\varphi^{(3)}_{2,ij}\varphi^{(1)}_{,jk}\varphi^{(1)}_{,ki}-\frac{1}{7}\left(\frac{1}{6}\epsilon^{ikl}\epsilon^{jmn}\epsilon^{lop}\varphi^{(1)}_{,ij}\varphi^{(1)}_{,km}v_{,on}^{(3)p}\right)-\frac{1}{7}\varphi^{(4)}_{5,ii}\varphi^{(1)}_{,ll} \nonumber \\[4pt]
    \mathcal{O}_2=&-\frac{1}{3}\varphi^{(3)}_{1,ij}\varphi^{(1)}_{,ij}\varphi^{(1)}_{,ll}+\frac{10}{21}\varphi^{(3)}_{2,ij}\varphi^{(1)}_{,ij}\varphi^{(1)}_{,ll}-\frac{1}{7}\varphi^{(4)}_{5,ii}\varphi^{(1)}_{,ll} \nonumber \\[4pt]
    \mathcal{O}_3=&-\frac{1}{3}\varphi^{(3)}_{1,ij}\varphi^{(2)}_{,ij}+\frac{10}{21}\varphi^{(3)}_{2,ij}\varphi^{(2)}_{,ij}-\frac{1}{7}\left(\frac{1}{2}\varphi^{(2)}_{,lj}\left(\epsilon^{jmn}v^{(3)n}_{,lm}+\epsilon^{lmn}v^{(3)n}_{,jm}\right)\right)\nonumber \\[4pt]
    \mathcal{O}_4=&\;\frac{39}{77}\varphi^{(4)}_{1,ij}\varphi^{(1)}_{,ij}+\frac{14}{33}\varphi^{(4)}_{2,ij}\varphi^{(1)}_{,ij}-\frac{20}{33}\varphi^{(4)}_{2,ij}\varphi^{(1)}_{,ij}-\frac{51}{539}\varphi^{(4)}_{3,ij}\varphi^{(1)}_{,ij}-\frac{1}{11}\varphi^{(4)}_{4,ij}\varphi^{(1)}_{,ij}-\nonumber\\[4pt]
    &\frac{1}{6}\left(\frac{1}{2}\varphi^{(1)}_{,lj}\left(\epsilon^{jmn}v^{(4)n}_{1,lm}+\epsilon^{lmn}v^{(4)n}_{1,jm}\right)\right)+\frac{5}{21}\left(\frac{1}{2}\varphi^{(1)}_{,lj}\left(\epsilon^{jmn}v^{(4)n}_{2,lm}+\epsilon^{lmn}v^{(4)n}_{2,jm}\right)\right)+\nonumber\\[4pt]
    &\frac{1}{14}\left(\frac{1}{2}\varphi^{(1)}_{,lj}\left(\epsilon^{jmn}v^{(4)n}_{3,lm}+\epsilon^{lmn}v^{(4)n}_{3,jm}\right)\right)\;,
\end{align}
and define them with an independent coefficient each. Following the arguments from \eqref{eq:lag_scalars_4}, the motivation to do this is to combine terms arising from the same order field. This exercise reduces the independent coefficients from $27\rightarrow14$, while in previous orders from $17\rightarrow15$, as shown in Eq.\eqref{eq:lag_sca_4_combined}. This gives us a sum total of 29 independent coefficients. We have shown that the momentum structures multiplying these coefficients have a linear map with the 29 operators from the usual bias expansion.
\vspace{10pt}\\
\underline{Discrepancy in matter kernel}
\vspace{5pt}\\
We noticed a discrepancy at fifth order, for the matter kernel as reported in \cite{Marinucci:2024add} and the one that we get from Eulerian bootstrap. After a careful analysis we notice that the kernel given in\,\cite{Marinucci:2024add} identically satisfies mass conservation but does not satisfy momentum conservation as given in Eq.\eqref{eq:mass_mom_conserv}. As shown in the attached mathematica notebook, we complete this procedure and obtain two more constraint relations of the form,
\begin{align}\label{eq:mm_5_const}
e_{14}+e_{11}-\frac{1}{2}e_{13}=0\;\;,\;\; e_{15}-\frac{3}{2}e_{11}+\frac{3}{2}e_{12}+\frac{3}{4}e_{13}=0\;,
\end{align}
where $e_{a}$ corresponds to the coefficient of the structures $\varphi^{(5)}_{a,ii}$ provided in\,\cite{Marinucci:2024add}. The explicit form for the structures given in Eq.\eqref{eq:mm_5_const} are given in the Mathematica notebook. This reduces the number of independent scalar contributions at fifth order from 15\,\cite{Marinucci:2024add} to 13. With this 13 we add the 8 independent coefficients coming from fourth order which makes the total number 21. This agrees with our result obtained from Eulerian bootstrap (Sec.\,\ref{sec:fifth_order}). This seems to suggest that one needs to take a closer look at the procedure for imposing momentum conservation as prescribed in \cite{DAmico:2021rdb} and \cite{Marinucci:2024add} to make sure they imply the same thing for the matter kernel. We have not indulged in such investigation and leave that for future work. 

\bibliographystyle{JHEP.bst}
\bibliography{references} 

\end{document}